\documentclass[aps,twocolumn,showpacs,preprintnumbers,amsmath,amssymb,prb,
superscriptaddress,floatfix]{revtex4}

\usepackage{graphicx}
\usepackage{color}

\begin{document}

\title{Spin-charge separation in one-dimensional fermion systems\\ beyond the Luttinger liquid theory}

\author{Thomas~L.~Schmidt}\affiliation{Department of Physics, Yale University, 217 Prospect Street, New Haven, Connecticut 06520, USA}

\author{Adilet~Imambekov}\affiliation{Department of Physics and Astronomy, Rice
University, Houston, Texas 77005, USA}

\author{Leonid~I.~Glazman}\affiliation{Department of Physics, Yale University, 217 Prospect Street, New Haven, Connecticut 06520, USA}

\date{\today}

\newcommand{\tpsi}{\tilde{\psi}}
\newcommand{\tV}{\tilde{V}}
\newcommand{\trho}{\tilde{\rho}}
\newcommand{\VJ}{\vec{J}}
\newcommand{\expct}[1]{\left\langle #1 \right\rangle}
\newcommand{\expcts}[1]{\langle #1 \rangle}
\renewcommand{\mod}{\textrm{mod }}
\newcommand{\etal}{\textit{et al.}}

\newcommand{\ket}[1]{\left| #1 \right\rangle}
\newcommand{\kets}[1]{| #1 \rangle}
\newcommand{\bra}[1]{\left\langle #1 \right|}
\newcommand{\bras}[1]{\langle #1 |}
\newcommand{\braket}[2]{{\left\langle #1 | #2 \right\rangle}}
\newcommand{\brakets}[2]{{\langle #1 | #2 \rangle}}

\renewcommand{\Re}{\operatorname{Re}} 
\renewcommand{\Im}{\operatorname{Im}} 

\begin{abstract}
We develop a nonperturbative zero-temperature theory for the dynamic response functions of interacting one-dimensional spin-1/2 fermions. In contrast to the conventional Luttinger liquid theory, we take into account the nonlinearity of the fermion dispersion exactly. We calculate the power-law singularities of the spectral function and the charge and spin density structure factors for arbitrary momenta and interaction strengths. The exponents characterizing the singularities are functions of momenta and differ significantly from the predictions of the linear Luttinger liquid theory. We generalize the notion of the spin-charge separation to the nonlinear spectrum. This generalization leads to phenomenological relations between threshold exponents and the threshold energy.
\end{abstract}

\maketitle

\section{Introduction}

One of the aspirations of condensed matter physics is to understand the physical properties of interacting many-body systems. A successful example is Landau's Fermi liquid theory\cite{nozieres_book} which provides a comprehensive framework for the description of repulsively interacting fermions. This theory is based on the concept of fermionic quasiparticles, which can be regarded as physical fermions surrounded by a cloud of particle-hole excitations. The quasiparticles carry the same quantum numbers as the physical fermions but generally have a different effective mass. Even if the interactions between the physical fermions are strong, phase space constraints near the Fermi surface limit the scattering rates of quasiparticles and they become stable towards low energies. In this limit, the system can be described by a theory of noninteracting fermionic quasiparticles.

Let us compare some properties of noninteracting Fermi gases and interacting Fermi liquids in three dimensions. In a Fermi gas, the occupation number $n(k) = \expcts{\psi^\dag_{k} \psi_{k}}{}$ which measures the number of physical fermions with momentum $k$, jumps from one to zero at the Fermi surface $|k| = k_F$. For a Fermi liquid the amplitude of the discontinuity at $|k| = k_F$ is reduced to a positive value $Z<1$, but remains nonzero. The so-called quasiparticle residue $Z$ is a measure of the overlap between the physical fermions and the quasiparticles. Another example is the spectral function $A(k,\omega)$, defined below in Eq.~(\ref{Adef}). For a Fermi gas with spectrum $\epsilon(k)$, the spectral function is $A(k,\omega) = \delta[\omega - \epsilon(k)]$. For a Fermi liquid, the $\delta$-peak at energies close to the Fermi level evolves into a symmetric Lorentzian peak with a width proportional to $(|k| - k_F)^2$.

The physical properties of interacting systems are drastically different in one dimension. The effect of interactions is nonperturbative. For example, an arbitrarily weak repulsion in one dimension leads to zero residue, $Z=0$, suggesting an absence of fermionic excitations. Instead, the elementary excitations are thought to be better represented by quantized waves of density obeying Bose statistics.\cite{tomonaga50,luttinger63} For low energies, the spectrum $\epsilon(k)$ of the physical fermions can be linearized around the two Fermi points $\pm k_F$. The interacting fermionic theory can then be mapped onto a theory of noninteracting bosons\cite{mattis65} and all correlation functions can be calculated exactly.\cite{haldane81} The universality class formed by gapless interacting one-dimensional systems at low energies is called Luttinger liquid (LL).\cite{haldane81_2} It is entirely characterized by the velocity of density waves $v$ and the Luttinger parameter $K$ which depends on the interaction strength.\cite{giamarchi03} The distinct nature of these systems compared to their higher-dimensional counterparts manifests itself in many observables. The spectral function $A(k,\omega)$, for example, instead of becoming a Lorentzian, displays asymmetric power-law divergencies.

The contrast between one-dimensional and higher-dimensional interacting systems becomes even stronger for spinful systems. The eigenmodes of a spinful LL are spin-carrying and charge-carrying density waves with linear dispersion.\cite{dzyaloshinskii74} In general, the velocity of these two types of excitations, $v_s$ and $v_c$ respectively, can be very different. The introduction of a physical fermion with charge and spin into the liquid leads to the formation of independent spin and charge density waves. This phenomenon is called spin-charge separation and can be probed in experiments.\cite{auslaender05,kim06,jompol09} When turning on the interactions, the $\delta$-function singularity in the spectral function of a noninteracting system, $A(q,\omega) = \delta(\omega - v_F q)$, splits into two power-law singularities at the threshold energies for spin and charge density waves,\cite{meden92,voit93} $A(q,\omega) \propto (\omega - v_s q)^{-\mu^s} (\omega - v_c q)^{-\mu^c}$ with exponents $\mu^{c,s}$ depending only on the Luttinger parameter.

The linearization of the generic spectrum of particles is the crucial simplification leading to the LL theory and it has a substantial impact on density excitations. For an LL, the charge and spin density structure factors defined below in Eq.~(\ref{DSFall}) have sharp peaks, $S(q,\omega) \propto \delta(\omega - v_c |q|)$ and $S^{zz}(q,\omega) \propto \delta(\omega - v_s |q|)$ respectively, because a linear spectrum entails a one-to-one correspondence between the energy and the momentum of density excitations. This is the reason why the eigenmodes of an LL are density excitations. Also note that spin and charge modes are entirely decoupled.

For a quadratic spectrum, in contrast, a density excitation of fixed momentum $q$ may have energies in a range of width proportional to $q^2$. The peak in the structure factor remains narrow in the sense that $q^2/(v_{c,s}q) \propto q$, as long as $q$ remains small compared to $k_F$. This agrees with the general idea of irrelevance of the curvature as a perturbation to the LL theory.\cite{haldane81} However, the form of the peak is far from being a simple broadening to a Lorentzian.\cite{pereira09} Moreover, a spectrum curvature leads to a coupling between the spin and charge modes.\cite{brazovskii93,vekua09,rizzi08} The charge density structure factor, for example, acquires a peak at energies $\omega \approx v_s q$  characteristic for the \emph{spinon} excitations.\cite{pereira10,teber07}

Along with the structure factors, the spectral function is also affected by the nonlinearity of the dispersion relation.\cite{khodas07_2,imambekov09,imambekov09_2} The energy domains in which the dynamic response functions deviate from the predictions of the LL theory expand with the particles' momenta tuned away from the Fermi points. It was shown, however, that even far away from the Fermi points there is a certain universality in the behavior of the response functions.\cite{imambekov09_2,imambekov09} The dynamic response functions at arbitrary wave vectors of single-species fermions or bosons were studied in detail.\cite{pustilnik03,pustilnik06,khodas07,cheianov08,imambekov08,pereira09} The case of spin-1/2 fermions, despite its practical importance, is studied far less.\cite{lante09,pereira10}

The experiments on momentum-resolved tunneling between 1D systems\cite{auslaender02,auslaender05} or between 1D and 2D systems,\cite{jompol09} as well as angle-resolved photoemission spectroscopy on quasi-1D systems,\cite{segovia99,claessen02,kim06,stewart08,wang09} provide a tool to measure the electron spectral function. Moreover, the density structure factor can be measured using the Coulomb drag\cite{pustilnik03,yamamoto06} or neutron scattering on spin chains.\cite{lake05} Some of these experiments may be interpreted, in a limited domain of wave vectors, in terms of spin-charge separation.\cite{claessen02,kim06,jompol09} Moreover, experiments using ultracold gases have been proposed\cite{recati03,kollath06} which could allow the observation of spin-charge separation in real space. The numerical calculation of dynamic response functions has become possible using time-dependent density matrix renormalization group techniques.\cite{benthien04,white08,barthel09,feiguin09,kokalj09,kohno10} None of the developed methods is limited to low energies and momenta, thus prompting the question of the nonlinear dispersion effects.

The main goal of this article is to present quantitative results for the dynamic response functions (spectral function and the density structure factors) for spinful 1D Fermi systems at arbitrary interaction strength. We extend the LL theory by taking into account the nonlinearity of the fermion spectrum exactly, and we obtain results for the dynamic response functions which are valid for arbitrary momenta. Moreover, we shall elucidate the fate of the spin-charge separation away from the Fermi points. Our results apply to a wide range of systems with gapless spectrum and will be used to track the evolution of the dynamic response functions all the way from the noninteracting to the strongly interacting limit.

We restrict our analysis to spin-$1/2$ systems at zero magnetic field, i.e., without spin polarization. Spin rotation symmetry then entails $SU(2)$-invariance of the spin degrees of freedom. On the other hand, spin-polarized one-dimensional systems are interesting in their own right and have been investigated theoretically\cite{yang01} as well as experimentally.\cite{liao10} A nonzero magnetic field breaks the $SU(2)$-symmetry and the Zeeman shift leads to different Fermi wavevectors for spin-up and spin-down fermions. This splits the peaks in the dynamic response functions, and for large magnetic fields spin and charge degrees of freedom become coupled even within the linear LL theory.\cite{rabello02} These complications do not arise in the absence of a magnetic field.

In order to calculate the dynamic response functions, it is convenient to translate the bosonic spin and charge modes into fermionic quasiparticles, spinons and holons. For a linear spectrum, the bosonic or fermionic languages may be used equally comfortably and both offer their particular benefits. The advantage of the former is the direct relation between the bosonic modes and the density response functions. On the other hand, the fermionic description connects to the well-known physics of the Fermi edge problem.\cite{mahan67,anderson67,nozieres69} For a nonlinear spectrum, the fermionic basis is superior because it avoids divergencies arising in the bosonic perturbation theory.\cite{samokhin98} It leads to a generalization of the quantum impurity model which was used previously to calculate dynamic response functions for spinless systems.\cite{imambekov09,imambekov09_2}

This article is organized as follows: in Sec.~\ref{sec:qual}, we present an overview of our results for the dynamic response functions at zero temperature and the spin-charge separation, and point out qualitatively the main differences between interacting systems with linear and nonlinear spectrum. In Sec.~\ref{sec:referm}, we rephrase the LL theory in the basis of fermionic spin and charge quasiparticles and reproduce the known results for the spectral function. We also discuss how the Hamiltonian changes in the presence of a nonlinear spectrum. In Sec.~\ref{sec:QI}, we present in detail the method for the calculation of threshold singularities of dynamic response functions for a nonlinear spectrum. We express the threshold exponents of the dynamic response functions in terms of scattering phase shifts and we calculate the spectral function at its edge of support for $|k| \to k_F$. In Sec.~\ref{sec:phen}, we construct phenomenological relations between these scattering phase shifts and the shape of the spinon spectrum $\epsilon_s(k)$ which are valid for arbitrary momenta in Galilean invariant system. In Secs.~\ref{sec:spin_exp} and \ref{sec:DSF}, we use these relations to calculate the spectral function $A(k,\omega)$ and density structure factors $S^{zz}(k,\omega)$, $S^{-+}(k,\omega)$ and $S(k,\omega)$ near their respective edges of support for arbitrary momenta. In Sec.~\ref{sec:holon}, we construct phenomenological relations fixing the exponents of correlation functions near the holon spectrum $\epsilon_c(k)$ in terms of its shape. In Sec.~\ref{sec:limits}, we apply our general theory to the limits of very strong and very weak interactions. For strong interactions, we reproduce the known results for the Hubbard model with infinite interaction. For weak interactions, we complement the phenomenological result by a perturbative calculation in the basis of free fermions in order to obtain $A(k,\omega)$ away from the spinon and holon spectra. In Sec.~\ref{sec:holon}, we estimate the width of the peak of $A(k,\omega)$ at the holon mass shell by investigating the decay rate of holons due to their interaction with spinons. Finally, in Sec.~\ref{sec:conclusion}, we present our conclusions.

\section{Qualitative picture and results}\label{sec:qual}

The LL theory is widely used to describe the low-energy properties of gapless one-dimensional interacting fermionic systems.\cite{haldane81_2} The restriction to low energies usually justifies a linearization of the spectrum of the physical fermions around the right and left Fermi points, $\epsilon(k) \approx v_F(\pm k - k_F)$. Within this approximation, the system remains exactly solvable even for nonzero interactions and can be cast into a linear theory of noninteracting bosonic fields.\cite{mattis65} These eigenmodes are collective density waves which correspond to many-particle excitations when expressed in terms of the physical fermions.

One of the notable features of spinful interacting systems is the spin-charge separation.\cite{voit93} The Hamiltonian of the interacting system splits into a sum of two commuting quadratic terms which act on different Hilbert spaces and describe the charge and spin degrees of freedom separately. For nonzero interactions, the velocities of these two types of excitations are different. The injection or extraction of a physical particle which carries both spin and charge thus leads to the formation of spin and charge density waves which separate in space.

Both the collective nature of the eigenmodes and the spin-charge separation are clearly observable in various dynamic response functions. The charge and spin density structure factors are defined as
\begin{align}\label{DSFall}
 S(k,\omega) &= \int dx dt e^{i \omega t - i k x} \expct{\rho_c(x,t) \rho_c(0,0)}{}, \notag \\
 S^{-+}(k,\omega) &= \int dx dt e^{i \omega t - i k x} \expct{ S^-(x,t) S^+(0,0)}{}, \notag \\
 S^{zz}(k,\omega) &= \int dx dt e^{i \omega t - i k x} \expct{ S^z(x,t) S^z(0,0)}{},
\end{align}
where $\rho_c(x)$ and $\vec{S}(x)$ denote the charge and spin density, respectively, and $S^\pm = S^x \pm i S^y$. The density structure factors measure the linear response of the system at momentum $k$ and energy $\omega$ to a perturbation which couples to the charge or spin density. The charge density perturbation, for instance, can be created by the absorption of a photon. For an LL, one finds $S(k,\omega) \propto \delta(\omega - v_c |k|)$ and $S^{zz}(k,\omega) = \frac{1}{2} S^{-+}(k,\omega) \propto \delta(\omega - v_s |k|)$. The Dirac-$\delta$ shape of these functions reflects the fact that charge and spin density waves are eigenmodes and thus have a sharp energy for a given momentum. Note that this is a consequence of the linearized spectrum of the physical fermions. Moreover, the functions demonstrate that charge and spin density waves propagate with velocities $v_c$ and $v_s$, respectively, which depend on the details of the interaction between the physical fermions.

\begin{figure}[t]
  \centering
  \includegraphics[width = 0.48 \textwidth]{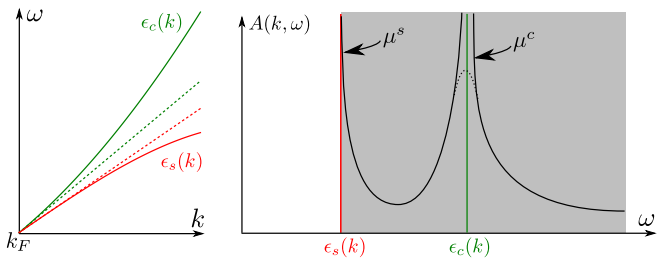}
\caption{(Color online) Spinon and holon spectra, $\epsilon_s(k)$ and $\epsilon_c(k)$, and spectral function $A(k,\omega)$ for momenta $k \geq k_F$. For repulsive interactions, a nonlinear fermion spectrum reduces $\epsilon_s(k)$ and increases $\epsilon_c(k)$ compared to the linear Luttinger liquid spectra. The spectral function has a power-law singularity at the spinon mass shell $\omega \approx \epsilon_s(k)$ with exponent $\mu^s$ and $\epsilon_s(k)$ is the edge of support of $A(k,\omega)$. The singularity at the holon mass shell $\omega \approx \epsilon_c(k)$ is smeared out (dotted line) away from $k_F$ for nonintegrable systems due to holon decay.}
  \label{fig:spectrumkF}
\end{figure}

The spectral function is defined in terms of the retarded Green's function by $A(k,\omega) = - \frac{1}{\pi} \Im G^\text{ret}(k,\omega)$ and can be written as
\begin{align}\label{Adef}
 A(k,\omega)  &= \frac{1}{\pi} \Re \int_0^\infty dt \int_{-\infty}^\infty dx e^{i \omega t - i k x} \notag \\
  &\times \expct{ \{ \psi_\sigma(x,t), \psi^\dag_\sigma(0,0) \} }{}.
\end{align}
The operator $\psi^\dag_\sigma(x)$ creates a physical fermion of spin $\sigma = \uparrow,\downarrow$. In the absence of a magnetic field, $SU(2)$-symmetry ensures that $A(k,\omega)$ is independent of $\sigma$. $A(k,\omega)$ represents a different type of dynamic response function which measures the response of the system to the addition of a physical particle or hole with momentum $k$ and energy $\omega$. This function can be determined experimentally, for instance, by measuring the momentum-resolved tunneling into LLs\cite{jompol09,auslaender05} or by photoemission spectroscopy.\cite{stewart08,kim06} For a noninteracting system, $A(k,\omega) = \delta[\omega- \epsilon(k)]$ defines the spectrum for single-particle excitations, $\epsilon (k)$. In the presence of interactions, customarily described by an LL, this function develops power-law singularities at the eigenenergies of spin and charge density waves,
\begin{align}\label{Agen}
 A[k,\omega \approx \epsilon_{c,s}(k)] \propto [ \omega - \epsilon_{c,s}(k) ]^{-\mu^{c,s}(k)}  .
\end{align}
The qualitative shape of $A(k,\omega)$ is shown in Fig.~\ref{fig:spectrumkF}. For an LL, the spectrum of left- and right-moving spin and charge density waves is linear, $\epsilon_{c,s}(k) = v_{c,s}(\pm k - k_F)$, and the exponents $\mu^{c,s}$ are $k$-independent. They only depend on the Luttinger parameter $K_c$ which encodes the interaction strength: for a noninteracting system $K_c = 1$, while repulsive interactions lead to $0 < K_c < 1$. For $\omega = v_c (k - k_F)$, the incoming particle leads to the formation of a charge density wave with energy $\omega$ while the spin density wave carries no energy. Similarly, for $\omega = v_s (k - k_F)$, the final state contains a spin density wave with energy $\omega$ and a charge density wave of zero energy. Away from these thresholds, the final state may contain multiple excitations of nonzero energy in the spin and charge sectors.

In the following, we shall refer to this conventional description as the \emph{linear} LL theory in order to emphasize the distinction to the case of nonlinear spectrum. Away from the Fermi points, the curvature of the physical spectrum $\epsilon(k)$ can no longer be neglected. It is convenient to refermionize the system and to express the spin and charge density waves in terms of fermionic quasiparticles, spinons and holons. Within the linear LL theory, the spinon and holon spectra $\epsilon_s(k)$ and $\epsilon_c(k)$ are linear and the quasiparticles are noninteracting. In contrast, for nonlinear $\epsilon(k)$ spinon and holon spectra become themselves nonlinear and interactions among spinons, among holons and between spinons and holons come into existence.

Throughout this article, we shall focus on the case of zero temperature and repulsive interactions between the physical fermions. Repulsion leads to $v_s<v_c$, as known from the linear LL theory. Let us discuss $A(k \geq k_F,\omega)$ for nonlinear spectrum. For momenta close to the Fermi points, $\epsilon_s(k)$ bends downwards away from the linear spectrum as depicted in Fig.~\ref{fig:spectrumkF}. The spinon spectrum $\epsilon_s(k)$ becomes the edge of support of the spectral function, i.e., $A(k,\omega) = 0$ for $- \epsilon_s(k) < \omega < \epsilon_s(k)$. Near the edge, for $\omega \approx \epsilon_s(k)$, $A(k,\omega)$ is given by Eq.~(\ref{Agen}) with a $k$-dependent exponent $\mu^s(k)$. This exponent coincides with the LL prediction in the limit $k \to k_F$. In order to calculate the spectral function away from $k_F$, we derive phenomenological expressions which yield the threshold exponent $\mu^s(k)$ at arbitrary $k$ in terms of the properties of $\epsilon_s(k)$. As the latter can be measured or calculated for many systems, this provides a useful relation between two independently measurable quantities.

On the other hand, as shown in Fig.~\ref{fig:spectrumkF}, the holon spectrum $\epsilon_c(k)$ bends upwards away from $k_F$. For generic nonintegrable systems, the singularity of the spectral function at $\omega \approx \epsilon_c(k)$ becomes smeared out because quasiparticle interactions allow a decay of holons. Therefore, the power-law behavior at the holon mass shell of the form~(\ref{Agen}) manifests itself only at $k \to k_F$. It turns out that for nonlinear $\epsilon_c(k)$, the exponent $\mu^c$ is different form the LL prediction even at $k \to k_F$. The reason is that the leading quadratic curvature of $\epsilon_c(k)$ introduces a new energy scale $(k - k_F)^2/(2m^*)$, where $m^*$ is the effective mass. The modified exponent only holds in an energy window of this width around $\epsilon_c(k)$. Since for a strictly linear spectrum, the width shrinks to zero, this does not contradict the linear LL theory. Thus for the charge mode the effect of the spectrum nonlinearity is similar to the spinless fermion case.\cite{imambekov09_2}

The strict spin-charge separation of the linear LL theory no longer exists once the band curvature is taken into account. The Hamiltonian does not consist of commuting spin and charge terms any more. The dynamic structure factors for the charge and spin density, $S(k,\omega)$ and $S^{zz}(k,\omega)$ respectively, cease to be $\delta$-functions.
Nevertheless, spin-charge separation continues to hold in a weaker sense: for arbitrary momentum $k$, the power-law singularity of the spectral function at its edge of support is determined by states where a spinon carries the entire energy $\epsilon_s(k)$ and has the velocity $\partial \epsilon_s(k)/\partial k$. The holon, in contrast, has zero energy and a strictly higher velocity $v_c$. Slightly away from the edge, the excess energy is used to generate additional low-energy particle-hole pairs in the holon sector, but no additional spinons. Moreover, spinon decay (in contrast to holon decay) remains forbidden by energy and momentum conservation. Therefore, the injection or extraction of a particle with momentum $k$ and energy near the threshold, $\omega \approx \epsilon_s(k)$, still forms spatially separating spin and charge density waves as in the linear LL theory. However, in contrast to the linear LL theory, this is no longer true for energies far away from the threshold.

The results for the spectral function and the density structure factors near the edges of support for arbitrary momenta are summarized in Table~\ref{tab:Exponents}.

\begin{table*}
  \centering
  \begin{tabular}{|c|c|c|l|}
   \hline
   $A(k,\omega \gtrless 0)$ & $(2n-1) k_F < k < (2n + 1) k_F$ & $\mu^s_{n,\pm}$
      & $\displaystyle 1 - \frac{1}{2} \left( -\frac{(2n + 1)
\sqrt{K_c}}{\sqrt{2}} + \frac{\delta^A_+ + \delta^A_-}{2\pi} \right)^2 -
\frac{1}{2} \left( \frac{1}{\sqrt{2 K_c}} - \frac{\delta^A_+ - \delta^A_-}{2\pi}
\right)^2 - m^2_\pm$ \\
   \hline
   $A(k,\omega \approx 0)$ & $k \approx (2n \pm 1) k_F  $ & $\mu^c_{n-}$
      & $\displaystyle -\frac{1}{2} - \frac{1}{4} \left( ( 2n + 1)^2 K_c  + \frac{1}{K_c} \right) + \frac{1}{\sqrt{2 K_c}} +  (2 n + 1) \sqrt{\frac{K_c}{2}}$ \\
   \hline
   $S(k,\omega)$ & $2n k_F < k < 2(n + 1) k_F$ & $\mu_{n}^{\text{DSF}}$
          & $\displaystyle \frac{1}{2} - \frac{1}{2} \left( \frac{2n
\sqrt{K_c}}{\sqrt{2}} + \frac{\delta^S_+ + \delta^S_-}{2\pi} \right)^2 -
\frac{1}{2} \left(\frac{\delta^S_+ - \delta^S_-}{2\pi} \right)^2$ \\
   \hline
  \end{tabular}
  \caption{Exponents for the spectral function $A(k,\omega)$ (see Fig.~\ref{fig:SpectralFunctionNLL} for notations) and the charge density structure factor $S(k,\omega)$ (see Fig.~\ref{fig:S} for notations) at the edge of support. The  exponents are determined in terms of the phase shifts $\delta^A_\pm = \Delta \delta_{\pm c}(k - 2 n k_F)$ and $\delta^S_\pm = \Delta \delta_{\pm c}[(2n + 1) k_F - k]$ calculated in Eq.~(\ref{phen_final}). Moreover, $m_{\pm} = (n+1/2 \pm 1/2)\, \mod 2$. The exponents and the edge of support for the spin structure factors $S^{-+}(k,\omega)$ and $S^{zz}(k,\omega)$ coincide with the ones for $S(k,\omega)$.}
  \label{tab:Exponents}
\end{table*}

\section{Refermionization of the Luttinger liquid theory}\label{sec:referm}

Let us start by recapitulating the results for the spectral function of interacting
fermions in one dimension using the conventional linear LL theory. This theory is universal in that it predicts the low-energy properties of microscopically very distinct systems using only very few measurable parameters.\cite{haldane81_2} One of its cornerstones is the realization that the elementary excitations of one-dimensional interacting fermion systems are collective bosonic charge and spin density waves.\cite{tomonaga50,mattis65} This is in stark contrast to higher dimensional systems, where the Fermi liquid theory\cite{nozieres_book} predicts that the elementary excitations are only weakly affected by interactions and remain fermionic. Spin and charge density waves can be encoded into the bosonic fields $\phi_\nu(x)$ ($\nu = s,c$) and the canonically conjugate fields $\theta_\nu(x)$, where\cite{giamarchi03}
\begin{align}
     [\phi_\nu(x), \partial_y \theta_\mu(y)] = i \pi \delta_{\mu\nu} \delta(x-y).
\end{align}
In the low-energy regime, the spectrum $\epsilon(k)$ of the physical fermions can be linearized around the two Fermi points, $\epsilon_{R,L}(k) \approx v_F(\pm k - k_F)$. For this spectrum, the kinetic energy becomes quadratic in $\phi_\nu$ and $\theta_\nu$. Moreover, the interaction energy is generally quadratic in the charge and spin densities $\rho_{c,s}$,
\begin{align}\label{rho_phi}
 \rho_{c,s}(x) = -\frac{\sqrt{2}}{\pi} \nabla \phi_{c,s}(x) .
\end{align}
Therefore, the interacting Hamiltonian, despite being quartic in fermionic operators, remains quadratic in the bosonic basis. It can be shown that the full Hamiltonian $H_0$ becomes a sum of commuting harmonic charge and spin terms, $H_0 = H_c + H_s$, which are given by\cite{giamarchi03} (using $\hbar = 1$)
\begin{align}\label{H_0}
 H_\nu = \frac{v_\nu}{2\pi} \int dx \left[ K_\nu (\nabla \theta_\nu)^2 + \frac{1}{K_\nu} (\nabla \phi_\nu)^2 \right]
\end{align}
for $\nu = c,s$. Interactions between the fermions lead to different velocities of charge and spin modes, $v_c \neq v_s$, and thus remove the degeneracy which is present in the noninteracting system. This leads to the spin-charge separation that has been observed in experiments:\cite{auslaender05,kim06,jompol09} once a spinful fermion is injected into the system, it creates spin and charge density waves propagating at different velocities, $v_s$ and $v_c$. Moreover, interactions determine the value of the constants $K_{c,s}$ which characterize the dynamical correlation functions and thermodynamic properties like the compressibility and the magnetic susceptibility. We shall assume that the interactions are repulsive. In the charge sector, this entails $0 < K_c < 1$, where $K_c = 1$ corresponds to noninteracting fermions. Furthermore, repulsive interactions lead to $v_s < v_c$. In addition to $H_0$, the spin part of the Hamiltonian will generally contain a sine-Gordon term
\begin{align}\label{Hg}
 H_g = \frac{2g}{(2 \pi a)^2} \int dx \cos[2 \sqrt{2} \phi_s(x)],
\end{align}
where $a$ is a short-distance cutoff. In terms of the original fermions, $H_g$ corresponds to spin-flip scattering $\propto \psi_{L\uparrow}^\dag \psi_{L\downarrow} \psi_{R\downarrow}^\dag \psi_{R\uparrow}$ and $g$ is proportional to the $2k_F$-component of the interaction potential. A renormalization group (RG) analysis shows that for repulsive interactions ($K_c < 1$) the interaction strength $g$ flows to zero as the bandwidth is reduced, so $H_g$ is irrelevant.\cite{solyom79} In the absence of a magnetic field, the system Hamiltonian must conserve $SU(2)$-symmetry, so the components $S^{-+}$ and $S^{zz}$ of the spin density structure factor in Eq.~(\ref{DSFall}) must coincide. For $g = 0$, this requirement leads to\cite{gogolin98,giamarchi03}
\begin{align}
    K_s = 1.
\end{align}

For energies small compared to $k_F^2/2m$, where $m$ is the bare mass of the fermions, all interaction processes will involve particles close to the two Fermi points $\pm k_F$. The physical fermion operators $\psi_\sigma$ ($\sigma=\uparrow,\downarrow$) can then be projected onto linearized bands of right-moving and left-moving fermions $\psi_{\alpha\sigma}(x)$ ($\alpha = R,L$) with momentum close to $\pm k_F$, i.e., $\psi_\sigma(x) = e^{i k_F x} \psi_{R\sigma}(x) + e^{-i k_F x} \psi_{L\sigma}(x)$. These are related to the bosonic fields $\theta_\nu$ and $\phi_\nu$ via the bosonization identity\cite{luther74,haldane81,giamarchi03}
\begin{align}\label{bos_iden}
 \psi_{\alpha\sigma}(x) \propto \frac{1}{\sqrt{2 \pi a}} e^{-i/\sqrt{2} [ \alpha \phi_c(x) - \theta_c(x) + \alpha \sigma \phi_s(x) -\sigma \theta_s(x) ]}.
\end{align}
The cutoff length $a$ is used to regularize the ultraviolet behavior of the theory. One can choose $a = 1/\Lambda$, where $\Lambda \ll k_F$ is the width of the linearized bands. In principle, the bosonization identity also contains Klein factors to ensure the correct fermionic anticommutation relations, e.g., $\{ \psi_{\alpha\sigma}(x), \psi^\dag_{\beta\tau}(y)\} = \delta_{\alpha\beta} \delta_{\sigma\tau} \delta(x-y)$.  However, we did not write them out explicitly because they commute with $\theta_\nu$ and $\phi_\nu$ and drop out whenever Eq.~(\ref{bos_iden}) is used to calculate expectation values of an operator which conserves charge and spin. The spectral function (\ref{Adef}) contains the expectation value $\expct{\{ \psi_{\alpha\sigma}^\dag(x,t), \psi_{\alpha\sigma}(0,0)\}}$ which obviously satisfies this requirement and the same is true for the other dynamic response functions we shall calculate.

The spectral function of a linear LL can be obtained by expressing the chiral fermions $\psi_{\alpha\sigma}$ in terms of $\phi_\nu$ and $\theta_\nu$ using Eq.~(\ref{bos_iden}) and then calculating the bosonic expectation values with respect to the Hamiltonian (\ref{H_0}). However, the bosonic language is not well suited to include band-curvature effects which generate anharmonic terms (in boson creation-annihilation operators) in the Hamiltonian. Bosonic perturbation theory in these anharmonic terms leads to divergences.\cite{samokhin98} The physical reason is that, e.g., the right-moving bosonic spin/charge excitations with momentum $k$ always have an energy $v_\nu k$ ($\nu = s,c$), which remains linear in $k$, so the boson velocity $v_\nu$ for right-movers is momentum-independent. Roughly speaking, bosons with different momenta thus have a long time to interact and this leads to a breakdown of perturbation theory. It turns out to be beneficial to rephrase the problem in a fermionic language by introducing left- and right-moving fermionic spin and charge quasiparticles, \emph{spinons} and \emph{holons}. The fermionic spectra will be curved, thereby avoiding the problem of the bosonic theory.

For spinless fermions, the transformation between the physical, interacting fermions and noninteracting fermionic quasiparticles can be performed directly using a unitary transformation\cite{rozhkov05} or via bosonization and subsequent refermionization.\cite{imambekov09_2} In the spinful case, we use the latter option. First, $H_0$ is diagonalized by the canonical scaling transformation
\begin{align}\label{ttheta}
 \tilde{\theta}_\nu &= \sqrt{K_\nu} \theta_\nu, \notag \\
 \tilde{\phi}_\nu &= \frac{\phi_\nu}{\sqrt{K_\nu}}.
\end{align}
Each of the Hamiltonians $H_\nu[\tilde{\theta}_\nu,\tilde{\phi}_\nu]$ ($\nu = c,s$) has the form of a free bosonic Hamiltonian with Luttinger parameters equal to unity and should therefore be representable as a noninteracting fermionic Hamiltonian by reversing the bosonization procedure. Indeed, the bosonization identity (\ref{bos_iden}) can be used to refermionize the theory by \emph{defining} new fermionic operators as
\begin{align}\label{referm}
 \tpsi_{\alpha\nu} \propto \frac{1}{\sqrt{2 \pi a}} e^{-i ( \alpha\tilde{\phi}_\nu - \tilde{\theta}_\nu )}
\end{align}
for $\alpha = R,L= +,-$ and $\nu = s,c$. Written in terms of spinons $\tpsi_{\alpha s}$ and holons $\tpsi_{\alpha c}$, the spin and charge parts of the LL Hamiltonian (\ref{H_0}) assume the form of free fermionic Hamiltonians with linear spectrum,
\begin{align}\label{H0}
 H_\nu = -i v_\nu \sum_{\alpha = R,L} \alpha \int dx : \tpsi^\dag_{\alpha \nu}(x) \nabla \tpsi_{\alpha\nu}(x):\ .
\end{align}
The colons denote normal ordering with respect to the ground state, which is given by filled Fermi seas: for $\alpha = R$ ($\alpha = L$) all states within the bandwidth $\Lambda$ with negative (positive) momentum are singly occupied whereas all other states are empty.

The bosonic fields $\tilde{\phi}_\nu$ and $\tilde{\theta}_\nu$ are related to the densities of the fermionic quasiparticles. In analogy to the conventional bosonization result,\cite{giamarchi03} the refermionization formula (\ref{referm}) leads to
\begin{align}
 \trho_{\alpha\nu}(x) &= :\tpsi^\dag_{\alpha\nu}(x) \tpsi_{\alpha\nu}(x): \notag \\
 &= -\frac{\alpha}{2\pi} \nabla [\alpha \tilde{\phi}_\nu(x) - \tilde{\theta}_\nu(x)].
\end{align}
This allows us to establish the relationship between the physical fermions and the spinons and holons. For this purpose, the physical fermions $\psi_{\alpha\sigma}$ are expressed in terms of the rescaled bosonic fields. Using (\ref{bos_iden}) and (\ref{ttheta}) it is straightforward to show that
\begin{align}\label{psi_tpsi}
 \psi_{\alpha\uparrow}(x)
&\propto
 \tpsi_{\alpha c}(x) F_{\alpha c}(x) \tpsi_{\alpha s}(x)
F_{\alpha s}(x), \notag \\
 \psi_{\alpha\downarrow}(x)
&\propto
 \tpsi_{\alpha c}(x) F_{\alpha c}(x) \tpsi^{\dag}_{\alpha s}(x)
F^\dag_{\alpha
 s}(x).
\end{align}
The Klein factors were discarded for the same reason as before. The unitary string operators $F_{\alpha\nu}(x)$ are functions of the quasiparticle densities and are given by
\begin{align}\label{string}
  F_{\alpha\nu}(x) = \exp\left\{ -i \alpha \int_{-\infty}^x dy \left[
      \delta_{+\nu} \trho_{\alpha \nu}(y) + \delta_{-\nu}
      \trho_{-\alpha\nu}(y) \right] \right\}.
\end{align}
where the effects of the interactions are contained in the $K_c$-dependent phase shifts
\begin{align}\label{phases}
  \frac{\delta_{+c}}{2\pi} &= 1 - \sqrt{\frac{1}{8 K_c}} - \sqrt{\frac{K_c}{8}}, \notag \\
  \frac{\delta_{-c}}{2\pi} &= \sqrt{\frac{1}{8 K_c}} - \sqrt{\frac{K_\nu}{8}} ,\notag \\
  \frac{\delta_{+s}}{2\pi} &= 1 - \frac{1}{\sqrt{2}}, \notag \\
  \frac{\delta_{-s}}{2\pi} &= 0.
\end{align}
For the phases $\delta_{\pm s}$, we used $K_s = 1$ which is a consequence of $SU(2)$-symmetry. As an example consider the creation of a right-moving spin-up fermion $\psi^\dag_{R\uparrow}$. According to Eq.~(\ref{psi_tpsi}), it corresponds to the creation of a holon $\smash{\tpsi^\dag_{Rc}}$ and a spinon $\smash{\tpsi^\dag_{Rs}}$ as well as of low-energy spin and charge density excitations $F^\dag_{Rc}, F^\dag_{Rs}$. Conversely, the creation of a spin-down particle corresponds to the annihilation of a spinon.

For a linear spectrum, the dynamics of the string operators (\ref{string}) is governed by the noninteracting Hamiltonian $H_0$ and becomes very simple. Hence, $A(k,\omega)$ can be calculated from the refermionized operators and one recovers the well-known LL result which is applicable for $k \approx \pm k_F$.\cite{meden92,voit93,gogolin98} Note that in the conventional approach to calculating $A(k,\omega)$, one first evaluates the correlators as functions of $x$ and $t$, and only then performs a Fourier transform. The latter step is a quite complicated problem in contour integration due to the existence of singularities in the integrand at $x=\pm v_\nu t$. Our method yields the threshold exponents in a much simpler way and provides a clear interpretation of $A(k,\omega)$ in terms of holons and spinons.

As an example, let us discuss the spectral function for $\omega < 0$ and $k \approx + k_F$. Expressing $\psi_{R\uparrow}(x)$ using Eqs.~(\ref{psi_tpsi}) and (\ref{string}) and Fourier transforming allow us to rewrite $A(k,\omega)$ as a convolution of correlation functions involving right- and left-moving spinons and holons. As $k \approx + k_F$, we measure the momentum relative to the right Fermi point, and use $q = k - k_F$. We then have
\begin{align}
 A(q,\omega) &= \frac{1}{2\pi} \int dx dt e^{i q x -i \omega t} \expct{ \psi_{R\uparrow}^\dag(x,t) \psi_{R\uparrow}(0,0)}{} \notag \\
&= \frac{1}{2\pi} \int
	\frac{d\omega_{Rc}}{2\pi} \frac{d q_{Rc}}{2\pi}
	\frac{d\omega_{Lc}}{2\pi} \frac{d q_{Lc}}{2\pi}
	\frac{d\omega_{Rs}}{2\pi} \frac{d q_{Rs}}{2\pi} \\
 &\times
	G_{Rc}(q_{Rc}, \omega_{Rc}) G_{Lc}(q_{Lc}, \omega_{Lc}) \notag \\
 &\times	
	G_{Rs}(q_{Rs}, \omega_{Rs}) G_{Ls}(q_{Ls}, \omega_{Ls}), \notag
\end{align}
where $q_{Ls} = q - q_{Rc} - q_{Lc} - q_{Rs}$ and $\omega_{Ls} = \omega - \omega_{Rc} - \omega_{Lc} - \omega_{Rs}$ due to energy and momentum conservation. We used the Fourier transforms of the correlation functions
\begin{align}\label{Galphanu_xt}
 G_{R\nu}(x,t) =& \Big\langle \tpsi^\dag_{R\nu}(x,t) e^{i \delta_{+\nu} \int^x dy \trho_{R\nu}(y,t)} \notag \\
 &\times e^{-i \delta_{+\nu} \int^0 dy \trho_{R\nu}(y,0)} \tpsi_{R\nu}(0,0)\Big\rangle, \\
 G_{L\nu}(x,t) =& \Big\langle e^{i \delta_{-\nu} \int^x dy \trho_{L\nu}(y,t)}
 	e^{-i \delta_{-\nu} \int^0 dy \trho_{L\nu}(y,0)} \Big\rangle. \notag
\end{align}
For linear spectrum, the time-dependence of the chiral left- and right-mover densities is simple, $\trho_{\alpha\nu}(x,t) = \trho_{\alpha\nu}(x - \alpha v_\nu t)$. These correlation functions can be calculated most easily in the bosonic basis, but a direct calculation in the fermionic basis is also feasible.\cite{rozhkov05} For the Fourier transforms, one finds
\begin{align}\label{Galphanu}
 G_{Rc}(q,\omega) &\propto \delta(\omega - v_c q) \theta(\omega + v_c q) (\omega + v_c q)^{\left(\frac{\delta_{+c}}{2\pi} - 1 \right)^2 - 1}, \notag \\
 G_{Lc}(q,\omega) &\propto \delta(\omega + v_c q) \theta(\omega - v_c q) (\omega - v_c q)^{\left(\frac{\delta_{-c}}{2\pi} \right)^2 - 1}, \notag \\
 G_{Rs}(q,\omega) &\propto \delta(\omega - v_s q) \theta(\omega + v_s q) (\omega + v_s q)^{\left(\frac{\delta_{+s}}{2\pi} - 1 \right)^2 - 1}, \notag \\
 G_{Ls}(q,\omega) &\propto \delta(\omega + v_s q) \delta(\omega - v_s q).
\end{align}
The arguments of the $\delta$-functions reflect the linear spectrum of the quasiparticles. The distinct form of $G_{Ls}(q,\omega)$ is a consequence of $\delta_{-s} = 0$. The Heaviside-$\theta$ functions appearing above have a simple interpretation in terms of the Fermi seas of spinons and holons. For right-movers (left-movers) all states with negative (positive) momenta are filled, thus only excitations with positive (negative) momenta can be created. The exponents occurring in these correlation functions can be interpreted in the context of the theoretical treatment of the Fermi edge singularity problem by Schotte and Schotte.\cite{schotte69}

The quadratic ``Anderson''-terms $\delta_{+c}^2$, $\delta_{-c}^2$, and $\delta_{+s}^2$ in the exponents of Eq.~(\ref{Galphanu}) are a consequence of the orthogonality catastrophe.\cite{anderson67} They reflect the shake-up of the right-moving and left-moving holons and of the right-moving spinons near the Fermi points, respectively, as a reaction to the introduction of the hole into the system. Left-moving spinons are not shaken up because $SU(2)$-symmetry leads to $\delta_{-s} = 0$. The exponents linear in $\delta_{+c}$ and $\delta_{+s}$, on the other hand, are ``Mahan'' terms\cite{mahan67} which arise because the injection of a right-moving hole leads to the creation of a right-moving holon and a right-moving spinon.

\begin{figure}[t]
  \centering
  \includegraphics[width = 0.48 \textwidth]{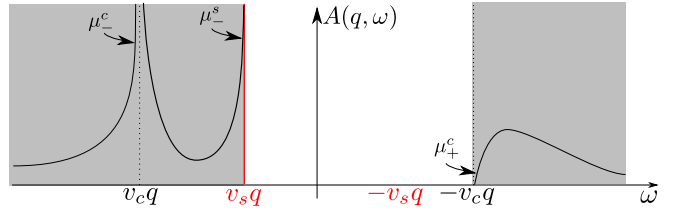}
\caption{(Color online) The spectral function
$A(q,\omega)$ for a linear Luttinger liquid with repulsive interactions along a cut for
fixed  $q = k - k_F < 0$. $A(q,\omega)$ is characterized by sharp power-law singularities at the holon and spinon mass shells $\omega = v_{c,s} q$ and at the inverted holon mass shell $\omega = -v_c q$.}
  \label{fig:SpectralFunctionLL}
\end{figure}

Consider $A(q,\omega)$ for $\omega \approx v_c q$. The singularity at this energy is generated by points in the integrand where $\omega \approx \omega_{Rc} = v_c q_{Rc}$. Due to momentum and energy conservation, the energies $\omega_{Lc}$ and $\omega_{Rs}$ as well as the momenta $q_{Lc}$ and $q_{Rs}$ will be small, so the functions $G_{Lc}$ and $G_{Rs}$ in the integrand lead to power-law singularities. One finds the exponent
\begin{align}\label{mu_mc_LL}
 \mu_{-}^{c} &= - 1 - \left[ \left(\frac{\delta_{-c}}{2\pi}\right)^2 - 1\right] - \left[ \left(\frac{\delta_{+s}}{2\pi} - 1\right)^2 - 1\right].
\end{align}
Similar arguments allow the calculation of the exponents at the spinon mass shell. Because $v_s < v_c$, the spectral function vanishes for $|\omega| < v_s |q|$. It is characterized by power-law singularities at the mass shells of spinons and holons, $A(q,\omega) \propto (\omega - v_{c,s} q)^{-\mu^{c,s}_-}$, with the exponents \cite{meden92,voit93}
\begin{align}\label{muLL}
  \mu^c_{-} &= \frac{1}{2} - \frac{1}{8} \left( K_c + \frac{1}{K_c} - 2 \right), \notag \\
  \mu^s_{-} &= \frac{1}{2} - \frac{1}{4} \left( K_c + \frac{1}{K_c} - 2 \right).
\end{align}
In addition, one finds a singularity at the inverted holon mass shell, $A(q,\omega) \propto (\omega + v_c q)^{-\mu^c_{+}}$, where
\begin{align}\label{muLL2}
   \mu^c_{+} &= - \frac{1}{8} \left( K_c + \frac{1}{K_c} - 2 \right).
\end{align}
The corresponding power-law at $\omega = -v_s q$ is suppressed because the particular form of $G_{Ls}(q_{Ls},\omega_{Ls})$ (\ref{Galphanu}) entails $q_{Ls} = \omega_{Ls} = 0$. This is a consequence of $K_s = 1$. It means physically that the right-moving hole with energy $\omega$ and momentum $k$ cannot excite left-moving spinons. Figure \ref{fig:SpectralFunctionLL} shows a cross-section of the spectral function $A(q,\omega)$ along a fixed $q = k - k_F < 0$.

\subsection{Corrections to the linear LL theory for nonlinear fermionic spectrum}

If the fermionic spectrum is not linear, two types of corrections have to be added to the Hamiltonian $H_0$ (\ref{H0}). First, the spectrum of the holons and spinons $\epsilon_{c,s}(k)$ will start to deviate from its linear behavior. In the case of the holons, the leading correction for $k \to k_F$ is quadratic,
\begin{align}\label{epsilonc}
 \epsilon_c(k) = v_c (k - k_F) + \frac{1}{2m^*} ( k - k_F)^2,
\end{align}
where $m^*$ is an effective mass which is related to the bare mass $m$ and the compressibility of the system and is generally positive, see Eq.~(\ref{mstar}). The spinon spectrum $\epsilon_s(k)$ also becomes nonlinear but its form is restricted by $SU(2)$-symmetry. The spin-up/spin-down symmetry of the physical fermions translates into particle-hole symmetry of the spinons. Therefore, the leading curvature is cubic,
\begin{align}\label{epsilons}
 \epsilon_s(k) = v_s (k - k_F) - \xi ( k - k_F)^3,
\end{align}
where in general $\xi > 0$. In addition to causing a band curvature of the spinon and holon spectra, a nonlinearity in the spectrum of physical fermions also leads to interactions between spinons and holons as well as within these two species. The different forms of holon and spinon mass shells (\ref{epsilonc})-(\ref{epsilons}) entail strikingly different consequences as far as the relevance of these interactions is concerned.

For the calculation of the spectral function $A(k,\omega)$ near $\omega \approx \epsilon_{c}(k)$, one needs to consider the scattering phase between a holon at momentum $k$ and holons and spinons at the Fermi edge. To lowest order in the interaction,\cite{pustilnik06} the scattering phase with left- and right-moving holons and spinons is proportional to $\tV^c_{\alpha\nu}(k-k_F)/(v_d - \alpha v_\nu)$, where $\tV^c_{\alpha\nu}(k-k_F)$ for $\alpha = R,L = +,-$ and $\nu = c,s$ is the corresponding interaction potential and $v_d = \partial \epsilon_c(k)/\partial k$ is the velocity of the holon with momentum $k$. For $k \to k_F$, the Hamiltonian of the system should reduce to that of a linear LL. In particular, this means that spinons and holons become noninteracting, i.e., $\tV^c_{\alpha\nu}(0) = 0$. Due to symmetry the expansion of the interaction potential starts with the quadratic term, $\tV^c_{\alpha\nu}(k-k_F) \propto (k - k_F)^2$ for $k \to k_F$. For a quadratic spectrum, $v_d - v_c \propto (k - k_F)$ for $k \to k_F$, whereas $v_d \pm v_s$ and $v_d + v_c$ remain finite in this limit. Therefore, all scattering phases vanish and interactions of the holon with other quasiparticles do not modify the spectral function for $k \to k_F$.

\begin{figure}[t]
  \centering
  \includegraphics[width = 0.48 \textwidth]{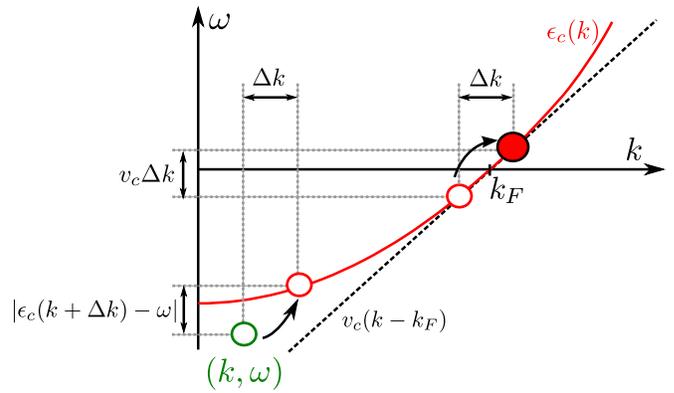}
\caption{(Color online) The injection of a hole with momentum $k$ and energy $\omega \approx \epsilon_c(k)$ leads to the formation of a holon on the mass shell and low-energy excitations around the holon Fermi edge. For clarity, the created low-energy spinons are not displayed. For this particular configuration, energy and momentum conservation lead to $v_c \Delta k = |\omega - \epsilon_c(k + \Delta k)|$. The impurity band (around $k$) and the low-energy band (around $k_F$) can be separated only if $\Delta k \ll k_F - k$. This is the case for $|\omega - \epsilon_c(k)| \ll (k - k_F)^2/(2m^*)$.}
  \label{fig:ImpurityBand}
\end{figure}

The quadratic form of the holon spectrum (\ref{epsilonc}), on the other hand, does lead to a change in $A(k,\omega)$. Let us focus again on $k < k_F$ and $\omega < \epsilon_c(k) < 0$. As depicted in Fig.~\ref{fig:ImpurityBand}, the injection of a hole with momentum $k$ and energy $\omega$ leads to the formation of a holon with shifted momentum $k + \Delta k$ on mass shell and low-energy spinons and holons near the Fermi points with total momentum $\Delta k$. If $\omega$ is close to the mass shell, i.e., $|\omega - \epsilon_c(k)| \ll |\epsilon_c(k) - v_c(k - k_F)| = (k - k_F)^2/(2m^*)$, energy and momentum conservation enforce $\Delta k \ll ( k_F - k)$. Therefore, the ``deep'' holon at momentum $k + \Delta k$ becomes well separated from excitations near the Fermi points and the respective regions of the spectrum cannot overlap, see Fig.~\ref{fig:ImpurityBand}. In this case, the averages over the fermion operator $\tpsi_{Rc}$ (which creates the deep holon at momentum $k$) and the string operators (which create the low-energy excitations near the Fermi points) in the definition (\ref{Galphanu_xt}) of $G_{Rc}(x,t)$ can be separated, similar to the case of spinless fermions.\cite{imambekov09_2} Then, one finds a modified exponent near the holon mass shell. For $\omega \approx \epsilon_c(k)$, $A(k,\omega) \propto [\omega - \epsilon_c(k)]^{-\mu^c_{0,-}}$, where
\begin{align}\label{mu_0_c}
 \mu^c_{0,-} &= 1 - \left(\frac{\delta_{-c}}{2\pi}\right)^2 -
 \left(\frac{\delta_{+s}}{2\pi}-1\right)^2 -
 \left(\frac{\delta_{+c}}{2\pi}\right)^2 \notag \\
&=
 - \frac{1}{2} - \frac{1}{4} \left( K_c + \frac{1}{K_c} \right) + \frac{1}{\sqrt{2 K_c}} + \frac{\sqrt{K_c}}{\sqrt{2}}.
\end{align}
This exponent can again be interpreted in the language of the Fermi edge problem: the quadratic exponents $\delta^2_{\pm c}$ and $\delta_{+s}^2$ are Anderson terms indicating a shake-up of the right- and left-moving holons as well as the right-moving spinons near the Fermi points, respectively. The linear exponent $\delta_{+s}$ can be interpreted as a Mahan term due to the creation of a spinon near the right Fermi point. In contrast to Eq.~(\ref{mu_mc_LL}), there is no Mahan term associated with $\delta_{+c}$ because, as explained above, the holon at momentum $k + \Delta k$ is in a different part of the spectrum than the Fermi point.

This exponent differs from the LL result (\ref{mu_mc_LL}). The exponent $\mu^c_{0,-}$ holds in a region of width $(k - k_F)^2/(2m^*)$ around $\epsilon_c(k)$. Beyond this window, the curvature of the spectrum becomes irrelevant and the exponent crosses over to the LL exponent $\mu^c_{-}$. Note that even for $K_c \to 1$, the exponent $\mu^c_{0,-} \to \sqrt{2} - 1$ differs from the LL prediction $\mu^c_{-} \to 1/2$.

For the calculation of the exponent near the spinon mass shell, $\omega \approx \epsilon_s(k)$, the interactions among spinons become important. This is a consequence of the cubic term in the spinon spectrum (\ref{epsilons}). Consider the scattering of a spinon with energy $\omega$ on low-energy left- and right-moving spinons and holons near the Fermi points. As previously, the lowest-order expansion of the respective interaction potentials for $k \to k_F$ is quadratic, $\tV_{\alpha\nu}(k-k_F) \propto (k - k_F)^2$. The velocity of the spinon is given by $v_d = \partial \epsilon_s(k)/\partial k$. However, in contrast to the quadratic holon spectrum, the leading curvature of the spinon spectrum $\epsilon_s(k)$ is cubic, so $v_d - v_s \propto (k - k_F)^2$. As a consequence, the scattering phase $\tV_{+s}(k-k_F)/(v_d - v_s)$ remains finite even in the limit $k \to k_F$. Hence, the scattering among spinons cannot be treated as a small perturbation. We shall investigate the scattering phase shifts for excitations near the spinon mass shell in the next section.

\section{Quantum impurity Hamiltonian}\label{sec:QI}

The investigation of the spectral function for momenta away from $\pm k_F$ necessitates a comment
about the momentum conservation when decomposing an injected physical particle or hole with momentum $k$ into a spinon-holon pair. Let us start from noninteracting fermions with Fermi momentum $k_F$. In terms of measurable quantities, $k_F$ can be defined as the smallest positive momentum $k$ for which $A(k,\omega = 0) \neq 0$. According to this definition, $k_F$ is defined by the singularities in the retarded Green's function, which according to Luttinger's theorem\cite{luttinger60,yamanaka97} are not shifted when the interactions are turned on. Therefore, the value of $k_F$ is not affected by interactions.

\begin{figure}[t]
  \centering
  \includegraphics[width = 0.48 \textwidth]{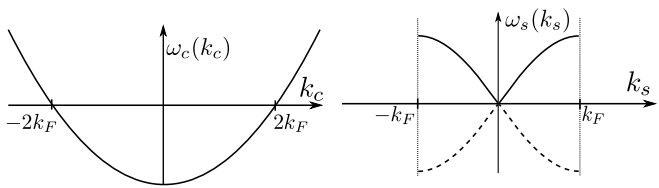}
  \caption{Holon spectrum $\omega_c(k_c)$ and spinon spectrum $\omega_s(k_s)$ in the noninteracting case. The holon spectrum has Fermi momentum $2 k_F$. The spinon spectrum is defined for $-k_F < k_s < k_F$ and the Fermi point is placed at $k_s = 0$. The spinon spectrum is particle-hole symmetric, so the dashed line indicates energies of spinon holes.}
  \label{fig:SpinonHolonSpectrum}
\end{figure}

The charge density of the physical fermions is related to $k_F$ by $\rho_c = 2 k_F / \pi$, where the factor $2$ results from the two spin orientations. The spinful physical fermions can be expressed in terms of two species, spinons and holons, of \emph{spinless} fermions. Since only the holons carry charge, their density must be equal to the physical charge density. Hence, the holon Fermi momentum $k_F^h = 2 k_F$. This has been found in the case of a generic strongly interacting system\cite{matveev07} as well as for integrable models at any interaction strength.\cite{carmelo91,essler05} In particular, the Bethe ansatz solutions for the Hubbard model at low filling in both the noninteracting limit and in the limit of infinite interaction lead to a parabolic holon spectrum $\omega_c(k_c)$ shown in Fig.~\ref{fig:SpinonHolonSpectrum}.

In the same context, it has also been shown that the spinon momentum is only defined up to multiples of $2 k_F$. This is well illustrated by the limit of strong repulsion.\cite{matveev07} In this limit, spinons live on a lattice provided by the holons. Therefore, we restrict the spinon spectrum $\omega_s(k_s)$ to momenta $-k_F < k_s < k_F$ and place the spinon Fermi level at zero momentum, $\omega_s(0) = 0$. For a physical particle with momentum $k$, the momentum conservation can then be expressed as\footnote{A slightly different, but compatible, decomposition of hole momentum into spinon and holon momenta has been chosen for the Hubbard model in [\onlinecite{essler05}]. The reason for the discrepancy is that in [\onlinecite{essler05}], the Fermi momentum of the spinons was placed at $\pm k_F$.}
\begin{align}
 k = k_c + k_s \pm k_F \quad \text{ for } k \lessgtr 0.
\end{align}
The spinon and holon mass shells $\epsilon_c(k)$ and $\epsilon_s(k)$ can be constructed from the spectra $\omega_c(k_c)$ and $\omega_s(k_s)$, respectively, by a combination of shifts and inversions.

The spectral function of interacting one-dimensional systems is generally characterized by power-law singularities. To explain the physical mechanism, let us focus briefly on the spinless case and consider a spinless hole with momentum $k < k_F$ tunneling into the system. If the energy is on the mass shell, the final state will contain a single ``deep'' hole with momentum around $k$ and energy $\epsilon(k)$. If the energy is close to, but not precisely at, the mass shell, the final state will contain the deep hole and, in addition, a number of particle-hole pairs, see Fig.~\ref{fig:ImpurityBand}. Since the energy available for the formation of these pairs is small, they will be located close to the Fermi points $\pm k_F$. It has been shown perturbatively\cite{pustilnik06,khodas07_2} that for a nonlinear excitation spectrum the deep hole can be regarded as separate from the particle-hole pairs at the Fermi points. The Hamiltonian can be projected onto three subbands, one containing the deep hole and two others containing excitations near the two Fermi points. The functional form of the spectral function is determined by how the deep hole interacts with the particle-hole pairs at the Fermi points. The physical mechanism is thus similar to the Fermi edge problem,\cite{mahan67,anderson67,schotte69} albeit in the present case with a mobile impurity instead of a static scattering potential.

The previous argument extends to spinful systems: in this case, power-law singularities occur at the spinon mass shell $\epsilon_s(k)$. Let us consider the case of a right-moving spinful hole with momentum $0 <k < k_F$ and energy close to $\epsilon_s(k)$. The final states giving rise to the power-law singularity will contain a deep spinon impurity with momentum $k_s = k - k_F =: k_d < 0$, a holon near its Fermi momentum and additional spin-carrying and charge-carrying particle-hole pairs at all Fermi points. Once more, the system can be projected onto narrow subbands containing the impurity and the Fermi points, respectively. This corresponds to decomposing the spinon operator by retaining only Fourier components close to zero and to $k_d$, $\tpsi_{Rs} \to \tpsi_{Rs} + e^{i k_d x} d$. The spectrum of the states near the Fermi points can be linearized and they are thus described by the LL Hamiltonian,
\begin{align}\label{MI_H0}
 H_0 = -i \sum_{\nu=c,s} v_\nu \sum_{\alpha = R,L} \alpha \int dx : \tpsi^\dag_{\alpha \nu}(x) \nabla \tpsi_{\alpha\nu}(x):\ .
\end{align}
Located on the spinon mass shell, the impurity has a velocity $v_d = \partial \epsilon_s(k)/\partial k$. The Hamiltonians containing its kinetic energy and the interaction between the mobile impurity and the subbands at the Fermi edges are given by
\begin{align}\label{MI_Hd}
  H_d &= \int dx\ d_s^\dag(x) [ \epsilon_{s}(k) - i v_d \nabla ]
  d_s(x),\\
H_{int}& = \int dx \sum_{\alpha \nu} \tV_{\alpha\nu}(k)
\trho_{\alpha \nu}(x) d_s^\dag(x) d_s(x), \label{MI_Hint}
\end{align}
where $\trho_{\alpha\nu}(x) = :\! \tpsi^\dag_{\alpha\nu}(x) \tpsi_{\alpha\nu}(x) :$ denotes the quasiparticle density in the bands around the Fermi edges, and the interaction constants $\tV_{\alpha\nu}(k)$ are yet to be determined.

\begin{figure}[t]
  \centering
  \includegraphics[width = 0.48 \textwidth]{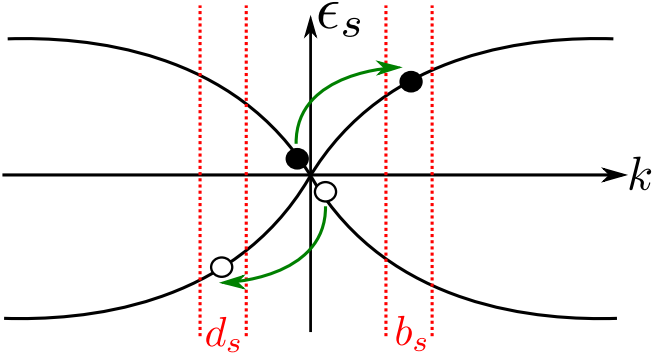}
\caption{(Color online) Spinon interaction process generated by the projection of the spin flip interaction in Eq.~(\ref{spinflip_general}): a spinon particle-hole pair near the left Fermi point scatters into a spinon hole at $k_s < 0$ and a spinon near $-k_s > 0$. The corresponding interaction operator is similar to the sine-Gordon term (\ref{Hg}) and, analogously, its amplitude flows to zero upon bandwidth reduction.}
  \label{fig:KondoTermSubbands}
\end{figure}

In addition, the interacting system Hamiltonian will generally contain spin-flip interaction terms of the form
\begin{align}\label{spinflip_general}
 \sum_{p,p',q} \psi^\dag_{\uparrow}(p) \psi_{\downarrow}(p + q) V(q) \psi^\dag_{\downarrow}(p') \psi_{\uparrow}(p' - q).
\end{align}
The projection of this term onto a reduced band structure with bands around the Fermi edges and around the impurity state leads to the following terms: (i) a density-density type interaction between particles near the Fermi edges.\cite{giamarchi03} Such a contribution merely renormalizes the Luttinger parameter $K_c$, as well as the spinon and holon velocities, $v_c$ and $v_s$; (ii) a density-density type interaction between the impurity and particles near either the left or right Fermi point. This leads to a term of the form (\ref{MI_Hint}) and can be absorbed into $\tV_{\alpha\nu}$; (iii) spin-flip interaction between particles at the two Fermi edges. The bosonization of these terms leads to a sine-Gordon term (\ref{Hg}) which vanishes logarithmically when reducing the width of the linearized bands around the Fermi points. In the SU(2)-symmetric case, $K_s$ approaches unity simultaneously with the reduction of $g$ in Eq.~\ref{Hg} to zero; (iv) spin-flip terms of the type $e^{-\sqrt{2} i ( \phi_s + \theta_s)} b_s^\dag d_s$ which transfer two left--moving spinons with $k \approx 0$ to the $d_s$-impurity band ($k_s<0$) and to its mirror image ($k\approx -k_d$, the ``$b_s$-band''). The process is illustrated in Fig.~\ref{fig:KondoTermSubbands}. These terms are the finite-$k$ counterparts of the sine-Gordon term in Eq.~(\ref{Hg}). Similar to the latter, these terms vanish in the limit of low energy of excitations.

Therefore, considering an interaction term of the form (\ref{MI_Hint}) is sufficient. The interacting Hamiltonian (\ref{MI_H0})-(\ref{MI_Hint}) can be diagonalized using a unitary transformation,
\begin{align}\label{U}
 U = \exp\left\{ - i \int_{-\infty}^\infty dx \sum_{\alpha \nu} \Delta \delta_{\alpha\nu} N_{\alpha\nu}(x) d^\dag_s(x) d_s(x) \right\},
\end{align}
where $N_{\alpha\nu} = \int^x dy \trho_{\alpha\nu}(y)$. The unitary operator $U$ is characterized by the phase shifts
\begin{align}\label{Deltadelta}
 \Delta \delta_{\alpha\nu} = \frac{\tV_{\alpha\nu}(k)}{v_d - \alpha v_\nu}.
\end{align}
This transformation removes the interaction term, $U^\dag ( H_0 + H_d + H_{int}) U = H_0 + H_d$. The $d_s$-operator in the rotated basis reads
\begin{align}
 U^\dag d_s(x) U = d_s(x) \exp\left\{ - i \int_{-\infty}^x dy \sum_{\alpha\nu} \Delta \delta_{\alpha\nu} \trho_{\alpha\nu}(y) \right\}.
\end{align}
One can see from Eq.~(\ref{string}) that $\Delta \delta_{\alpha\nu}$ simply adds to the phase $\delta_{\alpha\nu}$ which resulted from the refermionization, to produce the total shift $\delta^*_{\alpha \nu} = \delta_{\alpha\nu} + \Delta \delta_{\alpha\nu}$. Power-law singularities at the spinon mass shell in all dynamic response functions will be characterized by the phases $\delta^*_{\alpha\nu}$. The hole sector ($\omega < 0$) of the spectral function at momentum $k$ is given by
\begin{align}
 A(k,\omega < 0) &= \frac{1}{2\pi} \int dt dx e^{i k x - i\omega t} \notag \\
 &\times \expcts{ \psi^\dag_{R\sigma}(x,t) \psi_{R\sigma}(0,0)}_{H_0 + H_d + H_{int}},
\end{align}
and is independent of $\sigma$. It can be calculated by representing the physical fermions as $\psi_{R\uparrow} \sim e^{i k x} \tpsi_{Rc} F_{R c} d_s F_{Rs}$. The correlation function of the impurity after application of the unitary transformation (\ref{U}) is
\begin{align}
 \expct{d_s^\dag(x,t) d_s(0,0)}_{H_d} &= e^{i \epsilon_s(k) t} \delta(x - v_d t),
\end{align}
and the correlation functions involving the string operators can be calculated by using the bosonized expressions. One finds that $A(k,\omega) \propto [\omega - \epsilon_s(k)]^{-\mu^s_{0,-}}$ for $\omega \approx \epsilon_s(k)$, with the exponent
\begin{align}\label{mu_s0-}
 \mu^s_{0,-} &= 1 - \left(\frac{\delta^*_{+c}}{2\pi} - 1\right)^2 -
 \left(\frac{\delta^*_{-c}}{2\pi}\right)^2 \notag \\
&-
 \left(\frac{\delta^*_{+s}}{2\pi}\right)^2 -
 \left(\frac{\delta^*_{-s}}{2\pi}\right)^2.
\end{align}
We have seen previously that the cubic spectrum of the spinons means that the phase shift $\Delta \delta_{+s}$ remains finite even for $k \to k_F$ and cannot be treated as a perturbation. Luckily however, $SU(2)$-symmetry can by used to fix $\delta^*_{\pm s} = \delta_{\pm s} + \Delta \delta_{\pm s}$ for arbitrary momenta. As we shall show in Sec.~\ref{sec:DSF}, the exponents of the power-law singularities in the spin correlation functions $S^{zz}(k,\omega)$ and $S^{-+}(k,\omega)$ can be calculated as functions of $\delta^*_{\pm s}$. In an $SU(2)$-invariant system both exponents have to coincide for all momenta and this is only the case for
\begin{align}
    \delta^*_{\pm s} = 0.
\end{align}
As we laid out after Eq.~(\ref{Galphanu}), the phase shifts $\delta^*_{\pm s}$ can be interpreted as the probability amplitude of a shake-up of the spinons near the Fermi points. Vanishing phases $\delta^*_{\pm s} = 0$ mean that the creation of the spinon $d_s$ due to the incoming hole does not lead to such a shake-up and thus to a formation of particle-hole pairs in the spinon sector. For $|\omega| \gtrsim |\epsilon_s(k)|$, any excess energy can only be used to create particle-hole pairs in the holon sector. In this important sense, spin-charge separation remains meaningful even for systems with band curvature.

\section{Dynamic correlation functions away from the Fermi points}\label{sec:phen}

\subsection{Phenomenology for the scattering phase shifts}

In the limit $k \to k_F$, the edge exponent can be calculated from Eq.~(\ref{mu_s0-}) by using $\Delta \delta_{\pm c} = 0$ in addition to the relation $\delta^*_{\pm s}=0$ which is valid at arbitrary $k$. It turns out that $\mu^s_{0,-}$, unlike $\mu^c_{0,-}$, coincides with the corresponding LL exponent
\begin{align}
 \mu^s_{0,-} = \mu^s_{-}.
\end{align}
Therefore, for $\omega \approx \epsilon_s(k)$, one finds $A(k,\omega) \propto [\omega - \epsilon_s(k)]^{-\mu^s_{-}}$ even in the presence of band curvature.

In this section, we shall extend this result to momenta $k$ away from the Fermi points. We argued that for repulsive interactions, the edge of support of $A(k \approx \pm k_F,\omega)$ coincides with the spinon mass shell $\omega \approx \epsilon_s(k)$, and at this edge $A(k,\omega)$ is characterized by a power-law singularity. The state responsible for this singularity contains a spinon on mass shell and a holon at the Fermi point, so the mobile impurity $d_s$ has the quantum numbers of a spinon. Since the edge exponents must be continuous as a function of $k$, this must still be true for momenta away from the Fermi points. Hence, in order to calculate the dynamic response functions in this region, we can still make use of the same mobile-impurity Hamiltonian but in contrast to the limit $k \to k_F$, we can no longer use the lowest-order expansion of the phase shifts $\Delta \delta_{\pm c}$. Instead, we shall derive phenomenological expressions which relate the phase shifts to measurable properties of the spinon spectrum $\epsilon_s(k)$. In addition to the possibility of directly measuring $\epsilon_s(k)$, this function can be evaluated numerically by well-developed routines. It can also be calculated exactly for integrable models.

Let us focus again on the case $k < k_F$ and investigate the spectral function in the vicinity of the spinon spectrum $\omega \approx \epsilon_s(k)$. The configuration responsible for the edge singularity contains a spinon impurity near $k_d = k - k_F$, a holon close to the Fermi point, as well as particle-hole pairs of spinons and holons at the Fermi edges. The effective Hamiltonian is given by Eqs.~(\ref{MI_H0})-(\ref{MI_Hint}). We express the quasiparticle densities $\trho_{\alpha \nu}(x)$ in terms of bosonic operators $\theta_\nu$ and $\phi_\nu$. The interaction between the impurity $d_s$ and the low-energy spinons $\trho_{\alpha s}$ vanishes because $\delta_{\pm s}^* = 0$. The Hamiltonian $H = H_0 + H_d + H_{int}$ can be written as
\begin{align}\label{Hint2}
 H_0 &= \frac{v_c}{2\pi} \int dx \left[ K_c (\nabla \theta_c)^2 + \frac{1}{K_c} (\nabla \phi_c)^2 \right] \notag \\
&+ \frac{v_s}{2\pi} \int dx \left[ (\nabla \theta_s)^2 + (\nabla \phi_s)^2 \right], \notag \\
 H_d &= \int dx\ d^\dag_s(x) \left[ \epsilon_s(k) - i v_d \nabla \right] d_s(x), \notag \\
 H_{int} &= \int dx \left[ V_R \nabla \frac{\theta_c - \phi_c}{2 \pi} - V_L
\nabla \frac{\theta_c + \phi_c}{2\pi} \right] d_s^\dag
 d_s,
\end{align}
and contains two parameters, $V_L(k)$ and $V_R(k)$, which describe the interaction between the impurity and holons at the Fermi points. The interaction term can be removed using a unitary transformation like Eq.~(\ref{U}) and leads to the following relations between phase shifts and interaction constants,
\begin{align}\label{Vdelta}
 \sqrt{K_c} (V_L + V_R) &= - \Delta\delta_{-c} (v_d + v_c) - \Delta\delta_{+c} (v_d - v_c), \notag \\
 \frac{V_L - V_R}{\sqrt{K_c}} &= - \Delta\delta_{-c} (v_d + v_c) + \Delta\delta_{+c}(v_d - v_c).
\end{align}
In order to fix $V_L$ and $V_R$, we need two relations. The first one can be derived by considering the response of the system to a uniform charge density variation. Since the spinon spectrum $\epsilon_s(k)$ is defined with respect to the chemical potential $\mu$, for fixed $k$ the shift in the edge position upon a variation of the density $\delta \rho_c$ is given by
\begin{align}\label{shift1}
 \delta E = \left[\frac{\partial \epsilon_s(k)}{\partial \rho_c} + \frac{\partial \mu}{\partial \rho_c} \right] \delta \rho_c = \left[\frac{\partial \epsilon_s(k)}{\partial \rho_c} + \frac{\pi v_c}{2 K_c} \right]  \delta \rho_c\ .
\end{align}
In the last equality, we used a phenomenological relation\cite{giamarchi03} between $K_c$ and the compressibility $\kappa = \partial \rho_c/\partial \mu = 2 K_c / (\pi v_c)$. A second way to calculate the same energy shift is to use the Hamiltonian (\ref{Hint2}). According to Eq.~(\ref{rho_phi}), a density variation leads to a finite expectation value $\expct{\nabla \phi_c}{} = - \pi \delta \rho_c / \sqrt{2}$. We use the Hamiltonian (\ref{Hint2}) to calculate the energy of a state containing a spinon impurity $d_s$ with momentum $k_d$ and a holon at the Fermi edge. Then, we investigate the change of this energy due to the density variation. In this case, the interaction term (\ref{Hint2}) shifts by
\begin{align}
 \delta E_{int} &= \frac{V_R + V_L}{2\pi} \expct{\nabla \phi_c}{} = - \frac{V_R + V_L}{2\sqrt{2}} \delta \rho_c.
\end{align}
The impurity momentum is given by $k_d = k - k_F$. As the density variation affects the Fermi momentum, $\delta k_F = \pi \delta \rho_c/2$, one obtains an energy shift in the operator $H_d$ by
\begin{align}
 \delta E_{d} = -\frac{\pi v_d \delta \rho_c}{2} = -\frac{\pi}{2} \frac{\partial \epsilon_s(k)}{\partial k}  \delta \rho_c.
\end{align}
Finally, the energy of the holon at the right Fermi point changes due to the shift of the chemical potential and leads to $\delta E_0 = (\partial \mu/\partial \rho_c) \delta \rho_c$. Comparing the shift (\ref{shift1}) to the shifts $\delta E_0 + \delta E_d + \delta E_{int}$ calculated using the effective Hamiltonian, one obtains the first relation,
\begin{align}\label{Vsum}
 -\frac{V_R + V_L}{2\sqrt{2}} = \frac{\partial \epsilon_s(k)}{\partial \rho_c} + \frac{\pi}{2} \frac{\partial \epsilon_s(k)}{\partial k}.
\end{align}
In order to obtain a second relation which fixes the difference $V_L - V_R$, we consider the effect of uniform Galilean boost of the system. Let us assume the liquid is moving at velocity $\delta u$ and the incoming particle has momentum $k$. In a reference frame moving with the liquid, the momentum of the injected particle is $k - m \delta u$, where $m$ is the bare mass of the physical fermions. The total change of the energy acquired by the liquid due to the boost is given by
\begin{align}\label{shift2}
 \delta E' &= \frac{k^2}{2m} - \epsilon_s(k) - \frac{(k - m \delta u)^2}{2m} + \epsilon_s(k - m \delta u) \notag \\
&=
 \delta u \left[ k - m \frac{\partial \epsilon_s(k)}{\partial k} \right].
\end{align}
Next, we calculate the same shift using the Hamiltonian (\ref{Hint2}). The effect of a boost with velocity $\delta u$ on the physical fermions is to shift the Fermi momentum of right- and left-movers, $k_F^{R,L} \to k_F \pm m \delta u$. This gives rise to a difference between right-mover and left-mover density, $\rho_R - \rho_L = 2 m \delta u/\pi$. This density difference corresponds to a finite expectation value $\expct{\nabla \theta_c}{} = \sqrt{2} m \delta u$. We use again the Hamiltonian (\ref{Hint2}) to calculate to energy of a state containing an impurity $d_s$ and a holon at the Fermi point. Now, we investigate how this energy changes due to a finite $\expct{\nabla \theta_c}{} = \sqrt{2} m \delta u$. In the interaction Hamiltonian $H_{int}$, one obtains
\begin{align}
 \delta E'_{int} = \frac{V_L - V_R}{2\pi} \expct{\nabla \theta_c} = \sqrt{2} m \delta u \frac{V_L - V_R}{2\pi}.
\end{align}
The energy shift due to the holon at the Fermi point is given by $\delta E'_0 = K_c v_c m \delta u = v_F m \delta u$. Finally, the momentum of the impurity $k_d$ changes due to the shift of $k_F$ and leads to $\delta E'_d = - m \delta u v_d$. Equating the shifts $\delta E'_0 + \delta E'_d + \delta E'_{int}$ with (\ref{shift2}) leads to
\begin{align}\label{Vdiff}
 \frac{V_L - V_R}{2\pi} = \frac{ k - k_F}{\sqrt{2} m}.
\end{align}
Equations~(\ref{Vsum}) and (\ref{Vdiff}) allow us to fix the interaction strengths $V_L$ and $V_R$ in terms of the derivatives of $\epsilon_s(k)$ with respect to the density and the momentum. These, in turn, can be related to the phase shifts using Eq.~(\ref{Vdelta}). The result is
\begin{align}\label{phen_final}
 \frac{\Delta \delta_{\pm c}(k)}{2\pi} = \pm \frac{\frac{k - k_F}{m \sqrt{K_c}}
\pm \sqrt{K_c} \left( \frac{2}{\pi} \frac{\partial \epsilon_s(k)}{\partial \rho_c}
+ \frac{\partial \epsilon_s(k)}{\partial k} \right)}{2 \sqrt{2} \left( \frac{\partial \epsilon_s(k)}{\partial k} \mp \frac{k_F}{m K_c}
 \right)},
\end{align}
and it is valid for all systems with Galilean-invariant microscopic interactions. These relations are valid for all $-k_F < k < k_F$. The knowledge of these phase shifts allows the calculation of all dynamic response functions for energies close to the spinon mass shell at arbitrary momenta. Note that for the case of the exactly solvable 1D Yang-Gaudin model,\cite{yang67,gaudin67} the phase shifts predicted by Eq.~(\ref{phen_final}) coincide with the exact results obtained using the Bethe ansatz (which can be obtained as a limiting case of 1D Hubbard model considered in [\onlinecite{essler10}]).

\subsection{Edge exponents of the spectral function}\label{sec:spin_exp}

\begin{figure}[t]
  \centering
  \includegraphics[width = 0.48 \textwidth]{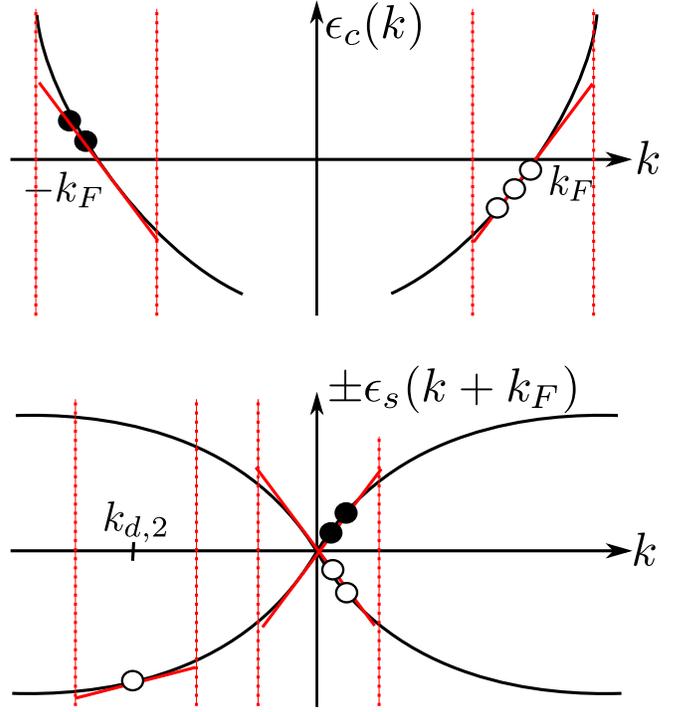}
\caption{(Color online) Mobile-impurity band structure for the calculation
  of $A(k,\omega)$ for $3 k_F < k < 5 k_F$ (i.e., $n = 2$) and 
  $\omega \approx |\epsilon_s(k - 4 k_F)|$. The final state contains a spinon
  with momentum $k_{d,2} = k - 5 k_F < 0$, a holon at its Fermi
  momentum as well as two particle-hole pairs in 
  the holon sector which absorb the extra momentum $4 k_F$. In this figure, we
  assume $m_- = 2$, so the spinon sector contains two additional
  particle-hole pairs. The Hamiltonian is projected onto narrow bands
  around these momenta. Within each band, the spectrum can be linearized.}
  \label{fig:bandstructure}
\end{figure}

\begin{figure}[t]
  \centering
  \includegraphics[width = 0.48 \textwidth]{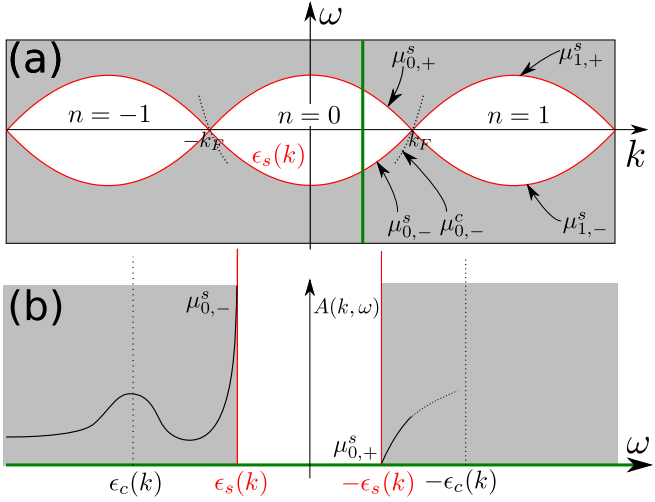}
\caption{(Color online) Spectral function $A(k,\omega)$ in the $(k,\omega)$-plane (a) and along a cross-section for fixed $0 < k < k_F$ (b). (a) $A(k,\omega)$ is nonzero in the shaded areas. For repulsive interactions, the edge of support is at the spinon mass shell $\omega = \pm |\epsilon_s(k)|$. At the edge for $\omega \gtrless 0$ in the momentum range $(2n - 1) k_F < k < (2 n + 1) k_F$, $A(k,\omega)$ has a power-law singularity characterized by the exponent $\mu^s_{n,\pm}$. A sharp power-law at the holon mass shell $\epsilon_c(k)$ (dotted lines) exists only for $k \approx (2n \pm 1) k_F$. The corresponding exponents $\mu_{n,-}^c$ are continuous at $\omega = 0$. (b) Compared to the Luttinger liquid case, the edge exponents is modified. The singularity at the holon mass shell $\epsilon_c(k)$ is smeared out because the band curvature leads in general to a finite holon lifetime.}
  \label{fig:SpectralFunctionNLL}
\end{figure}

The phase shifts $\delta^*_{\alpha\nu}$ allow us to derive the edge exponents of $A(k,\omega)$ for momenta $k$ away from $\pm k_F$. In the following, we shall focus on a general momentum interval $(2 n - 1) k_F < k < (2 n + 1) k_F$ (with integer $n$) and energies near the spinon mass shell, $\omega \approx \epsilon_s(k - 2 n k_F)$ and its shadow bands, $\omega \approx - \epsilon_s(k - 2n k_F)$, see Fig.~\ref{fig:SpectralFunctionNLL}. Similar to Sec.~\ref{sec:QI}, for $\omega < 0$ injecting a hole into the system will create a spinon impurity with momentum near $k_{d,n} = k - ( 2n + 1) k_F$ on its mass shell as well as a holon at approximately the Fermi momentum. The remaining momentum $2 n k_F$ must be given to additional spinon and holon particle-hole excitations near the Fermi edges, see Fig.~\ref{fig:bandstructure}. In terms of the physical fermions, the most general configuration of additional particle-hole pairs reads as follows,
\begin{align}\label{nparticles}
 \Phi_{2nk_F} := (\psi_{R\uparrow})^{n_{R\uparrow}}
 (\psi_{R\downarrow})^{n_{R\downarrow}}
 (\psi_{L\uparrow})^{n_{L\uparrow}}
 (\psi_{L\downarrow})^{n_{L\downarrow}}\ .
\end{align}
All $n_{\alpha\sigma}$ are integer numbers. Positive values correspond to powers of the annihilation operators $\psi_{\alpha\sigma}$, whereas negative numbers denote powers of the creation operators $\psi^\dag_{\alpha\sigma}$. Since spin and charge of the incoming hole are absorbed by the deep spinon at momentum $k_{d,n}$ and the holon at the Fermi point, the total charge and total spin of these additional excitations must vanish. This means
\begin{align}\label{nsum}
 n_{R\uparrow} + n_{R\downarrow} + n_{L\uparrow} + n_{L\downarrow} &= 0, \notag \\
 n_{R\uparrow} - n_{R\downarrow} + n_{L\uparrow} - n_{L\downarrow} &= 0.
\end{align}
From these equations, one finds $n_{L\uparrow} = - n_{R\uparrow}$ and $n_{L\downarrow} = - n_{R\downarrow}$. As an additional requirement, the excess momentum $2 n k_F$ must be accommodated. This leads to the constraint $n_{R\uparrow} + n_{R\uparrow} - n_{L\uparrow} - n_{L\downarrow} = 2 n$ and therefore to $n_{R\uparrow} + n_{R\downarrow} = n$. This leaves one free parameter $m_-$, which must satisfy the selection rule
\begin{align}\label{selm_minus}
 m_- \equiv n (\mod 2).
\end{align}
The general solution for integer $n_{\alpha\sigma}$ reads
\begin{align}
 n_{R\uparrow} &= \frac{n + m_-}{2}, \notag \\
 n_{R\downarrow} &= \frac{n - m_-}{2}.
\end{align}
Therefore, the state (\ref{nparticles}) can be written as
\begin{align}\label{nparticles2}
  \Phi_{2 n k_F} = (\psi^\dag_{R\uparrow} \psi_{R\downarrow} \psi_{L\uparrow} \psi^\dag_{L\downarrow})^{-(n+m_-)/2}
 (\psi^\dag_{L\downarrow} \psi_{R\downarrow})^n\ .
\end{align}
Physically, the first term comes about as a result of spin-flip scattering. Indeed, the fermionic representation of the sine-Gordon term (\ref{Hg}) produces exactly this type of scattering. The second term comes about due to scattering between the right and left Fermi points and absorbs the excess momentum $2 n k_F$. The spectral function can be calculated by decomposing the fermionic operators according to
\begin{align}
 \psi_{R\uparrow} = e^{i k x} \tpsi_{Rc} F_{Rc} d_s F_{Rs} \times \Phi_{2n k_F}
\end{align}
and bosonizing $\Phi_{2 n k_F}$ and $\tpsi_{Rc}$ using Eqs.~(\ref{bos_iden}) and (\ref{referm}), respectively. One finds that in the interval $(2 n - 1) k_F < k < (2 n + 1) k_F$, it has a power-law singularity $A(k,\omega) \propto [ \omega + |\epsilon_s(k - 2 n k_F)|]^{-\mu^s_{n,-}}$ with the exponent
\begin{align}\label{mu_n}
 \mu^s_{n,-}
&=
 1 - \frac{1}{2} \left( -\frac{(2n + 1)
\sqrt{K_c}}{\sqrt{2}} + \frac{\Delta \delta_{+c} + \Delta \delta_{-c}}{2\pi} \right)^2 \notag \\
&-
\frac{1}{2} \left( \frac{1}{\sqrt{2 K_c}} - \frac{\Delta \delta_{+c} - \Delta \delta_{-c}}{2\pi}
\right)^2 - m_-^2\ ,
\end{align}
where $\Delta \delta_{\pm c} \equiv \Delta \delta_{\pm c}(k - 2 n k_F)$ is evaluated for momenta on the main spinon branch, $-k_F < k - 2n k_F < k_F$. Since the selection rule requires integer $m_-$, the leading exponent is given by $m_- = 0$ for even $n$, and $m_- = \pm 1$ for odd $n$.

A similar line of reasoning can be applied to calculate the edge exponents $\mu^s_{n,+}$ for $(2n -1)k_F < k < (2n+1)k_F$ and $\omega > 0$ (see Fig.~\ref{fig:SpectralFunctionNLL}). The configuration with this combination of momentum and energy contains a spinon with momentum $k_d = k - (2n - 1) k_F > 0$ as well as a holon near the Fermi point. Similar to the previous case, the remaining momentum $2(n-1) k_F$ must be absorbed by additional particle-hole excitations of the form (\ref{nparticles}). Stipulating again charge and spin neutrality leads to a modified selection rule $m_+ \equiv n + 1 (\mod 2)$. As a result, one obtains $A(k,\omega) \propto [ \omega - |\epsilon_s(k - 2 n k_F)|]^{-\mu^s_{n,+}}$ where $\mu^s_{n,+}$ is similar to Eq.~(\ref{mu_n}),
\begin{align}\label{mu_np}
 \mu^s_{n,+}
&=
 1 - \frac{1}{2} \left( -\frac{(2n + 1)
\sqrt{K_c}}{\sqrt{2}} + \frac{\Delta \delta_{+c} + \Delta \delta_{-c}}{2\pi} \right)^2 \notag \\
&-
\frac{1}{2} \left( \frac{1}{\sqrt{2 K_c}} - \frac{\Delta \delta_{+c} - \Delta \delta_{-c}}{2\pi}
\right)^2 - m_+^2\ ,
\end{align}
except for the different selection rule for $m_+$.

Using the formula for the phase shifts (\ref{phen_final}), it is thus possible to express the exponents (\ref{mu_n}) and (\ref{mu_np}) entirely in terms of the spinon spectrum $\epsilon_s(k)$ and the Luttinger parameters. This establishes a relation between two distinct sets of observable quantities, exponents and spectra, which applies to a large range of systems and which can in principle be checked in experiments.

As a consequence of the symmetry of the spinon spectrum $\epsilon_s(k) = \epsilon_s(-k)$, the edge exponents also satisfy the $k \to -k$ symmetry, $\mu^s_{n,\pm}(k) = \mu^s_{-n,\pm}(-k)$. Moveover, since the phase shifts $\Delta \delta_{\pm c}$ vanish for $k \to k_F$, one can also verify that the edge exponents change continuously when crossing a Fermi point,
\begin{align}
 \mu^s_{n,\pm}[(2n+1)k_F - 0] &= \mu^s_{n+1,\mp}[(2n+1) k_F +0].
\end{align}
The final result\cite{schmidt09_3} for the edge exponents of $A(k,\omega)$ in all regions of the $(k,\omega)$-plane is shown in Table \ref{tab:Exponents}. The positions of the edges and the corresponding notations for the exponents are shown in Fig.~\ref{fig:SpectralFunctionNLL}.

\subsection{Charge and spin density structure factors}\label{sec:DSF}

The charge density structure factor is defined as
\begin{align}\label{DSF}
 S(k,\omega) = \int dx dt e^{i \omega t - i k x} \expct{\rho_c(x,t) \rho_c(0,0)}{},
\end{align}
where $\rho_c(x) = \sum_\sigma \psi^\dag_\sigma(x) \psi_\sigma(x)$ denotes the charge density. In the case of a linear spectrum, spin and charge density waves with momentum $k$ have a uniquely defined energy $v_{s,c} k$. One of the predictions of the linear LL theory is that even for nonzero interactions, these are stable excitations. Therefore, one finds $S_{LL}(k,\omega) = 2 K_c |k| \delta(\omega - v_c |k|)$ where the effect of interactions is limited to a renormalization of the velocity $v_c \neq v_F$ and the prefactor is fixed by the $f$-sum rule.\cite{nozieres_book}

\begin{figure}[t]
  \centering
  \includegraphics[width = 0.48 \textwidth]{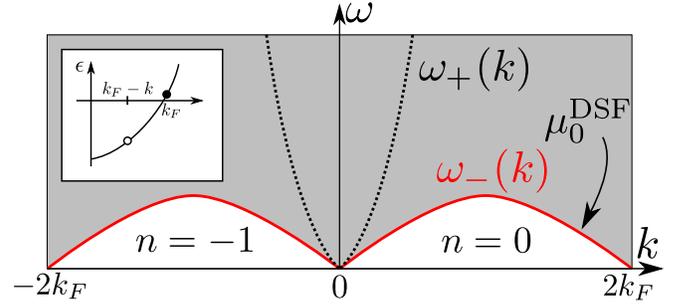}
\caption{(Color online) Density structure factor $S(k,\omega)$. For noninteracting systems $S(k,\omega)$ is constant for $\omega_-(k) < \omega < \omega_+(k)$ and vanishes otherwise. For nonzero interactions, a power-law singularity with exponent $\mu_n^\text{DSF}$ forms at the lower threshold $\omega_-(k)$. \emph{In the inset:} Particle-hole excitation with momentum $k$ giving rise to the singularity at $\omega_-(k)$.}
  \label{fig:S}
\end{figure}

In the case of finite band curvature, a one-to-one relation between momentum and energy of particle-hole pairs no longer exists. Let us illustrate the consequences for the noninteracting case. The fermion spectrum is given by $\epsilon(k) = (k^2 - k_F^2)/(2m)$ and it turns out that for $0 < k < 2 k_F$, $S(k,\omega)$ is nonvanishing only in the interval $\omega_-(k) < \omega < \omega_+(k)$, where $\omega_\pm(k) = v_F k \pm k^2/(2m)$, see Fig.~\ref{fig:S}. The density excitation of lowest energy $\omega_-$ for given momentum $k$ contains a particle at the Fermi level and a hole at $k_F - k$ (see inset of Fig.~\ref{fig:S}). The upper threshold $\omega_+$ reflects the highest energy such an excitation can have. It corresponds to a particle at $k_F + k$ and a hole at the Fermi level.

For spinless weakly interacting systems, a power-law singularity develops at $\omega = \omega_-(k)$ but the structure factor still vanishes below it, $S(k,\omega) = 0$ for $\omega < \omega_-(k)$.\cite{pustilnik06} The upper threshold, on the contrary, no longer exists since density excitations can give away excess energy to create other particle-hole excitations. Instead, for $\omega \gg \omega_+(k)$, the density structure factor decays as a power-law. A schematic picture of the domain of support of $S(k,\omega)$ in the $(k,\omega)$-plane is shown in Fig.~\ref{fig:S}.

For spinful systems, the charge density structure factor describes responses in the charge sector. Its analog in the spin sector is the spin structure factor. Since the calculations of both functions are identical and the invariance of the spin structure under spin rotations allows us to fix the phase shift $\delta^*_{\pm s}$, we shall henceforth focus on the latter.

In the absence of a magnetic field, the system Hamiltonian is $SU(2)$-invariant. One of the consequences of this symmetry is identical power laws in different components of the spin structure factor. In particular, the functions
\begin{align}\label{SpinCorrelations}
 S^{-+}(k,\omega) &= \int dx dt e^{i \omega t - i k x} \expct{ S^-(x,t) S^+(0,0)}{}, \notag \\
 S^{zz}(k,\omega) &= \int dx dt e^{i \omega t - i k x} \expct{ S^z(x,t) S^z(0,0)}{}
\end{align}
must be identical even in the presence of band curvature. In terms of the physical fermions, the spin density is given by $\vec{S}(x) = \frac{1}{2} \sum_{\sigma,\sigma'} \psi_\sigma(x) \vec{\tau}_{\sigma\sigma'} \psi_{\sigma'}(x)$, where $\vec{\tau}$ denotes the vector of Pauli matrices.

To be specific, let us first investigate $S^{-+}(k,\omega)$ in the $n=0$ band, i.e., for $0 < k < 2k_F$. Similar to the dynamic structure factor $S(k,\omega)$, the spin structure factor will vanish below a threshold energy $\omega_-(k)$ and have a power-law singularity at this threshold. The excitation which is responsible for the singularity contains a spinon impurity at momentum $k$ as well as a spinon hole with small momentum. Therefore, we can project the spinon operator on two bands using $\tpsi_{Rs}(x) \to \tpsi_{Rs}(x) + e^{i k x} d_s(x)$. For $S^+_R(x)$, one finds
\begin{align}\label{SpR}
 S^+_R(x)
&=
 \psi_{R\uparrow}^\dag(x) \psi_{R\downarrow}(x)\\
&\propto
 e^{-i k x} d_s^\dag F^\dag_{Rs} \tpsi^\dag_{Rs} F^\dag_{Rs} \notag \\
&\propto
 e^{-i k x} d_s^\dag \exp\left\{ i \left(\frac{\delta_{+s}}{2\pi} - \frac{1}{\sqrt{2}} \right) [\tilde{\theta}_s(x) - \tilde{\phi}_s(x) ] \right\}. \notag
\end{align}
For the $z$-component of the spin density, we use
\begin{align}
 S^z_R(x)
&= \frac{1}{2} \left[ \psi^\dag_{R\uparrow}(x) \psi_{R\uparrow}(x) - \psi^\dag_{R\downarrow}(x) \psi_{R\downarrow}(x) \right] \notag \\
&=
 \psi^\dag_{R\uparrow}(x) \psi_{R\uparrow}(x)  - \frac{1}{2} \rho_{Rc}(x),
\end{align}
where $\rho_{Rc} = \psi^\dag_{R\uparrow} \psi_{R\uparrow} + \psi^\dag_{R\downarrow} \psi_{R\downarrow}$ denotes the right-moving charge density. If the spectrum of the physical fermions is linear, the spin-charge separation ensures that $\rho_{Rc}$ does not involve spinon operators. For a nonlinear spectrum, this strict separation does not apply any more and $\rho_{Rc}$ may indeed contain spinon operators. This fact will become important in the calculation of the charge density structure factor. Since all terms lead to the same series of exponents, it is sufficient to only project the first term and use
\begin{align}\label{SRz}
 S^z_R(x)
&\propto
 e^{-i k x} d_s^\dag F^\dag_{Rs} \tpsi_{Rs} F_{Rs} + \text{h.c.} \\
&\propto
 e^{-i k x} d_s^\dag \exp\left\{ i \left(\frac{\delta_{+s}}{2\pi} + \frac{1}{\sqrt{2}} \right) [\tilde{\theta}_s(x) - \tilde{\phi}_s(x) ] \right\} \notag \\
&+ \text{h.c.} \notag
\end{align}
The ensuing calculation of the exponent of $S^{-+}$ is performed using the mobile impurity Hamiltonian (\ref{MI_H0})-(\ref{MI_Hint}) and general phase shifts $\delta^*_{\alpha\nu} = \delta_{\alpha\nu} + \Delta \delta_{\alpha\nu}$ and results in the leading edge exponent $\mu^{-+}_{0,0}$.

In analogy to the previous section, the marginally irrelevant spin-flip scattering (\ref{Hg}) may create a final state with identical energy and momentum but which contains additional particle-hole pairs and thus produces subleading exponents. To calculate these, the operators $S_R^+$ and $S_R^z$ must be multiplied by the operator $\Phi_{2 n k_F}$ (\ref{nparticles2}) for $n = 0$. A state arising from $|m|$ spin-flip scattering events will contribute the exponent (for $m  \in \mathbb{Z}$)
\begin{align}\label{mu0m_1}
 \mu^{-+}_{0,m} &= 1 - \frac{1}{2} \left( \frac{ 2 m + 1}{\sqrt{2}} - \frac{\delta^*_{+s} + \delta^*_{-s}}{2\pi} \right)^2 \\
 &- \frac{1}{2} \left(\frac{1}{\sqrt{2}} - \frac{\delta^*_{+s} - \delta^*_{-s}}{2\pi} \right)^2 \notag \\
 &- \frac{1}{2} \left(\frac{\Delta \delta_{+c} + \Delta \delta_{-c}}{2\pi}\right)^2
 - \frac{1}{2} \left(\frac{\Delta \delta_{+c} - \Delta \delta_{-c}}{2\pi}\right)^2.\notag
\end{align}
where $\delta^*_{\pm s}$ is evaluated at momentum $k \in [0, 2 k_F]$ and $\Delta \delta_{\pm c}$ at momentum $k_F - k$. An analogous calculation yields the leading and subleading exponents for $S^{zz}$,
\begin{align}\label{mu0m_2}
 \mu^{zz}_{0,m} &= 1 - \frac{1}{2} \left( \frac{ 2 m - 1}{\sqrt{2}} - \frac{\delta^*_{+s} + \delta^*_{-s}}{2\pi} \right)^2 \\
 &- \frac{1}{2} \left(\frac{1}{\sqrt{2}} + \frac{\delta^*_{+s} - \delta^*_{-s}}{2\pi} \right)^2 \notag \\
 &- \frac{1}{2} \left(\frac{\Delta \delta_{+c} + \Delta \delta_{-c}}{2\pi}\right)^2
 - \frac{1}{2} \left(\frac{\Delta \delta_{+c} - \Delta \delta_{-c}}{2\pi}\right)^2.\notag
\end{align}
It should be emphasized that each of the exponents in Eqs.~(\ref{mu0m_1})-(\ref{mu0m_2}) actually generates an infinite ``ladder'' of exponents differing by an integer power, and Eqs.~(\ref{mu0m_1})-(\ref{mu0m_2}) give only the most divergent exponent. Indeed, when we calculate, e.g., the exponent of $S^{-+}(k,\omega)$, the expected behavior of $m$th term is
\begin{align}
    S^{-+}(k,\omega) \propto [\omega - |\epsilon_s(k_F-k)|]^{-\mu_{0,m}^{-+}} R[\omega - |\epsilon_s(k_F-k)|],
\end{align}
where $R$ is at most logarithmically divergent for $\omega - |\epsilon_s(k_F-k)| \to 0$. A Taylor expansion of this function generates an infinite ladder of exponents. $SU(2)$-symmetry requires that these full sets of exponents generated by different $\mu_{0,m}^{-+}$ and $\mu_{0,m}^{zz}$ should coincide. This is a weaker requirement compared to coincidence of sets of $\mu_{0,m}^{-+}$ and  $\mu_{0,m}^{zz}$. However, it turns out that both these constraints lead to the same requirement
\begin{align}
 \delta^*_{-s} = \delta^*_{+s} = 0.
\end{align}
By construction, this equality holds for arbitrary momenta. Note that in order to fix both phase shifts $\delta^*_{\pm s}$, a comparison of only the leading $(m = 0)$ exponents is not sufficient.

In the momentum interval $2 n k_F < k < 2 (n + 1) k_F$, a spinon impurity will be created at momentum $k - 2 n k_F$ and the second spinon near zero momentum. As in the calculation for $A(k,\omega)$, the excess momentum $2 n k_F$ is accommodated by scattering across the Fermi points. A number $|m|$ of spin-flip events will lead to a selection rule of the form (\ref{selm_minus}). If we focus only on the leading $(m = 0)$ exponent, we can conclude that the spin structure factor near its edge of support is given by $S^{-+}(k,\omega) \propto S^{zz}(k,\omega) \propto \{ \omega - |\epsilon_s[(2 n + 1) k_F - k]|\}^{-\mu_n^\text{SSF}}$ where
\begin{align}\label{mu_SSF}
 \mu_n^\text{SSF} &= \frac{1}{2} - \frac{1}{2} \left( \frac{2n
\sqrt{K_c}}{\sqrt{2}} + \frac{\Delta \delta_{+c} + \Delta \delta_{-c}}{2\pi} \right)^2 \notag \\
&-
\frac{1}{2} \left(\frac{\Delta \delta_{+c} - \Delta \delta_{-c}}{2\pi} \right)^2,
\end{align}
with $\Delta \delta_{\pm c} \equiv \Delta \delta_{\pm c}[(2n + 1) k_F - k]$. Here, the phase shifts are related to the spinon spectrum by Eq.~(\ref{phen_final}).

For the calculation of the charge density structure factor $S(k,\omega)$, the operator $\rho_c(x)$ needs to be examined. For a nonlinear spectrum of the physical fermions, this operator will contain spinon operators. Therefore, $\rho_c$ can create a state which contains a spinon with momentum $k$ on its mass shell and additional spinons and holons near the Fermi points. For a given momentum $k$, this state is the one with the least energy, so the edge of support of $S(k,\omega)$ coincides with the one for $S^{-+}(k,\omega)$ and $S^{zz}(k,\omega)$. Therefore, we project $\rho_c$ onto two subbands using again $\tpsi_{Rs}(x) \to \tpsi_{Rs}(x) + e^{i k x} d_s(x)$. Because the respective terms in the projection of the operator $\rho_c(x)$ are identical to the terms in the projection of $S^z_R(x)$ in Eq.~(\ref{SRz}), it follows that for $2 n k_F < k < 2 (n + 1) k_F$, near the edge of support $S(k,\omega) \propto \{ \omega - |\epsilon_s[(2 n + 1) k_F - k]|\}^{-\mu_n^\text{DSF}}$, and
\begin{align}\label{mu_DSF}
 \mu_n^\text{DSF} = \mu_n^\text{SSF}.
\end{align}
An overview of the edge exponents of the charge and spin density structure factors in the different regions of the $(k,\omega)$-plane is contained in Table~\ref{tab:Exponents}.\cite{schmidt09_3}

For small momenta $k \ll k_F$, the spin and charge density structure factors were investigated recently within the Abelian\cite{teber07} and non-Abelian\cite{pereira10} bosonization approaches. In the former approach, the nonlinear spectrum of fermions produces terms which mix the spin and charge fields in the bosonized Hamiltonian. The perturbation theory in these terms developed in Ref.~[\onlinecite{teber07}] corroborates one of our conclusions (and of Ref.~[\onlinecite{pereira10}]): the edge of support of $S(k,\omega)$ actually coincides with that of $S^{zz}(k,\omega)$. For repulsive interactions, it is located at the spinon mass shell, i.e., $\omega = v_s k$ for small momenta. However, the second-order bosonic self-energy actually diverges at this edge. As a consequence, it was not possible in Ref.~[\onlinecite{teber07}] to investigate the shape of the edge singularity. Despite this drawback, the Abelian bosonization provides an easy access to $S(k,\omega)$ and $S^{zz}(k,\omega)$ away from the singularities, for instance to the high-frequency tails for $\omega\gg v_c k$. These ``off-shell'' tails were studied in Ref.~[\onlinecite{teber07}]. The broadening of the peaks in $S(k,\omega)$ near the spinon and holon mass shells was addressed in Ref.~[\onlinecite{pereira10}]. Its conclusions regarding the threshold exponents agree with the limit $k\to 0$ of our results.

\section{Holon edge singularities}\label{sec:holon}

So far, we have only discussed those power law singularities which occur at the edges of support of a dynamical correlation function. We argued that in the case of repulsive interaction, the spinon-holon excitation with the lowest possible energy for a given momentum contains a holon at its Fermi edge, whereas the remaining momentum and the whole energy are carried by the spinon. Therefore, the exponents at the edges of support are characterized by the phase shifts $\Delta \delta_{\pm c}$ produced by the interaction of a spinon impurity with low-energy holons.

For energies above the edge of support it is also possible to create spinon-holon excitations which contain a spinon at its Fermi point and give the entire energy to a holon. For a generic system, such an excitation will not be stable since energy and momentum conservation allow the decay of a holon through the creation of spinon pairs. This will lead to a broadening of the threshold. One important exception is the case of integrable models, where holon excitations may be stable. But also the cases of very weak and very strong interaction allow long-lived holon excitations. In this section, we shall therefore develop the theory for holon edge singularities assuming the holon is stable.

Let us denote the holon threshold in the spectral function by $\epsilon_c(k)$. By definition, the configuration giving rise to this threshold contains a spinon at the Fermi point as well as a holon excitation of energy $\epsilon_c(k)$. A first step towards the calculation of the corresponding threshold exponents is generalizing Sec.~\ref{sec:QI} to calculate the phase shifts caused by such a holon impurity. The derivation is very similar to the spinless case.\cite{imambekov09} The Hamiltonians $H_0$ and $H_d$ describing, respectively, the LLs at the Fermi points and the holon impurity are given by
\begin{align}\label{MI_Hdc}
 H_0 &= \frac{v_c}{2\pi} \int dx \left[ K_c (\nabla \theta_c)^2 + \frac{1}{K_c} (\nabla \phi_c)^2 \right] \notag \\
&+ \frac{v_s}{2\pi} \int dx \left[ (\nabla \theta_s)^2 + (\nabla \phi_s)^2 \right], \notag \\
H_d &= \int dx\ d_c^\dag(x) [ \epsilon_c(k) - i v_d \nabla ] d_c(x),
\end{align}
where $v_d = \partial \epsilon_c(k)/\partial k$. Interactions between the holon impurity $d_c$ and the spinons at the Fermi points do not lead to phase shifts because of $SU(2)$-symmetry: a density-density interaction of the type $\trho_{\alpha s}(x) d_c^\dag(x) d_c(x)$, being linear in spinon density would violate spin-up/spin-down symmetry. Therefore, the interaction term only contains holon-holon interaction and becomes in bosonized form
\begin{align}\label{MI_Hintc}
 H_{int} = \int dx \left[ V'_R \nabla \frac{\theta_c - \phi_c}{2 \pi} - V'_L
\nabla \frac{\theta_c + \phi_c}{2\pi} \right] d^\dag_c
 d_c.
\end{align}
Removing the interaction term by means of a unitary transformation as in the spinon case leads to phase shifts. In order to distinguish them from the spinon phase shifts, we label them $\chi_{\pm}$. In analogy to Eq.~(\ref{Vdelta}), these phase shifts are determined by the equations
\begin{align}\label{Vdelta_c}
 (V'_L \mp V'_R)K^{\mp 1/2}_c &= - \chi_{-} (v_d + v_c) \pm \chi_{+} (v_d - v_c).
\end{align}
In order to express the interaction potentials $V'_{L,R}$ in terms of measurable quantities, we consider again the variations of the energy of the system with respect to a variation of the density and to a Galilean boost using the Hamiltonian $H_0 + H_d + H_{int}$. A uniform density variation $\delta \rho_c$ leads to a nonzero $\expcts{\nabla \phi_c} = - \pi \delta \rho_c/\sqrt{2}$ and thus shifts $H_{int}$ by
\begin{align}
 \delta E_{int} = -\frac{V'_R + V'_L}{2\sqrt{2}} \delta \rho_c.
\end{align}
The density variation also causes a change in the holon Fermi momentum but this does not lead to an energy shift in $H_0$ and $H_d$. Calculating the same shift in energy using the definition of $\epsilon_c(k)$ [see Eq.~(\ref{shift1})] and equating both expressions leads to
\begin{align}
 - \frac{V'_{R} + V'_{L}}{2\sqrt{2}} = \frac{\partial \epsilon_c(k)}{\partial \rho_c} + \frac{\pi v_c}{2 K_c}.
\end{align}
Next, we consider again the energy shift due to a uniform change of momentum. For a system moving at velocity $\delta u$, this leads to a finite $\expct{\nabla \theta_c}{} = \sqrt{2} m \delta u$. Equating the corresponding shift in $H_{int}$ with the shift calculated from the definition of $\epsilon_c(k)$ [see Eq.~(\ref{shift2})] leads to the second relation,
\begin{align}
 \frac{V'_{L}-V'_{R}}{2\pi} = \frac{1}{\sqrt{2}} \left( \frac{k}{m} - \frac{\partial \epsilon_c}{\partial k} \right).
\end{align}
The phase shifts can now be determined from Eq.~(\ref{Vdelta_c}). The result is
\begin{align}\label{xi}
 \frac{\chi_{\pm}(k)}{2\pi} &= \frac{1}{2 (\pm \frac{\partial \epsilon_c}{\partial k} - v_c)}
 \bigg\{ \frac{1}{\sqrt{2 K_c}} \left[ \frac{k}{m} - \frac{\partial \epsilon_c}{\partial k} \right] \notag \\
&\pm \sqrt{\frac{K_c}{2}} \left[ \frac{2}{\pi} \frac{\partial \epsilon_c}{\partial \rho_c} + \frac{v_c}{K_c} \right] \bigg\}.
\end{align}
This result is therefore a direct generalization of the relation derived for the spinless case in Ref.~[\onlinecite{imambekov09}]. The symmetry of the holon edge $\epsilon_c(k) = \epsilon_c(-k)$ leads to the symmetry $\chi_{\pm}(k) = - \chi_{\mp}(-k)$. Note that in contrast to the spinon momentum, the holon momentum is not bounded. Therefore, the relations (\ref{xi}) hold for arbitrary $k$.

The phase shifts (\ref{xi}) reproduce the correct universal phase shifts in the vicinity of $k \to k_F$. Close to $k_F$, one can expand $\epsilon_c(k) = v_c(k - k_F) + (k - k_F)^2/(2m^*)$, where $m^*$ is an effective mass which will generally be different from the bare mass $m$ of the physical particles. Using this form of $\epsilon_c(k)$ it may easily be checked that
\begin{align}
 \chi_-(k \to k_F) = \delta_{-c},
\end{align}
where $\delta_{-c}$ is defined in Eq.~(\ref{phases}). In order to verify the correct behavior for $\chi_+$, it is necessary to determine the effective mass $m^*$. This can be done by bosonizing the Hamiltonian with quadratic spectrum and considering the response of the system to a density variation. The result is similar to the spinless case \cite{pereira06} and reads as follows for a Galilean-invariant system,
\begin{align}\label{mstar}
 \frac{1}{m^*}
&=
 \frac{\sqrt{2 K_c}}{2 \pi} \frac{\partial v_c}{\partial \rho_c} +
 \frac{1}{2 m \sqrt{2 K_c}}.
\end{align}
Using this effective mass, it can be shown that $\chi_+(k \to k_F) = \delta_{+c}$.

Power-law singularities of the spectral function may appear at $\epsilon_c(k)$ as well as at the shifted holon lines $\epsilon_c(k - 2 n k_F)$ for $n \in \mathbb{Z}$. The configurations giving rise to singularities at $\omega \approx \epsilon_c(k - 2 n k_F)$ contain a holon which carries almost the entire energy $\omega$ of the incoming particle (or hole), as well as $|n|$ additional low-energy particle-hole pairs with total momentum $2 n k_F$. The calculation of these edge exponents is analogous to the calculation in Sec.~\ref{sec:spin_exp}. The fermion annihilation operator is projected as
\begin{align}
 \psi_{\uparrow} = e^{i k x} d_c F_{Rc} \tpsi_{Rs} F_{Rs} \times \Phi_{2 n k_F},
\end{align}
where $\Phi_{2 n k_F}$ is given by Eq.~(\ref{nparticles2}). The edge exponents can now be found by bosonizing $\Phi_{2 n k_F}$ and $\tpsi_{Rs}$ using Eqs.~(\ref{bos_iden}) and (\ref{referm}).

Assuming the holon energy is well-defined, the spectral function displays a power-law behavior near the holon spectrum, $A(k,\omega) \propto [ \omega - \epsilon_c(k - 2 n k_F)]^{-\mu^c_{n,-}}$ with
\begin{align}\label{mu_c_n}
 \mu^c_{n,-}
&=
 \frac{1}{2} - \frac{1}{2} \left[n \sqrt{2 K_c} - \frac{\chi_{+} + \chi_{-}}{2\pi} \right]^2 \notag \\
&- \frac{1}{2} \left[ \frac{\chi_{+} - \chi_{-}}{2\pi}\right]^2 - m_-(m_- + 1).
\end{align}
Note that $m_-$ has to satisfy the selection rule in Eq.~(\ref{selm_minus}), so the leading exponent is always reached for $m_-(m_- + 1) = 0$. The edge positions and the labels of the exponents in the different regions of the $(k,\omega)$-plane are illustrated in Fig.~\ref{fig:AHubbard}.

\section{Limiting cases}\label{sec:limits}

\subsection{Strongly interacting fermions}\label{sec:Strong}

For strong repulsive interactions, it becomes increasingly difficult for fermions in one dimension to pass each other. The excitations can be separated into charge and spin parts and the spin part can be modeled using a Heisenberg Hamiltonian, $H_s = J \sum_j \vec{S}_j \cdot \vec{S}_{j+1}$, where $\vec{S}_j$ denotes the spin density on site $j$ of the lattice. A strong finite-range interaction leads to an exponential\cite{matveev04_2,matveev04,matveev07,matveev07_2} suppression of $J$ and, as a consequence, the spinon spectrum becomes almost flat, $|\epsilon_s(k)| \lessapprox J$. In this case, the phenomenological phase shifts defined in Eq.~(\ref{phen_final}) are
\begin{align}
 \frac{\Delta \delta_{\pm c}(k)}{2\pi} = - \frac{\sqrt{K_c}}{2\sqrt{2}} \frac{ k - k_F}{k_F}.
\end{align}
The edge exponents $\mu^s_{n,\pm}$ in the range $(2 n -1) k_F < k < (2 n + 1) k_F$ can then be determined from Eqs.~(\ref{mu_n}) and (\ref{mu_np}) using the proper selection rules for $m_\pm$. They are given by
\begin{align}
 \mu^s_{n \text{ odd},-} &= \mu^s_{n \text{ even},+} = - \frac{K_c}{4} \left(\frac{k}{k_F}\right)^2 - \frac{1}{4 K_c}, \notag \\
 \mu^s_{n \text{ odd},+} &= \mu^s_{n \text{ even},-} = 1 - \frac{K_c}{4} \left(\frac{k}{k_F}\right)^2 - \frac{1}{4 K_c}.
\end{align}
For $k \to k_F$, this reproduces the universal exponents (\ref{muLL}) of the LL theory.

In general, the holon branch of excitations is broadened by possible decay via the creation of pairs of spinons. However, in the limit $J\to 0$, we expect the interactions between the holon and spinon branches to vanish.\cite{fiete07} In this limit Eqs.~(\ref{xi}) and (\ref{mu_c_n}) are applicable. The resulting exponents depend on details of the interaction which ultimately defines the dependence of the holon spectrum on momentum and density. Below we consider the special case of a Hubbard model, in which the existence of holon mode is protected by integrability, and the interaction becomes strong at low electron filling factors.

The Hubbard model is described by the Hamiltonian
\begin{align}
 H_\text{Hubbard} =  -t \sum_{n\sigma} \left[ \psi^\dag_{n\sigma} \psi_{(n+1)\sigma} + \text{h.c.} \right] \notag \\
 + (U/2) \sum_n \psi^\dag_{n\uparrow} \psi_{n\uparrow} \psi^\dag_{n\downarrow} \psi_{n\downarrow},
\end{align}
where $\psi_{n\sigma}$ annihilates a fermion of spin $\sigma = \uparrow,\downarrow$ at the $n$th lattice site, $t$ is the hopping amplitude between neighboring sites and $U$ is the on-site interaction strength. For large interaction, double occupancy of a single lattice site is suppressed and the Hubbard Hamiltonian can be mapped onto a $t$-$J$ model,\cite{chao78} where the $J$-term describes a coupling between neighboring spins with an exchange coupling $J = 4 t^2/U$. For large $U$, the spinon spectrum $\epsilon_s(k)$ collapses. Moreover, $U \to \infty$ ensures that double occupancy of a single lattice site is forbidden and the charge degrees of freedom of the spinful interacting fermions behave largely like spinless noninteracting fermions of the doubled density. Therefore, the charge part can be regarded as noninteracting particles with Fermi momentum $2k_F$ and hence Fermi velocity $v_c = 2 v_F$. Going back to the continuum case, this corresponds to a Luttinger parameter $K_c = v_F/v_c = 1/2$. The edge exponents become
\begin{align}\label{mu_YG}
 \mu^s_{n \text{ odd},\mp} &= \mu^s_{n \text{ even},\pm} = \mp \frac{1}{2} - \frac{1}{8} \left(\frac{k}{k_F}\right)^2.
\end{align}
They coincide with known results determined from the finite-size corrections\cite{essler10} of the exact solution of the Hubbard model using the Bethe ansatz and from the factorization of the exact solution into charge and spin parts.\cite{carmelo92,carmelo04,carmelo08,carmelo08_2}

\begin{figure}[t]
  \centering
  \includegraphics[width = 0.48 \textwidth]{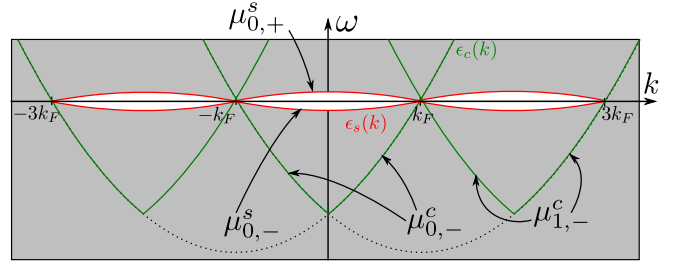}
\caption{(Color online) Position of power-law singularities in the spectral function $A(k,\omega)$ for the strongly interacting Hubbard model. At the main holon branch $\epsilon_c(k)$, power-law singularities have an exponent $\mu_{0,-}^c$ regardless of the sign of $\omega$. Weaker power-laws with exponents $\mu_{n,-}^c$ can be found at the shifted holon mass shell $\epsilon_c(k + 2 n k_F)$ for $n \in \mathbb{Z}$.}
  \label{fig:AHubbard}
\end{figure}

Being an integrable model, it is expected that the Hubbard model has stable holon excitations and that the power-laws in the spectral function at the holon mass shell are resolved. The exponents are given by Eq.~(\ref{mu_c_n}), were the phase shifts $\chi_\pm(k)$ have to be determined using the holon spectrum $\epsilon_c(k)$ for the strongly interacting Hubbard model. From Eq.~(\ref{mstar}), one finds that the effective mass for $K_c = 1/2$ is equal to the bare mass, $m^* = m$. Moreover, for $U \to \infty$, the charge sector behaves like noninteracting spinless fermions with Fermi momentum $2k_F$. Hence, the holon energy as a function of the holon momentum $k_c$ is given by $\omega_c(k_c) = k_c^2/(2m) - (2k_F)^2/(2m)$, see Fig.~\ref{fig:AHubbard}. This is also confirmed in the Bethe ansatz solution. For $k > 0$, the edge position in the spectral function is given by $\epsilon_c(k) = \omega_c(k + k_F)$. The symmetry $\epsilon_c(k) = \epsilon_c(|k|)$ then fixes the threshold for arbitrary $k$. This threshold position has been found numerically in [\onlinecite{penc96}]. Using Eq.~(\ref{xi}), one readily finds $\chi_\pm/(2\pi) = 1/4$ independent of $k$. Therefore, for $\omega \approx \epsilon_c(k)$, we find $A(k,\omega) \propto [\omega - \epsilon_c(k)]^{-\mu^c_{0,-}}$ with
\begin{align}
 \mu^c_{0,-} = \frac{3}{8}.
\end{align}
In the limit $k \to k_F$, this reproduces the universal result Eq.~(\ref{mu_0_c}). A schematic picture of the spectral function for the strongly interacting Hubbard model is shown in Fig.~\ref{fig:AHubbard}.

\subsection{Weakly interacting fermions}\label{sec:Weak}
\newcommand{\ret}{\text{ret}}

\begin{figure}[t]
  \centering
  \includegraphics[width = 0.48 \textwidth]{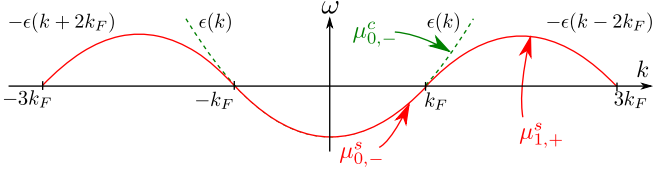}
\caption{(Color online) Singular lines in the spectral function $A(k,\omega)$ for a weakly interacting systems with spectrum $\epsilon(k) = (k^2 - k_F^2)/(2m)$. Up to second order in the interaction strength, singularities appear at the mass shell $\omega \approx \epsilon(k)$ for all $k$ as well as at $\omega \approx -\epsilon(k \mp 2k_F)$ for $k \gtrless \pm k_F$. In the configurations giving rise to the singularities, the energy is carried by a spinon along the red (solid) line or by a holon along the green (dashed) line.}
  \label{fig:SpectralFunctionWeaklyInteracting}
\end{figure}

For noninteracting fermions with dispersion $\epsilon(k)$, $A(k,\omega) = \delta[\omega - \epsilon(k)]$. As we have seen in the previous sections, for $|\omega| \geq |\epsilon(k)|$ interactions generally turn the singularity at the mass shell into a power-law. Moreover, for $|k| > k_F$, the edge of support no longer coincides with the mass shell but, in the limit of vanishingly small interaction, with the shifted and inverted threshold $\omega_{th} = - \epsilon(k \pm 2k_F)$, see Fig.~\ref{fig:SpectralFunctionWeaklyInteracting}. In this section, we shall calculate the spectral function using perturbation theory in the interaction.

For this purpose, let us start from the definition of the spectral function in terms of the retarded Green's function, $A(k,\omega) = -\frac{1}{\pi} \Im G^\ret(k,\omega)$. Due to $SU(2)$-symmetry, all correlation functions are independent of the spin orientation, so the spin index was dropped. The exact retarded Green's function of the interacting system can be expressed in terms of the retarded self-energy as
\begin{align}
 G^\ret(k,\omega) = \frac{1}{\omega - \epsilon(k) - \Sigma^\ret(k,\omega)}.
\end{align}
The self-energy $\Sigma^\ret(k,\omega)$ will be calculated perturbatively using the Hamiltonian $H = H_0 + H_{int}$, where
\begin{align}\label{Hint1}
 H_0 =& \sum_{k}\sum_{\sigma=\uparrow,\downarrow} \epsilon(k) \psi^\dag_\sigma(k) \psi_\sigma(k), \\
 H_{int} =& \frac{1}{2L} \sum_{k_1,k_2,k_3} \sum_{\sigma,\tau=\uparrow,\downarrow} \Big[ \notag \\
& \psi^\dag_\sigma(k_1) \psi^\dag_\tau(k_2) V_1(k_3) \psi_\tau(k_2 + k_3) \psi_\sigma(k_1 - k_3) \notag \\
-& \psi^\dag_\sigma(k_1) \psi^\dag_\tau(k_2) V_2(k_3) \psi_\sigma(k_2 + k_3) \psi_\tau(k_1 - k_3) \Big]. \notag
\end{align}
The free fermion spectrum is $\epsilon(k) = ( k^2 - k_F^2)/(2m)$, and $\psi_\sigma$ ($\psi^\dag_\sigma$) denotes the annihilation (creation) operator for a physical fermion of spin $\sigma = \uparrow,\downarrow$. These operators obey the fermionic commutation relations, $\{ \psi_\sigma(k), \psi^\dag_{\tau}(k') \} = \delta_{\sigma\tau} \delta_{kk'}$. For system length $L$, the momentum $k = 2\pi n/L$ ($n \in \mathbb{Z}$) is quantized due to the periodic boundary conditions, but we shall take the limit $L \to \infty$ in the following. The term $H_{int}$ is the most general two-particle interaction term allowed by $SU(2)$-symmetry and translation invariance. In terms of the charge density $\rho_c(x)$ and the spin density $\vec{S}(x)$ defined in Sec.~\ref{sec:DSF}, one can write it as
\begin{align}
 H_{int} &= \frac{1}{2} \int dx dy [ U_\rho(x-y) \rho_c(x) \rho_c(y) \notag \\
 &+ U_\sigma(x-y) \vec{S}(x) \cdot \vec{S}(y) ].
\end{align}
The interaction potentials of charge and spin densities are related to $V_{1,2}$ by $V_1 = U_\rho - U_\sigma/4$ and $V_2 = -U_\sigma/2$.

The self-energy will be calculated perturbatively in the interaction strengths $V_{1,2}$. Separating real and imaginary parts of the self-energy, the spectral function reads
\begin{align}\label{ASigmaRet}
 A(k,\omega) = \frac{1}{\pi} \frac{ - \Im \Sigma^\ret }{[\omega - \epsilon(k) - \Re \Sigma^\ret]^2 + [\Im \Sigma^\ret]^2}.
\end{align}
The first-order contribution to $\Sigma^\text{ret}(k,\omega)$ can be accessed most conveniently by calculating the first-order self-energy $\Sigma(k, i\omega_n)$ in imaginary time and then performing an analytic continuation $i \omega_n \to \omega + i \delta$ in order to translate this to the retarded self-energy. The only contributions in the first order are the well-known Hartree and Fock terms. The result is energy-independent and reads
\begin{align}
 \Sigma^{\ret(1)}(k) &= - \int_{k-k_F}^{k+k_F} \frac{dq}{2\pi} [ V_1(q) - 2 V_1(0) \notag \\
&-2 V_2(q) + V_2(0)].
\end{align}
Because the first-order term is real, it merely leads to a shift of the edge position. As the energy $\omega$ is measured with respect to the chemical potential, the new edge position is shifted to
\begin{align}
 \epsilon'(k) = \epsilon(k) - \Sigma^{\ret(1)}(k) + \Sigma^{\ret(1)}(k_F),
\end{align}
and the spectral function up to first order in the interaction strength remains a $\delta$-function, $A^{(1)}(k,\omega) = \delta[\omega - \epsilon'(k)]$.

The second-order self-energy could in principle also be calculated using the imaginary-time Green's functions. In view of the calculation of the power-laws in the spectral function, however, our primary interest is in its imaginary part. The most convenient way to calculate the imaginary part of the self-energy to second order is provided by Fermi's Golden Rule,
\begin{align}\label{ImSigma2}
 - \Im \Sigma^{\ret(2)}(k,\omega) = 2 \pi \sum_f |\bra{f} H_{int} \ket{i}|^2 \delta(\omega - \epsilon_f).
\end{align}
The initial state $\ket{i}$ has energy and momentum $\omega$ and $k$, respectively. The sum over final states $\ket{f}$ is over a complete basis of the Hilbert space and $\epsilon_f$ denotes the energy of the state $\ket{f}$. The $\delta$-function reflects energy conservation.

\begin{figure}[t]
  \centering
  \includegraphics[width = 0.2 \textwidth]{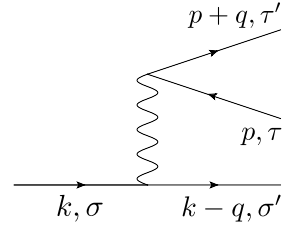}
  \caption{Feynman diagram for the imaginary part of the second order self-energy. The straight lines denote fermions with momentum and spin quantum numbers. The wiggly line depicts the interaction with momentum exchange $q$.}
  \label{fig:FGR}
\end{figure}

Let us investigate first the particle sector and assume $k > k_F$. In this case, the initial state is given by $\ket{i} = \psi^\dag_\sigma(k) \ket{FS}$, where $\ket{FS}$ denotes the Fermi sea of spin-up and spin-down particles where all states with $|k| < k_F$ are filled. The most general final state yielding a nonzero matrix element is given by $\ket{f} = \psi^\dag_{\sigma'}(k-q) \psi^\dag_{\tau'}(p+q) \psi_{\tau}(p) \ket{FS}$, so Eq.~(\ref{ImSigma2}) measures the decay probability of a single particle with momentum $k$ and energy $\omega$ (not necessarily on mass shell) via the creation of a particle-hole pair with momentum $q$. The physical process is depicted in Fig.~\ref{fig:FGR}.

Let us first calculate (\ref{ImSigma2}) for $k > k_F$ for energies slightly above the mass shell, $\omega \gtrapprox \epsilon(k)$. In this case, energy and momentum conservation imply that either $q \gtrapprox k + k_F$ or $q \lessapprox 0$. Both processes are identical since they can be mapped onto each other by exchanging the momenta $p+q$ and $k-q$ of the outgoing particles. Physically, for the quadratic spectrum $\epsilon(k)$, a right-moving particle with $k > k_F$ and $\omega > \epsilon(k)$ can only decay by creating a particle-hole pair near the opposite Fermi point. For small $\omega - \epsilon(k)$,
\begin{equation}
 - \Im \Sigma^{\ret(2)}(k,\omega)
    =
 \frac{2 U^2_\text{eff}(k+k_F)}{\pi} \theta[\omega - \epsilon(k)] \frac{\omega - \epsilon(k)}{(v +v_F)^2},
\end{equation}
where we used $v = k/m$ and $v_F = k_F/m$. For interaction potentials $V_{1,2}(q)$ which do not vary appreciably on the scale $\Delta q = m [\omega - \epsilon(k)]/(k + k_F)$ near $q = k + k_F$ and $q = 0$, the effective interaction vertex is given by
\begin{align}\label{PT_defU}
 U^2_\text{eff}(k) &= [V_1(0) + V_2(k)]^2 +  [V_2(0) + V_1(k)]^2  \notag \\
&-  [V_1(0) + V_2(k)]  [V_2(0) + V_1(k)].
\end{align}
Next, let us focus on $k > k_F$ but energies slightly below the mass shell, $\omega \lessapprox \epsilon(k)$. In this case, energy and momentum conservation are satisfied for $q \lessapprox k - k_F$ and $q \gtrapprox 0$. As previously, both types of processes are related by exchanging the momenta on the outgoing lines. The result reads
\begin{equation}\label{PT_ImSigma}
 - \Im \Sigma^{\ret(2)}(k,\omega)
=
 \frac{2 U^2_\text{eff}(k - k_F)}{\pi} \theta[\epsilon(k)-\omega] \frac{|\omega - \epsilon(k)|}{(v -v_F)^2}.
\end{equation}
The spectral function near the mass shell can now be obtained from Eq.~(\ref{ASigmaRet}). Due to the analytic structure of the retarded Green's function, the real part can be estimated from Kramers-Kronig relations. Close to the mass shell, these predict a logarithmic divergence $\Re \Sigma^{\ret(2)}(k,\omega) \propto U^2_\text{eff} [\omega - \epsilon(k)] \ln \{[\omega - \epsilon(k)]/\epsilon_F \}$ where $\epsilon_F$ is a high-energy cutoff of the order of the Fermi energy. This indicates an expected breakdown of the perturbation theory in a narrow vicinity of the spectrum. For a sufficiently weak interaction and at fixed $(k-k_F)^2/m$ there are domains of energy $\omega$ where Eqs.~(\ref{PT_defU}) and (\ref{PT_ImSigma}) are valid, while $\Re\Sigma^{\ret(2)}$ can be dispensed with in comparison with $|\omega-\epsilon(k)|$. In these domains, the spectral function reads
\begin{align}\label{WI_holon}
 A(k,\omega) &= \frac{2 U^2_\text{eff}(k+k_F)}{\pi^2(v +v_F)^2} \frac{\theta[\omega - \epsilon(k)]}{\omega - \epsilon(k)} \notag \\
&+
 \frac{2 U^2_\text{eff}(k - k_F)}{\pi^2(v -v_F)^2}  \frac{\theta[\epsilon(k)-\omega]}{|\omega - \epsilon(k)|}.
\end{align}
Therefore, comparing this to the general structure $A(k,\omega) \propto [\omega-\epsilon(k)]^{-\mu_{0,-}^c}$, perturbation theory predicts an exponent $\mu_{0,-}^c = 1$ in the vicinity of the mass shell. The singular lines of the spectral function in the weakly interacting case are depicted in Fig.~\ref{fig:SpectralFunctionWeaklyInteracting}.

Next, we calculate the spectral function for $k_F < k < 3 k_F$ near the edge of support, $\omega \approx \omega_{th} = -\epsilon(k - 2k_F)$. It turns out that for small $\omega - \omega_{th}$, due to energy and momentum conservation, the only allowed momentum exchange must satisfy $q \lessapprox k - k_F$. To lowest order in $\omega - \omega_{th}$, one finds
\begin{align}
 - \Im \Sigma^{\ret(2)}(k,\omega)
=
 \frac{V^2_\text{eff}}{\pi} \theta[\omega - \omega_{th}] \frac{\omega - \omega_{th}}{(v -v_F)^2}
\end{align}
with the effective interaction vertex,
\begin{align}\label{Veff_th}
 V_\text{eff}^2
&= [V_1(k-k_F) + V_2(k-k_F)]^2.
\end{align}
The same caveats as previously about the applicability of perturbation theory apply also near the threshold. In the range of applicability, one finds the spectral function from Eq.~(\ref{ASigmaRet}),
\begin{align}\label{WI_threshold}
 A(k,\omega) &= \frac{V^2_\text{eff}}{\pi^2 m^2 (v -v_F)^6} \theta(\omega -\omega_{th}) (\omega -\omega_{th}).
\end{align}
Therefore, the threshold exponent predicted by perturbation theory is $\mu_{1,+}^s = -1$.

The second-order self-energy can be calculated for $-k_F < k < k_F$ in a similar fashion. In this case, the initial state reads $\ket{i} = \psi_\sigma(k)\ket{FS}$ while the final state is given by $\ket{f} = \psi^\dag_{\tau}(p) \psi_{\tau'}(p+q) \psi_{\sigma'}(k-q) \ket{FS}$. The calculation of the matrix elements $\bra{f} H_{int} \ket{i}$ is performed as previously but now leads to different integration ranges for the momenta $p$ and $q$. For $\omega \approx \epsilon(k) < 0$, a hole with momentum $k$ can relax by creating a low-energy particle-hole pair near either the left or right Fermi point by transferring the momentum $k - k_F$ or $k + k_F$, respectively. For $|k| < k_F$, the spectral function near the mass shell reads
\begin{align}\label{WI_mu}
 A(k,\omega) &= \left[ \frac{U^2_\text{eff}(k+k_F)}{\pi^2(v +v_F)^2} + \frac{U^2_\text{eff}(k-k_F)}{\pi^2(v -v_F)^2} \right] \frac{\theta[\epsilon(k)-\omega]}{|\omega - \epsilon(k)|}.
\end{align}
This corresponds to an exponent $\mu_{0,-}^s = 1$. For $|k| < k_F$, the mass shell in the limit of vanishing interaction coincides with the edge of support. However, note that second-order perturbation theory is insufficient to explain the finite values of $A(|k|<k_F,\omega)$ near the shadow band $\omega \approx -\epsilon(k) > 0$. The final state in this region contains an additional particle-hole pair with momentum $\pm 2 k_F$ and is thus not captured by the second-order calculation. Note also that unlike the case of spinless fermions, here $U_\text{eff}(0)\neq 0$, indicating problems with the perturbation theory in the vicinity of Fermi points; we have to require $|k-k_F|/m \gg U_\text{eff}$. We note that the difference between the holon and spinon velocities, $v_c-v_s\propto U_\text{eff}$, so the latter requirement may also be viewed as the condition that the spectrum curvature term in the fermion energy is appreciable, $(k-k_F)^2/m\gg (v_c-v_s)|k-k_F|$. Because the particle-hole pair in the final state has to respect this constraint, perturbation theory can only be used to evaluate $A(k,\omega)$ for energies $\omega$ away from the true singularities, $|\omega - \epsilon(k)|, |\omega - \omega_{th}| \gg (v_c - v_s) |k - k_F|$.

In order to calculate the exponents in closer vicinity of the thresholds, we use the phenomenological relations, Eqs.~(\ref{phen_final}) and (\ref{xi}). First, let us explain how to decompose a physical fermion into spinons and holons in the weakly interacting limit. We use the Bethe ansatz solution of the Yang-Gaudin model\cite{yang67,gaudin67} in the limit of zero interaction and compare the thresholds $\epsilon(k)$ and $\omega_{th}$ to its spinon and holon spectra.\cite{essler05} For $k < k_F$, the singularity at $\omega \approx \epsilon(k)$ is created by configurations which contain a holon at its Fermi point while the energy $\epsilon(k)$ is carried by a spinon on mass shell. For $k > k_F$, the excitations at $\omega \approx \omega_{th}$ are spinon excitations in the same sense. Hence, the general observation that for repulsive interactions the edge of support of the spectral function corresponds to spinon excitations continues to hold in the noninteracting limit. In contrast, for $k > k_F$, the threshold at $\omega \approx \epsilon(k)$ corresponds to a holon excitation with energy $\epsilon(k)$, while the spinon rests at its Fermi point. The singular lines of $A(k,\omega)$ in different regions of the $(k,\omega)$-plane along with the nature of the respective excitations are depicted in Fig.~\ref{fig:SpectralFunctionWeaklyInteracting}.

Away from $k_F$, the physical spectrum $\epsilon(k)$ is only weakly affected by the interactions, so we can calculate the exponents in this region using the noninteracting spinon and holon spectra. For $0<k<2k_F$, excluding again a domain $|k-k_F|\approx mU_\text{eff}(0)$ around the Fermi points, we can use $\epsilon_s(k \gtrless k_F) = v_F( k - k_F) \mp (k - k_F)^2/(2m)$ and calculate the exponents from the phenomenological relations Eq.~(\ref{phen_final}). For weak interactions, the Luttinger parameter can be expanded as $K_c = 1 - \delta K_c$, where $\delta K_c \to 0^+$. Everywhere at $|k|<k_F$, except the narrow vicinities of the Fermi points, $|k\pm k_F|\lessapprox mU_\text{eff}$, the threshold exponent is given by $\mu_{0,-}^s \approx 1 - O(\delta K_c^2)$ and thus is compatible with the perturbative result (\ref{WI_mu}). For $k > k_F$ and $\omega \approx \omega_{th}$, the phenomenology yields the exponent $\mu_{1,+}^s = O(\delta K_c)$ at the threshold, different from the exponent Eq.~(\ref{WI_threshold}) valid away from the threshold.

Similarly, the holon spectrum at $|k-k_F|\gg mU_\text{eff}$ can also be approximated by its noninteracting limit, $\epsilon_c(k > k_F) = v_F(k - k_F) + ( k -k_F)^2/(2m)$. A calculation of the phenomenological phase shifts and the exponent using Eqs.~(\ref{xi}) and (\ref{mu_c_n}) then leads to $\mu_{0,-}^c = 1/2 - O(\delta K_c^2)$. Note that this coincides with the exponent at the holon mass shell predicted for $k \approx k_F$ by the linear LL theory, see Eq.~(\ref{muLL}). But it is different from the perturbative exponent $\mu_{0,-}^c =1$ in Eq.~(\ref{WI_holon}) which is valid away from the mass shell.

We conclude that the lowest-order perturbation theory performed here is only able to predict the behavior of $A(k,\omega)$ away from the true edge. Closer to the edge, the nonperturbative spin-charge separation becomes important and the correct exponents can be derived using the phenomenological relations.

\section{Holon relaxation}\label{sec:relax}

The spinon and holon excitations of the linear LL theory are eigenstates of the Hamiltonian $H_0$ [see Eq.~(\ref{H_0})] and are thus predicted to be stable. As a consequence of the infinite spinon and holon lifetimes, the singularities of the spectral function and other dynamic response functions at the spinon and holon mass shells are characterized by true power laws.

In the case of nonlinear spectrum, bosonization leads to spinon-spinon and holon-holon interactions, as well as to spinon-holon interactions. This may give rise to finite lifetimes for holons (we assume that spinons remain the lowest-energy excitations). In the spectral function, such a finite lifetime generally leads to smearing of the singularities\cite{mahan90} and is thus directly measurable.

For spinless fermions, the effects of a nonlinear spectrum on the particle and hole lifetimes were analyzed in [\onlinecite{khodas07_2}]. For weakly interacting fermions with a generic short-ranged repulsive interaction potential $V(k)$, it was shown that the decay rate of a particle with momentum $k \approx k_F$ scales as $\Gamma \propto (k - k_F)^{8}$. The exponent here comes from a limitation on the phase space available for the decay combined with the $q^2$ momentum dependence of the effective interaction at small momentum transfer $q$. An important exception is the case of integrable models,\cite{cheon98,cheon99,sutherland04} where these calculations predict infinite lifetime for all momenta. Away from the limit of weak interactions, the rate is limited by the phase space argument only, yielding the decay rate scaling\cite{pereira09} $\Gamma \propto (k - k_F)^4$.

A similar phase space argument applied to the decay of a holon with creation of two spinons would lead to $\Gamma\propto |k-k_F|$, possibly contradicting to the notion of a well-defined holon branch at $k \to k_F$. However, as we show below, the decay rate of a holon with small (measured from the Fermi point) momentum must scale to zero with $k \to k_F$ faster than $|k - k_F|^3$.

In order to elucidate the possible decay processes for holons, it is convenient to start again from a description in terms of refermionized quasiparticles. The band curvature of the physical fermions leads to interactions between the quasiparticles. Away from the Fermi points, it is advantageous to classify the interaction processes by their relevance in the RG sense and to consider all possible interaction operators which are allowed by $SU(2)$-symmetry and Galilean invariance. Due to its built-in $SU(2)$-symmetry, non-Abelian bosonization\cite{gogolin98} is a convenient tool to achieve this. Expressed using the left- and right-moving holon densities $J_\alpha(x)$ and spinon densities $\VJ_\alpha(x)$ ($\alpha = L,R$), the Hamiltonian (\ref{H_0}) of the linear LL reads $H_0 = H_c + H_s$, where
\begin{align}
    H_c &= 2\pi v_c \int dx [ J_R^2(x) + J_L^2(x) ], \notag \\
    H_s &= \frac{2\pi v_s}{3} \int dx [ \VJ_R^2(x) + \VJ_L^2(x) ].
\end{align}
The operators $J_\alpha(x)$ are related to the physical charge density by $\rho_c(x) = 2 \sqrt{K_c} \expct{J_L(x) + J_R(x)}$. This Hamiltonian emerges at the low-energy RG fixed point and is valid in the narrow-band limit. The leading correction for increased bandwidth is an interaction between left-moving and right-moving spin densities,\cite{gogolin98}
\begin{align}
    H_g = - 2 \pi v_s g \int dx\ \VJ_R(x) \cdot \VJ_L(x).
\end{align}
Note that when expressed in terms of the Abelian spinon fields $\tilde{\phi}_s$ and $\tilde{\theta}_s$, the operator $H_g$ generates the sine-Gordon term (\ref{Hg}). The band curvature of the physical fermions leads to interaction operators which are cubic in spin and charge densities,
\begin{align}\label{Hekz}
    H_\eta &= \frac{4\pi^2}{3} \int dx  \left[ \eta_- (J_R^3 + J_L^3) - \eta_+ (J_R^2 J_L + J_L^2 J_R) \right], \notag \\
    H_\kappa &= \frac{4\pi^2}{3} \int dx \Big[ \kappa_- (J_R \VJ_R^2 + J_L \VJ_L^2) \notag \\
    &+ \kappa_+ ( J_R \VJ_L^2 + J_L \VJ_R^2) \Big], \notag \\
    H_\zeta &= \frac{4\pi^2 \zeta}{3} \int dx\ (J_L + J_R) \VJ_R \cdot \VJ_L.
\end{align}
Note that these operators represent all cubic terms which are compatible with $SU(2)$-symmetry. In particular, this symmetry prohibits terms linear in the vector operators $\VJ_\alpha(x)$. Interaction operators containing quartic and higher-order terms in $\VJ_\alpha(x)$ and $J_\alpha(x)$ do exist but their contribution is subleading for small bandwidths.

The prefactors $g, \zeta, \kappa_\pm$ and $\eta_\pm$ can be fixed phenomenologically by relating them to other observable quantities. For this purpose, let us investigate the variation of the interaction operators as a response to a uniform variation of the charge density $\rho_c(x) \to \rho_c(x) + \delta \rho_c$. This variation shifts $J_\alpha(x) \to J_\alpha(x) + \delta J$, where $\delta J = \delta \rho_c / ( 4 \sqrt{K_c})$, and thus creates the following additions to the Hamiltonian,
\begin{align}
    \delta H_\eta &= \frac{4\pi^2}{3} \delta J \Big[ (3 \eta_- - \eta_+) \int dx \left( J_R^2 + J_L^2 \right) \notag \\
    &- 4 \eta_+ \int dx\ J_L J_R \Big],\notag \\
    \delta H_\kappa &= \frac{4\pi^2}{3} \delta J (\kappa_- + \kappa_+) \int dx (\VJ_R^2 + \VJ_L^2),\notag \\
    \delta H_\zeta &= \frac{4\pi^2 \zeta}{3} (2 \delta J) \int dx\ \VJ_R \cdot \VJ_L.
\end{align}
Combined with $H_c$, the Hamiltonian $\delta H_\eta$ leads to a renormalization of $v_c$ and $K_c$.\cite{pereira06} The shifts $\delta H_{\kappa}$ and $ \delta H_\zeta$ modify $H_s$ and $H_g$, respectively, and thus renormalize the parameters $v_s$ and $v_s g$. Expressing the density variation as a variation of chemical potential using $\partial \rho_c/\partial \mu = 2 K_c / ( \pi v_c)$,\cite{giamarchi03} one finds\cite{pereira10}
\begin{align}\label{kappa_phen}
    \kappa_- + \kappa_+ &= \frac{v_c}{\sqrt{K_c}} \frac{\partial v_s}{\partial \mu}, \\
    \zeta &= - \frac{3}{2} \frac{v_c }{\sqrt{K_c}} \frac{\partial(v_s g)}{\partial \mu}. \label{zeta_phen}
\end{align}
In a similar way, the difference $\kappa_- - \kappa_+$ can be related to the mass $m$ of the physical fermions by considering a charge current variation of the Galilean-invariant system. One finds,\cite{nayak01,pereira10}
\begin{align}
    \kappa_- - \kappa_+ = \frac{1}{m \sqrt{K_c}}.
\end{align}

It is known that upon a bandwidth reduction $g$ flows logarithmically to zero.\cite{gogolin98} Assuming the initial bandwidth to be of order $k_F$, for a smaller bandwidth of order $k$ the effective coupling constant will flow to $g(k) = 1/\ln[k_F / (k-k_F)]$. As the chemical potential $\mu$ is proportional to $k_F$, the derivative $\partial g/\partial \mu \propto - g^2/k_F$. The derivative $\partial v_s/\partial \mu$, on the other hand, remains finite for small bandwidths. Therefore, in leading logarithmic approximation, $\partial g/\partial\mu$ can be neglected and the coupling constants $\kappa_\pm$ and $\zeta$ can be related as
\begin{align}\label{zeta_kappa_phen}
    \zeta \approx -\frac{3}{2} g \frac{v_c}{\sqrt{K_c}} \frac{\partial v_s}{\partial \mu} = -\frac{3}{2} g ( \kappa_- + \kappa_+).
\end{align}

For repulsive interactions, the excitation of lowest energy for a given momentum $k$ is a spinon with energy $\epsilon_s(k)$. Due to the shape of the spinon spectrum (see Fig.~\ref{fig:SpectralFunctionNLL}), the absolute value of the velocity $\partial \epsilon_s(k)/\partial k$ reaches its maximum $v_s$ near the Fermi points. As a consequence, spinon relaxation by creation of low-energy spinons is ruled out by energy and momentum conservation. Similarly, because $v_c > v_s \geq \partial \epsilon_s(k)/\partial k$, decay of spinons by the creation of holons is also forbidden.  Therefore, spinon excitations are stable.

Holons, on the other hand, can relax via the creation of low-energy spinons. Let us investigate the decay of an initial state $\ket{i} = \ket{k}_c \ket{0}_s$ which contains an additional holon with momentum above the Fermi edge and no spinon excitations. Relaxation of the holon to a momentum $k' < k$ can happen via the creation of two spinon density excitations with momenta $q_L < 0$ and $q_R > 0$. This final state will be labeled $\ket{f} = \ket{k'}_c \ket{q_L,q_R}_s$. For momenta $k$ close to the Fermi point, momentum and energy conversation for this process read
\begin{align}
    k &= k' + q_R + q_L, \notag \\
    v_c k &= v_c k' + v_s (q_R - q_L),
\end{align}
and have nontrivial solutions ($k \neq k'$) for $v_c > v_s$.

The holon lifetimes associated with this decay channel can be calculated using Fermi's Golden Rule. Two combinations of operators from the interaction terms (\ref{Hekz}) have a nonzero matrix element between the states $\ket{i}$ and $\ket{f}$. To first order in the interaction, $\bra{f} H_\zeta \ket{i}$ is the only such term. To second order, only $\bra{f} H_g H_\kappa \ket{i}$ and $\bra{f} H_\kappa H_g \ket{i}$ are nonzero.

The calculation of these matrix elements is greatly simplified by the fact that the initial state $\ket{i}$ contains no excitation in the spin sector and thus corresponds to an $SU(2)$-invariant singlet state. Therefore, one can use $\VJ_\alpha(x) \cdot \VJ_\beta(x) \ket{0}_s = 3 J^z_\alpha(x) J^z_\beta(x) \ket{0}_s$ for $\alpha, \beta \in \{L, R\}$. Moreover, all interaction Hamiltonians $H_g, H_\eta, H_\kappa$ and $H_\zeta$ conserve the total spin $\vec{S} = \int dx [ \VJ_R(x) + \VJ_L(x) ]$. This can be verified by calculating the commutators using the $SU(2)$-Kac-Moody algebra for the operators $\VJ_\alpha(x)$.\cite{gogolin98} Therefore, the spin sector remains in a singlet state even if acted on by the interaction Hamiltonians. It means that only the $z$-components of the spin operators, $J^z_\alpha(x)$, are needed for the calculation of the holon lifetime. A normal-mode expansion allows one to represent the Fourier transform $J_\alpha^z(p)$ in terms of bosonic operators $b_p$,
\begin{align}\label{Jbosons}
    J_\alpha^z(p)
&=
    \sqrt{\frac{L |p|}{4 \pi}} \left[ \theta(\alpha p) b^\dag_{p} + \theta(-\alpha p) b_{-p}  \right],
\end{align}
where $L$ is the system length and $\alpha = R,L = +,-$. The operators $b_q$ satisfy the commutation relations $[b_q, b^\dag_{q'}] = \delta_{qq'}$. Using the creation operators $b^\dag_q$, the spin part of the final state $\ket{f}$ can be written as $\ket{q_L,q_R}_s = b^\dag_{q_L} b^\dag_{q_R} \ket{0}_s$. For the calculation of the holon lifetime, the following spinon matrix elements are needed,
\begin{align}\label{mat:spinon}
 _s\!\bra{q} J_\alpha^z(p) \ket{0}_s
&= \sqrt{\frac{L|q|}{4 \pi}} \theta(\alpha q) \delta_{p,q}.
\end{align}
For the expectation values of holon operators, on the other hand, it is convenient to retain the description in terms of fermionic quasiparticles where $J_\alpha(x) = \frac{1}{2} \trho_{\alpha c}(x) = \frac{1}{2} \tpsi^\dag_{\alpha c}(x) \tpsi_{\alpha c}(x)$. The charge part of the initial and final states can be expressed as $\ket{k}_c = \tpsi^\dag_{Rc}(k) \ket{FS}_c$ and $\ket{k'}_c = \tpsi^\dag_{Rc}(k') \ket{FS}_c$, respectively, where $\ket{FS}_c$ denotes the Fermi sea of holons. Then, one finds the matrix element,
\begin{align}\label{mat:holon}
 _c\!\bra{k'} J_R(p) \ket{k}_c
&=
 \frac{1}{2}\delta_{k',p+k}.
\end{align}

The first-order matrix element $T_\zeta = \bra{f} H_\zeta \ket{i}$ can be calculated by using these matrix elements. One finds
\begin{align}\label{T_zeta}
    T_\zeta
    &=
        \frac{2\pi^2 \zeta}{L^2} \sum_{p,p'} \ _s\!\bra{q_L} J_L^z(p) \ket{0}_s \ _s\!\bra{q_R} J_R^z(p) \ket{0}_s \notag \\
    &\times _c\!\bra{k'} J_R(-p-p') \ket{k}_c \notag \\
    &= \frac{\pi \zeta}{2 L} \delta_{k-k'-q_L-q_R} \sqrt{|q_L q_R|}.
\end{align}
The matrix elements $\bra{f} H_g \ket{i}$ and $\bra{f} H_\eta \ket{i}$ vanish because $H_g$ and $H_\eta$ do not couple spinons and holons. The remaining first-order matrix element $\bra{f} H_\kappa \ket{i} = 0$ because it contains only terms of the form $\VJ_\alpha^2(x)$, which do not create spinons on opposite branches.

To the second order, cross-terms of the operators $H_g$ and $H_\kappa$ may couple the same initial and final states as above. Since these are second-order terms, the matrix elements can be calculated using the $S$-matrix expansion\cite{mahan90}
\begin{align}
    T_{\kappa g} = \bra{f} H_\kappa \frac{1}{E - H_0} H_g \ket{i}, \notag \\
    T_{g \kappa} = \bra{f} H_g \frac{1}{E - H_0} H_\kappa \ket{i},
\end{align}
where $H_0 = H_c + H_s$ is the noninteracting Hamiltonian. The energy $E$ denotes the energy of the initial state, $E = v_c (k - k_F)$. As Fermi's Golden Rule imposes energy conservation, it ultimately becomes equal to the energy of the final state. After Fourier transforming $H_g$ and $H_\kappa$, going over to bosonic operators using Eq.~(\ref{Jbosons}) and using the spinon and holon matrix elements (\ref{mat:spinon})-(\ref{mat:holon}), one finds that
\begin{align}\label{T_kappa_g}
     T_{\kappa g}
&=
 \frac{3 \pi g}{4 L} (\kappa_- + \kappa_+) \delta_{k-k'-q_L-q_R}\sqrt{|q_L q_R|}, \notag \\
    T_{g \kappa}
 &=
    0.
\end{align}
Other second-order terms exist but they contain higher powers of $q_L$ and $q_R$ and are therefore subleading compared to $T_{\kappa g}$ for holon momenta $k$ near the Fermi points. According to Fermi's Golden Rule the lifetime is
\begin{align}\label{FGR}
    \Gamma = 2\pi \sum_{\ket{f}} |T_\zeta  + T_{\kappa g} |^2 \delta(\epsilon_f - \epsilon_i),
\end{align}
where $\epsilon_i$ and $\epsilon_f$ are the energies of the initial state $\ket{i}$ and the final state $\ket{f}$, respectively. The sum over all final states $\ket{f}$ translates to a summation over the momenta $q_L < 0$, $q_R > 0$ and $k' \in [k_F, k]$. It can be seen from Eqs.~(\ref{T_zeta}) and (\ref{T_kappa_g}) that each of the decay channels taken individually would lead to a decay rate $\Gamma \propto (k - k_F)^3$. However, Fermi's Golden Rule (\ref{FGR}) contains the square of the sum of the probability amplitudes $T_{\zeta}$ and $T_{\kappa g}$. The prefactors of both amplitudes are related according to Eq.~(\ref{zeta_kappa_phen}) and one finds $T_{\zeta} + T_{\kappa g} = 0$. Therefore, the decay rate vanishes up to terms proportional to $g^2 (k-k_F)^3$, in the calculation of $\Gamma$ performed in the second order\footnote{Our conclusion here differs from the one of Pereira and Sela in [\onlinecite{pereira10}], see Eq.~(14) therein.} of $g=1/\ln[k_F/(k-k_F)]$. Retaining in Eq.~(\ref{zeta_phen}) the derivative $\partial g/\partial\mu\propto g^2/\epsilon_F$ exceeds the accuracy of our calculation. It is not clear if the evaluation of $\Gamma$ to order $g^4$ would yield zero. Possibly, in that order the distinction between integrable and non-integrable systems emerges.

In the limit of weak backscattering, $V(2k_F) \ll V(0) \ll v_F$, the universal logarithmic dependence for $g(k-k_F)$ is reached only at very low energies, while its bare value $g \propto V(2k_F)/v_F$ is applicable as long as $[V(2k_F)/v_F]\ln[k_F/(k- k_F)]\ll 1$. In that case, Eqs.~(\ref{zeta_phen}) and (\ref{FGR}) yield
\begin{align}
 \Gamma \propto \epsilon_F \frac{V(0)}{v_F} \left[\frac{V(2k_F)}{v_F}\right]^2 \left[\frac{k-k_F}{k_F}\right]^3
\end{align}
This estimate should be viewed as the result of perturbation theory in $V(2k_F)$ in the basis of well-defined holon and spinon modes with linear spectrum, which sets a limit on holon momenta, $k - k_F \lesssim mV(0)$ (we also used $v_c-v_s\sim V(0)$ in the derivation). Curiously, the latter estimate for $\Gamma$ at the limit of its applicability, $k - k_F \approx mV(0)$, matches the relaxation rate of a spinful fermion evaluated, in the basis of free fermions, by the perturbation theory with respect to the entire interaction.\cite{karzig10}

\section{Conclusion}\label{sec:conclusion}

In conclusion, we have investigated spinful one-dimensional interacting Fermi systems at zero temperature beyond the low-energy regime and calculated their dynamic response functions for arbitrary momenta near the edges of support in the $(k,\omega)$-plane. Away from the Fermi points, the nonlinearity of the fermion spectrum becomes important and the physical properties can no longer be explained by the linear Luttinger liquid theory. In particular, we shed light on the meaning of spin-charge separation away from the low-energy limit.

The Luttinger liquid theory is based on the assumption of a linear fermionic spectrum. The eigenmodes are spin and charge density waves and the theory can be formulated in terms of noninteracting bosonic fields. Refermionizing these fields in Eq.~(\ref{referm}), we introduced fermionic quasiparticles, spinons and holons, which constitute a convenient basis even away from the low-energy regime. In contrast to the linear LL theory, a curvature of the spectrum of the physical fermions leads to interactions between spinons and holons. For repulsive interactions and small $|k| - k_F$, spinons are the lowest-energy excitations. By continuity, we expect that for generic repulsive potentials, the spinon spectrum $\epsilon_s(k)$ remains the edge of support for the spectral function for arbitrary momenta.

We found that the spin-charge separation exists also for nonlinear spectrum but in a weaker sense than in a linear LL. If a particle with arbitrary momentum $k$ and energy $\omega \approx \epsilon_s(k)$ tunnels into the system, it creates a single spinon with energy close to $\epsilon_s(k)$, momentum close to $k$, and velocity $v_d = \partial \epsilon_s(k)/\partial k$. In addition, it creates low-energy holon excitations with velocity $v_c > v_d$ and momenta near the Fermi points, but no additional spinon excitations. The created spinon separates in space from the charge excitations due to its different velocity. This is reminiscent of the conventional spin-charge separation in linear LLs. However, in contrast to the linear LL theory, such a decoupling only survives for energies close to $\epsilon_s(k)$.

The separation in the momenta of the spinon and the low-energy excitations allows us to introduce an effective mobile impurity model as a tool for the evaluation of measurable dynamic response functions, such as the spectral function $A(k,\omega)$ and the charge and spin density structure factors, $S(k,\omega)$ and $S^{zz}(k,\omega)$, respectively. In analogy to the Fermi edge problem, the created spinon acts as a mobile impurity and causes a shake-up of the Fermi seas of holons. The threshold exponents of $A(k,\omega)$ can therefore be expressed for arbitrary momenta and interaction strengths in terms of scattering phase shifts $\Delta \delta_{\pm c}(k)$, see Eqs.~(\ref{mu_n}) and (\ref{mu_np}). In the vicinities of the Fermi points, $\Delta \delta_{\pm c}$ depend only on the Luttinger parameter $K_c$, see Eq.~(\ref{phases}), which leads to universal values of the exponents. For arbitrary momenta and Galilean invariant systems, we related the phase shifts to another set of measurable properties given by the dependence of $\epsilon_s(k)$ on $k$ and the charge density $\rho_c$, see Eq.~(\ref{phen_final}). Using similar considerations, we calculated the threshold behavior of the spin and charge density structure factors $S(k,\omega)$ and $S^{zz}(k,\omega)$, see Eqs.~(\ref{mu_SSF}) and (\ref{mu_DSF}). These results are summarized in Table~\ref{tab:Exponents}. The general evolution of the spectral function with increasing interaction was considered by the analysis of the limits of weak and strong interaction. In the former one, we found $A(k,\omega)$ perturbatively, see Sec.~\ref{sec:Weak}. In the latter limit, we utilized the phenomenological relations to calculate the threshold properties of $A(k,\omega)$ from the exactly known spectrum, see Sec.~\ref{sec:Strong}.

Unlike spinons, the holon excitations do decay as a consequence of the nonlinear spectrum of the fermions. The corresponding lifetime is long for holons with momenta near the Fermi points, see Sec.~\ref{sec:relax}. For integrable models, the holon spectrum $\epsilon_c(k)$ is well-defined for arbitrary momenta. We determined the exponents of the spectral function for $\omega \approx \epsilon_c(k)$ by deriving phenomenological expressions relating the corresponding scattering phase shifts to properties of $\epsilon_c(k)$, see Sec.~\ref{sec:holon}.

\acknowledgments The authors thank F.~Essler, A.~Lamacraft, R.~G.~Pereira, and E.~Sela for discussions. The authors acknowledge support by
the NSF DMR Grant No.~0906498, the Texas Norman Hackerman Advanced Research Program under Grant No.~01889, the Alfred P.~Sloan Foundation, and the Swiss National Science Foundation.

\bibliography{paper}

\begin{thebibliography}{81}
\expandafter\ifx\csname natexlab\endcsname\relax\def\natexlab#1{#1}\fi
\expandafter\ifx\csname bibnamefont\endcsname\relax
  \def\bibnamefont#1{#1}\fi
\expandafter\ifx\csname bibfnamefont\endcsname\relax
  \def\bibfnamefont#1{#1}\fi
\expandafter\ifx\csname citenamefont\endcsname\relax
  \def\citenamefont#1{#1}\fi
\expandafter\ifx\csname url\endcsname\relax
  \def\url#1{\texttt{#1}}\fi
\expandafter\ifx\csname urlprefix\endcsname\relax\def\urlprefix{URL }\fi
\providecommand{\bibinfo}[2]{#2}
\providecommand{\eprint}[2][]{\url{#2}}

\bibitem[{\citenamefont{Nozi\`eres}(1997)}]{nozieres_book}
\bibinfo{author}{\bibfnamefont{P.}~\bibnamefont{Nozi\`eres}},
  \emph{\bibinfo{title}{Theory of interacting {F}ermi systems}}
  (\bibinfo{publisher}{Addison-Wesley}, \bibinfo{address}{Reading, MA},
  \bibinfo{year}{1997}).

\bibitem[{\citenamefont{Tomonaga}(1950)}]{tomonaga50}
\bibinfo{author}{\bibfnamefont{S.}~\bibnamefont{Tomonaga}},
  \bibinfo{journal}{Prog. Theor. Phys.} \textbf{\bibinfo{volume}{5}},
  \bibinfo{pages}{544} (\bibinfo{year}{1950}).

\bibitem[{\citenamefont{Luttinger}(1963)}]{luttinger63}
\bibinfo{author}{\bibfnamefont{J.~M.} \bibnamefont{Luttinger}},
  \bibinfo{journal}{J. Math. Phys.} \textbf{\bibinfo{volume}{4}},
  \bibinfo{pages}{1154} (\bibinfo{year}{1963}).

\bibitem[{\citenamefont{Mattis and Lieb}(1965)}]{mattis65}
\bibinfo{author}{\bibfnamefont{D.~C.} \bibnamefont{Mattis}} \bibnamefont{and}
  \bibinfo{author}{\bibfnamefont{E.~H.} \bibnamefont{Lieb}},
  \bibinfo{journal}{J. Math. Phys.} \textbf{\bibinfo{volume}{6}},
  \bibinfo{pages}{304} (\bibinfo{year}{1965}).

\bibitem[{\citenamefont{Haldane}(1981{\natexlab{a}})}]{haldane81}
\bibinfo{author}{\bibfnamefont{F.~D.~M.} \bibnamefont{Haldane}},
  \bibinfo{journal}{J. Phys. C: Solid State Phys.}
  \textbf{\bibinfo{volume}{14}}, \bibinfo{pages}{2585}
  (\bibinfo{year}{1981}{\natexlab{a}}).

\bibitem[{\citenamefont{Haldane}(1981{\natexlab{b}})}]{haldane81_2}
\bibinfo{author}{\bibfnamefont{F.~D.~M.} \bibnamefont{Haldane}},
  \bibinfo{journal}{Phys. Rev. Lett.} \textbf{\bibinfo{volume}{47}},
  \bibinfo{pages}{1840} (\bibinfo{year}{1981}{\natexlab{b}}).

\bibitem[{\citenamefont{Giamarchi}(2003)}]{giamarchi03}
\bibinfo{author}{\bibfnamefont{T.}~\bibnamefont{Giamarchi}},
  \emph{\bibinfo{title}{Quantum Physics in One Dimension}}
  (\bibinfo{publisher}{Clarendon Press}, \bibinfo{address}{Oxford},
  \bibinfo{year}{2003}).

\bibitem[{\citenamefont{Dzyaloshinskii and Larkin}(1974)}]{dzyaloshinskii74}
\bibinfo{author}{\bibfnamefont{I.~E.} \bibnamefont{Dzyaloshinskii}}
  \bibnamefont{and} \bibinfo{author}{\bibfnamefont{A.~I.}
  \bibnamefont{Larkin}}, \bibinfo{journal}{Sov. Phys. JETP}
  \textbf{\bibinfo{volume}{38}}, \bibinfo{pages}{202} (\bibinfo{year}{1974}).

\bibitem[{\citenamefont{Auslaender et~al.}(2005)\citenamefont{Auslaender,
  Steinberg, Yacoby, Tserkovnyak, Halperin, Baldwin, Pfeiffer, and
  West}}]{auslaender05}
\bibinfo{author}{\bibfnamefont{O.~M.} \bibnamefont{Auslaender}},
  \bibinfo{author}{\bibfnamefont{H.}~\bibnamefont{Steinberg}},
  \bibinfo{author}{\bibfnamefont{A.}~\bibnamefont{Yacoby}},
  \bibinfo{author}{\bibfnamefont{Y.}~\bibnamefont{Tserkovnyak}},
  \bibinfo{author}{\bibfnamefont{B.~I.} \bibnamefont{Halperin}},
  \bibinfo{author}{\bibfnamefont{K.~W.} \bibnamefont{Baldwin}},
  \bibinfo{author}{\bibfnamefont{L.~N.} \bibnamefont{Pfeiffer}},
  \bibnamefont{and} \bibinfo{author}{\bibfnamefont{K.~W.} \bibnamefont{West}},
  \bibinfo{journal}{Science} \textbf{\bibinfo{volume}{308}},
  \bibinfo{pages}{88} (\bibinfo{year}{2005}).

\bibitem[{\citenamefont{Kim et~al.}(2006)\citenamefont{Kim, Koh, Rotenberg, Oh,
  Eisaki, Motoyama, Uchida, Tohyama, Maekawa, Shen et~al.}}]{kim06}
\bibinfo{author}{\bibfnamefont{B.~J.} \bibnamefont{Kim}},
  \bibinfo{author}{\bibfnamefont{H.}~\bibnamefont{Koh}},
  \bibinfo{author}{\bibfnamefont{E.}~\bibnamefont{Rotenberg}},
  \bibinfo{author}{\bibfnamefont{S.-J.} \bibnamefont{Oh}},
  \bibinfo{author}{\bibfnamefont{H.}~\bibnamefont{Eisaki}},
  \bibinfo{author}{\bibfnamefont{N.}~\bibnamefont{Motoyama}},
  \bibinfo{author}{\bibfnamefont{S.}~\bibnamefont{Uchida}},
  \bibinfo{author}{\bibfnamefont{T.}~\bibnamefont{Tohyama}},
  \bibinfo{author}{\bibfnamefont{S.}~\bibnamefont{Maekawa}},
  \bibinfo{author}{\bibfnamefont{Z.-X.} \bibnamefont{Shen}},
  \bibnamefont{et~al.}, \bibinfo{journal}{Nature Physics}
  \textbf{\bibinfo{volume}{2}}, \bibinfo{pages}{397} (\bibinfo{year}{2006}).

\bibitem[{\citenamefont{Jompol et~al.}(2009)\citenamefont{Jompol, Ford,
  Griffiths, Farrer, Jones, Anderson, Ritchie, Silk, and Schofield}}]{jompol09}
\bibinfo{author}{\bibfnamefont{Y.}~\bibnamefont{Jompol}},
  \bibinfo{author}{\bibfnamefont{C.~J.~B.} \bibnamefont{Ford}},
  \bibinfo{author}{\bibfnamefont{J.~P.} \bibnamefont{Griffiths}},
  \bibinfo{author}{\bibfnamefont{I.}~\bibnamefont{Farrer}},
  \bibinfo{author}{\bibfnamefont{G.~A.~C.} \bibnamefont{Jones}},
  \bibinfo{author}{\bibfnamefont{D.}~\bibnamefont{Anderson}},
  \bibinfo{author}{\bibfnamefont{D.~A.} \bibnamefont{Ritchie}},
  \bibinfo{author}{\bibfnamefont{T.~W.} \bibnamefont{Silk}}, \bibnamefont{and}
  \bibinfo{author}{\bibfnamefont{A.~J.} \bibnamefont{Schofield}},
  \bibinfo{journal}{Science} \textbf{\bibinfo{volume}{325}},
  \bibinfo{pages}{597} (\bibinfo{year}{2009}).

\bibitem[{\citenamefont{Meden and Sch\"onhammer}(1992)}]{meden92}
\bibinfo{author}{\bibfnamefont{V.}~\bibnamefont{Meden}} \bibnamefont{and}
  \bibinfo{author}{\bibfnamefont{K.}~\bibnamefont{Sch\"onhammer}},
  \bibinfo{journal}{Phys. Rev. B} \textbf{\bibinfo{volume}{46}},
  \bibinfo{pages}{15753} (\bibinfo{year}{1992}).

\bibitem[{\citenamefont{Voit}(1993)}]{voit93}
\bibinfo{author}{\bibfnamefont{J.}~\bibnamefont{Voit}}, \bibinfo{journal}{Phys.
  Rev. B} \textbf{\bibinfo{volume}{47}}, \bibinfo{pages}{6740}
  (\bibinfo{year}{1993}).

\bibitem[{\citenamefont{Pereira et~al.}(2009)\citenamefont{Pereira, White, and
  Affleck}}]{pereira09}
\bibinfo{author}{\bibfnamefont{R.~G.} \bibnamefont{Pereira}},
  \bibinfo{author}{\bibfnamefont{S.~R.} \bibnamefont{White}}, \bibnamefont{and}
  \bibinfo{author}{\bibfnamefont{I.}~\bibnamefont{Affleck}},
  \bibinfo{journal}{Phys. Rev. B} \textbf{\bibinfo{volume}{79}},
  \bibinfo{pages}{165113} (\bibinfo{year}{2009}).

\bibitem[{\citenamefont{Brazovskii et~al.}(1993)\citenamefont{Brazovskii,
  Matveenko, and Nozi\`eres}}]{brazovskii93}
\bibinfo{author}{\bibfnamefont{S.}~\bibnamefont{Brazovskii}},
  \bibinfo{author}{\bibfnamefont{F.}~\bibnamefont{Matveenko}},
  \bibnamefont{and}
  \bibinfo{author}{\bibfnamefont{P.}~\bibnamefont{Nozi\`eres}},
  \bibinfo{journal}{JETP Lett.} \textbf{\bibinfo{volume}{58}},
  \bibinfo{pages}{796} (\bibinfo{year}{1993}).

\bibitem[{\citenamefont{Vekua et~al.}(2009)\citenamefont{Vekua, Matveenko, and
  Shlyapnikov}}]{vekua09}
\bibinfo{author}{\bibfnamefont{T.}~\bibnamefont{Vekua}},
  \bibinfo{author}{\bibfnamefont{S.~I.} \bibnamefont{Matveenko}},
  \bibnamefont{and} \bibinfo{author}{\bibfnamefont{G.~V.}
  \bibnamefont{Shlyapnikov}}, \bibinfo{journal}{JETP Letters}
  \textbf{\bibinfo{volume}{90}}, \bibinfo{pages}{289} (\bibinfo{year}{2009}).

\bibitem[{\citenamefont{Rizzi et~al.}(2008)\citenamefont{Rizzi, Polini,
  Cazalilla, Bakhtiari, Tosi, and Fazio}}]{rizzi08}
\bibinfo{author}{\bibfnamefont{M.}~\bibnamefont{Rizzi}},
  \bibinfo{author}{\bibfnamefont{M.}~\bibnamefont{Polini}},
  \bibinfo{author}{\bibfnamefont{M.~A.} \bibnamefont{Cazalilla}},
  \bibinfo{author}{\bibfnamefont{M.~R.} \bibnamefont{Bakhtiari}},
  \bibinfo{author}{\bibfnamefont{M.~P.} \bibnamefont{Tosi}}, \bibnamefont{and}
  \bibinfo{author}{\bibfnamefont{R.}~\bibnamefont{Fazio}},
  \bibinfo{journal}{Phys. Rev. B} \textbf{\bibinfo{volume}{77}},
  \bibinfo{pages}{245105} (\bibinfo{year}{2008}).

\bibitem[{\citenamefont{Pereira and Sela}(2010)}]{pereira10}
\bibinfo{author}{\bibfnamefont{R.~G.} \bibnamefont{Pereira}} \bibnamefont{and}
  \bibinfo{author}{\bibfnamefont{E.}~\bibnamefont{Sela}},
  \bibinfo{journal}{Phys. Rev. B} \textbf{\bibinfo{volume}{82}},
  \bibinfo{pages}{115324} (\bibinfo{year}{2010}).

\bibitem[{\citenamefont{Teber}(2007)}]{teber07}
\bibinfo{author}{\bibfnamefont{S.}~\bibnamefont{Teber}},
  \bibinfo{journal}{Phys. Rev. B} \textbf{\bibinfo{volume}{76}},
  \bibinfo{pages}{045309} (\bibinfo{year}{2007}).

\bibitem[{\citenamefont{Khodas et~al.}(2007{\natexlab{a}})\citenamefont{Khodas,
  Pustilnik, Kamenev, and Glazman}}]{khodas07_2}
\bibinfo{author}{\bibfnamefont{M.}~\bibnamefont{Khodas}},
  \bibinfo{author}{\bibfnamefont{M.}~\bibnamefont{Pustilnik}},
  \bibinfo{author}{\bibfnamefont{A.}~\bibnamefont{Kamenev}}, \bibnamefont{and}
  \bibinfo{author}{\bibfnamefont{L.~I.} \bibnamefont{Glazman}},
  \bibinfo{journal}{Phys. Rev. B} \textbf{\bibinfo{volume}{76}},
  \bibinfo{pages}{155402} (\bibinfo{year}{2007}{\natexlab{a}}).

\bibitem[{\citenamefont{Imambekov and
  Glazman}(2009{\natexlab{a}})}]{imambekov09}
\bibinfo{author}{\bibfnamefont{A.}~\bibnamefont{Imambekov}} \bibnamefont{and}
  \bibinfo{author}{\bibfnamefont{L.~I.} \bibnamefont{Glazman}},
  \bibinfo{journal}{Phys. Rev. Lett.} \textbf{\bibinfo{volume}{102}},
  \bibinfo{pages}{126405} (\bibinfo{year}{2009}{\natexlab{a}}).

\bibitem[{\citenamefont{Imambekov and
  Glazman}(2009{\natexlab{b}})}]{imambekov09_2}
\bibinfo{author}{\bibfnamefont{A.}~\bibnamefont{Imambekov}} \bibnamefont{and}
  \bibinfo{author}{\bibfnamefont{L.~I.} \bibnamefont{Glazman}},
  \bibinfo{journal}{Science} \textbf{\bibinfo{volume}{323}},
  \bibinfo{pages}{228} (\bibinfo{year}{2009}{\natexlab{b}}).

\bibitem[{\citenamefont{Pustilnik et~al.}(2003)\citenamefont{Pustilnik,
  Mishchenko, Glazman, and Andreev}}]{pustilnik03}
\bibinfo{author}{\bibfnamefont{M.}~\bibnamefont{Pustilnik}},
  \bibinfo{author}{\bibfnamefont{E.~G.} \bibnamefont{Mishchenko}},
  \bibinfo{author}{\bibfnamefont{L.~I.} \bibnamefont{Glazman}},
  \bibnamefont{and} \bibinfo{author}{\bibfnamefont{A.~V.}
  \bibnamefont{Andreev}}, \bibinfo{journal}{Phys. Rev. Lett.}
  \textbf{\bibinfo{volume}{91}}, \bibinfo{pages}{126805}
  (\bibinfo{year}{2003}).

\bibitem[{\citenamefont{Pustilnik et~al.}(2006)\citenamefont{Pustilnik, Khodas,
  Kamenev, and Glazman}}]{pustilnik06}
\bibinfo{author}{\bibfnamefont{M.}~\bibnamefont{Pustilnik}},
  \bibinfo{author}{\bibfnamefont{M.}~\bibnamefont{Khodas}},
  \bibinfo{author}{\bibfnamefont{A.}~\bibnamefont{Kamenev}}, \bibnamefont{and}
  \bibinfo{author}{\bibfnamefont{L.~I.} \bibnamefont{Glazman}},
  \bibinfo{journal}{Phys. Rev. Lett.} \textbf{\bibinfo{volume}{96}},
  \bibinfo{pages}{196405} (\bibinfo{year}{2006}).

\bibitem[{\citenamefont{Khodas et~al.}(2007{\natexlab{b}})\citenamefont{Khodas,
  Pustilnik, Kamenev, and Glazman}}]{khodas07}
\bibinfo{author}{\bibfnamefont{M.}~\bibnamefont{Khodas}},
  \bibinfo{author}{\bibfnamefont{M.}~\bibnamefont{Pustilnik}},
  \bibinfo{author}{\bibfnamefont{A.}~\bibnamefont{Kamenev}}, \bibnamefont{and}
  \bibinfo{author}{\bibfnamefont{L.~I.} \bibnamefont{Glazman}},
  \bibinfo{journal}{Phys. Rev. Lett.} \textbf{\bibinfo{volume}{99}},
  \bibinfo{pages}{110405} (\bibinfo{year}{2007}{\natexlab{b}}).

\bibitem[{\citenamefont{Cheianov and Pustilnik}(2008)}]{cheianov08}
\bibinfo{author}{\bibfnamefont{V.~V.} \bibnamefont{Cheianov}} \bibnamefont{and}
  \bibinfo{author}{\bibfnamefont{M.}~\bibnamefont{Pustilnik}},
  \bibinfo{journal}{Phys. Rev. Lett.} \textbf{\bibinfo{volume}{100}},
  \bibinfo{pages}{126403} (\bibinfo{year}{2008}).

\bibitem[{\citenamefont{Imambekov and Glazman}(2008)}]{imambekov08}
\bibinfo{author}{\bibfnamefont{A.}~\bibnamefont{Imambekov}} \bibnamefont{and}
  \bibinfo{author}{\bibfnamefont{L.~I.} \bibnamefont{Glazman}},
  \bibinfo{journal}{Phys. Rev. Lett.} \textbf{\bibinfo{volume}{100}},
  \bibinfo{pages}{206805} (\bibinfo{year}{2008}).

\bibitem[{\citenamefont{Lante and Parola}(2009)}]{lante09}
\bibinfo{author}{\bibfnamefont{V.}~\bibnamefont{Lante}} \bibnamefont{and}
  \bibinfo{author}{\bibfnamefont{A.}~\bibnamefont{Parola}},
  \bibinfo{journal}{Phys. Rev. B} \textbf{\bibinfo{volume}{80}},
  \bibinfo{pages}{195113} (\bibinfo{year}{2009}).

\bibitem[{\citenamefont{Auslaender et~al.}(2002)\citenamefont{Auslaender,
  Yacoby, de~Picciotto, Baldwin, Pfeiffer, and West}}]{auslaender02}
\bibinfo{author}{\bibfnamefont{O.~M.} \bibnamefont{Auslaender}},
  \bibinfo{author}{\bibfnamefont{A.}~\bibnamefont{Yacoby}},
  \bibinfo{author}{\bibfnamefont{R.}~\bibnamefont{de~Picciotto}},
  \bibinfo{author}{\bibfnamefont{K.~W.} \bibnamefont{Baldwin}},
  \bibinfo{author}{\bibfnamefont{L.~N.} \bibnamefont{Pfeiffer}},
  \bibnamefont{and} \bibinfo{author}{\bibfnamefont{K.~W.} \bibnamefont{West}},
  \bibinfo{journal}{Science} \textbf{\bibinfo{volume}{295}},
  \bibinfo{pages}{825} (\bibinfo{year}{2002}).

\bibitem[{\citenamefont{Segovia et~al.}(1999)\citenamefont{Segovia, Purdie,
  Hengsberger, and Baer}}]{segovia99}
\bibinfo{author}{\bibfnamefont{P.}~\bibnamefont{Segovia}},
  \bibinfo{author}{\bibfnamefont{D.}~\bibnamefont{Purdie}},
  \bibinfo{author}{\bibfnamefont{M.}~\bibnamefont{Hengsberger}},
  \bibnamefont{and} \bibinfo{author}{\bibfnamefont{Y.}~\bibnamefont{Baer}},
  \bibinfo{journal}{Nature} \textbf{\bibinfo{volume}{402}},
  \bibinfo{pages}{504} (\bibinfo{year}{1999}).

\bibitem[{\citenamefont{Claessen et~al.}(2002)\citenamefont{Claessen, Sing,
  Schwingenschl\"ogl, Blaha, Dressel, and Jacobsen}}]{claessen02}
\bibinfo{author}{\bibfnamefont{R.}~\bibnamefont{Claessen}},
  \bibinfo{author}{\bibfnamefont{M.}~\bibnamefont{Sing}},
  \bibinfo{author}{\bibfnamefont{U.}~\bibnamefont{Schwingenschl\"ogl}},
  \bibinfo{author}{\bibfnamefont{P.}~\bibnamefont{Blaha}},
  \bibinfo{author}{\bibfnamefont{M.}~\bibnamefont{Dressel}}, \bibnamefont{and}
  \bibinfo{author}{\bibfnamefont{C.~S.} \bibnamefont{Jacobsen}},
  \bibinfo{journal}{Phys. Rev. Lett.} \textbf{\bibinfo{volume}{88}},
  \bibinfo{pages}{096402} (\bibinfo{year}{2002}).

\bibitem[{\citenamefont{Stewart et~al.}(2008)\citenamefont{Stewart, Gaebler,
  and Jin}}]{stewart08}
\bibinfo{author}{\bibfnamefont{J.~T.} \bibnamefont{Stewart}},
  \bibinfo{author}{\bibfnamefont{J.~P.} \bibnamefont{Gaebler}},
  \bibnamefont{and} \bibinfo{author}{\bibfnamefont{D.~S.} \bibnamefont{Jin}},
  \bibinfo{journal}{Nature} \textbf{\bibinfo{volume}{454}},
  \bibinfo{pages}{744} (\bibinfo{year}{2008}).

\bibitem[{\citenamefont{Wang et~al.}(2009)\citenamefont{Wang, Alvarez, Allen,
  Mo, He, Jin, Mandrus, and H\"ochst}}]{wang09}
\bibinfo{author}{\bibfnamefont{F.}~\bibnamefont{Wang}},
  \bibinfo{author}{\bibfnamefont{J.~V.} \bibnamefont{Alvarez}},
  \bibinfo{author}{\bibfnamefont{J.~W.} \bibnamefont{Allen}},
  \bibinfo{author}{\bibfnamefont{S.-K.} \bibnamefont{Mo}},
  \bibinfo{author}{\bibfnamefont{J.}~\bibnamefont{He}},
  \bibinfo{author}{\bibfnamefont{R.}~\bibnamefont{Jin}},
  \bibinfo{author}{\bibfnamefont{D.}~\bibnamefont{Mandrus}}, \bibnamefont{and}
  \bibinfo{author}{\bibfnamefont{H.}~\bibnamefont{H\"ochst}},
  \bibinfo{journal}{Phys. Rev. Lett.} \textbf{\bibinfo{volume}{103}},
  \bibinfo{pages}{136401} (\bibinfo{year}{2009}).

\bibitem[{\citenamefont{Yamamoto et~al.}(2006)\citenamefont{Yamamoto, Stopa,
  Tokura, Hirayama, and Tarucha}}]{yamamoto06}
\bibinfo{author}{\bibfnamefont{M.}~\bibnamefont{Yamamoto}},
  \bibinfo{author}{\bibfnamefont{M.}~\bibnamefont{Stopa}},
  \bibinfo{author}{\bibfnamefont{Y.}~\bibnamefont{Tokura}},
  \bibinfo{author}{\bibfnamefont{Y.}~\bibnamefont{Hirayama}}, \bibnamefont{and}
  \bibinfo{author}{\bibfnamefont{S.}~\bibnamefont{Tarucha}},
  \bibinfo{journal}{Science} \textbf{\bibinfo{volume}{313}},
  \bibinfo{pages}{204} (\bibinfo{year}{2006}).

\bibitem[{\citenamefont{Lake et~al.}(2005)\citenamefont{Lake, Tennant, Frost,
  and Nagler}}]{lake05}
\bibinfo{author}{\bibfnamefont{B.}~\bibnamefont{Lake}},
  \bibinfo{author}{\bibfnamefont{D.~A.} \bibnamefont{Tennant}},
  \bibinfo{author}{\bibfnamefont{C.~D.} \bibnamefont{Frost}}, \bibnamefont{and}
  \bibinfo{author}{\bibfnamefont{S.~E.} \bibnamefont{Nagler}},
  \bibinfo{journal}{Nature Materials} \textbf{\bibinfo{volume}{4}},
  \bibinfo{pages}{329} (\bibinfo{year}{2005}).

\bibitem[{\citenamefont{Recati et~al.}(2003)\citenamefont{Recati, Fedichev,
  Zwerger, and Zoller}}]{recati03}
\bibinfo{author}{\bibfnamefont{A.}~\bibnamefont{Recati}},
  \bibinfo{author}{\bibfnamefont{P.~O.} \bibnamefont{Fedichev}},
  \bibinfo{author}{\bibfnamefont{W.}~\bibnamefont{Zwerger}}, \bibnamefont{and}
  \bibinfo{author}{\bibfnamefont{P.}~\bibnamefont{Zoller}},
  \bibinfo{journal}{Phys. Rev. Lett.} \textbf{\bibinfo{volume}{90}},
  \bibinfo{pages}{020401} (\bibinfo{year}{2003}).

\bibitem[{\citenamefont{Kollath and Schollw\"ock}(2006)}]{kollath06}
\bibinfo{author}{\bibfnamefont{C.}~\bibnamefont{Kollath}} \bibnamefont{and}
  \bibinfo{author}{\bibfnamefont{U.}~\bibnamefont{Schollw\"ock}},
  \bibinfo{journal}{New Journal of Physics} \textbf{\bibinfo{volume}{8}},
  \bibinfo{pages}{220} (\bibinfo{year}{2006}).

\bibitem[{\citenamefont{Benthien et~al.}(2004)\citenamefont{Benthien, Gebhard,
  and Jeckelmann}}]{benthien04}
\bibinfo{author}{\bibfnamefont{H.}~\bibnamefont{Benthien}},
  \bibinfo{author}{\bibfnamefont{F.}~\bibnamefont{Gebhard}}, \bibnamefont{and}
  \bibinfo{author}{\bibfnamefont{E.}~\bibnamefont{Jeckelmann}},
  \bibinfo{journal}{Phys. Rev. Lett.} \textbf{\bibinfo{volume}{92}},
  \bibinfo{pages}{256401} (\bibinfo{year}{2004}).

\bibitem[{\citenamefont{White and Affleck}(2008)}]{white08}
\bibinfo{author}{\bibfnamefont{S.~R.} \bibnamefont{White}} \bibnamefont{and}
  \bibinfo{author}{\bibfnamefont{I.}~\bibnamefont{Affleck}},
  \bibinfo{journal}{Phys. Rev. B} \textbf{\bibinfo{volume}{77}},
  \bibinfo{pages}{134437} (\bibinfo{year}{2008}).

\bibitem[{\citenamefont{Barthel et~al.}(2009)\citenamefont{Barthel,
  Schollw\"ock, and White}}]{barthel09}
\bibinfo{author}{\bibfnamefont{T.}~\bibnamefont{Barthel}},
  \bibinfo{author}{\bibfnamefont{U.}~\bibnamefont{Schollw\"ock}},
  \bibnamefont{and} \bibinfo{author}{\bibfnamefont{S.~R.} \bibnamefont{White}},
  \bibinfo{journal}{Phys. Rev. B} \textbf{\bibinfo{volume}{79}},
  \bibinfo{pages}{245101} (\bibinfo{year}{2009}).

\bibitem[{\citenamefont{Feiguin and Huse}(2009)}]{feiguin09}
\bibinfo{author}{\bibfnamefont{A.~E.} \bibnamefont{Feiguin}} \bibnamefont{and}
  \bibinfo{author}{\bibfnamefont{D.~A.} \bibnamefont{Huse}},
  \bibinfo{journal}{Phys. Rev. B} \textbf{\bibinfo{volume}{79}},
  \bibinfo{pages}{100507} (\bibinfo{year}{2009}).

\bibitem[{\citenamefont{Kokalj and Prelov\ifmmode~\check{s}\else
  \v{s}\fi{}ek}(2009)}]{kokalj09}
\bibinfo{author}{\bibfnamefont{J.}~\bibnamefont{Kokalj}} \bibnamefont{and}
  \bibinfo{author}{\bibfnamefont{P.}~\bibnamefont{Prelov\ifmmode~\check{s}\else
  \v{s}\fi{}ek}}, \bibinfo{journal}{Phys. Rev. B}
  \textbf{\bibinfo{volume}{80}}, \bibinfo{pages}{205117}
  (\bibinfo{year}{2009}).

\bibitem[{\citenamefont{Kohno}(2010)}]{kohno10}
\bibinfo{author}{\bibfnamefont{M.}~\bibnamefont{Kohno}},
  \bibinfo{journal}{Phys. Rev. Lett.} \textbf{\bibinfo{volume}{105}},
  \bibinfo{pages}{106402} (\bibinfo{year}{2010}).

\bibitem[{\citenamefont{Yang}(2001)}]{yang01}
\bibinfo{author}{\bibfnamefont{K.}~\bibnamefont{Yang}}, \bibinfo{journal}{Phys.
  Rev. B} \textbf{\bibinfo{volume}{63}}, \bibinfo{pages}{140511}
  (\bibinfo{year}{2001}).

\bibitem[{\citenamefont{an~Liao et~al.}(2010)\citenamefont{an~Liao, Rittner,
  Paprotta, Li, Partridge, Hulet, Baur, and Mueller}}]{liao10}
\bibinfo{author}{\bibfnamefont{Y.}~\bibnamefont{an~Liao}},
  \bibinfo{author}{\bibfnamefont{A.~S.~C.} \bibnamefont{Rittner}},
  \bibinfo{author}{\bibfnamefont{T.}~\bibnamefont{Paprotta}},
  \bibinfo{author}{\bibfnamefont{W.}~\bibnamefont{Li}},
  \bibinfo{author}{\bibfnamefont{G.~B.} \bibnamefont{Partridge}},
  \bibinfo{author}{\bibfnamefont{R.~G.} \bibnamefont{Hulet}},
  \bibinfo{author}{\bibfnamefont{S.~K.} \bibnamefont{Baur}}, \bibnamefont{and}
  \bibinfo{author}{\bibfnamefont{E.~J.} \bibnamefont{Mueller}},
  \bibinfo{journal}{Nature} \textbf{\bibinfo{volume}{467}},
  \bibinfo{pages}{567} (\bibinfo{year}{2010}).

\bibitem[{\citenamefont{Rabello and Si}(2002)}]{rabello02}
\bibinfo{author}{\bibfnamefont{S.}~\bibnamefont{Rabello}} \bibnamefont{and}
  \bibinfo{author}{\bibfnamefont{Q.}~\bibnamefont{Si}},
  \bibinfo{journal}{Europhysics Letters} \textbf{\bibinfo{volume}{60}},
  \bibinfo{pages}{882} (\bibinfo{year}{2002}).

\bibitem[{\citenamefont{Mahan}(1967)}]{mahan67}
\bibinfo{author}{\bibfnamefont{G.~D.} \bibnamefont{Mahan}},
  \bibinfo{journal}{Phys. Rev.} \textbf{\bibinfo{volume}{163}},
  \bibinfo{pages}{612} (\bibinfo{year}{1967}).

\bibitem[{\citenamefont{Anderson}(1967)}]{anderson67}
\bibinfo{author}{\bibfnamefont{P.~W.} \bibnamefont{Anderson}},
  \bibinfo{journal}{Phys. Rev. Lett.} \textbf{\bibinfo{volume}{18}},
  \bibinfo{pages}{1049} (\bibinfo{year}{1967}).

\bibitem[{\citenamefont{Nozi\`eres and de~Dominicis}(1969)}]{nozieres69}
\bibinfo{author}{\bibfnamefont{P.}~\bibnamefont{Nozi\`eres}} \bibnamefont{and}
  \bibinfo{author}{\bibfnamefont{C.~T.} \bibnamefont{de~Dominicis}},
  \bibinfo{journal}{Phys. Rev.} \textbf{\bibinfo{volume}{178}},
  \bibinfo{pages}{1097} (\bibinfo{year}{1969}).

\bibitem[{\citenamefont{Samokhin}(1998)}]{samokhin98}
\bibinfo{author}{\bibfnamefont{K.~V.} \bibnamefont{Samokhin}},
  \bibinfo{journal}{J. Phys. Condens. Matter} \textbf{\bibinfo{volume}{10}},
  \bibinfo{pages}{L533} (\bibinfo{year}{1998}).

\bibitem[{\citenamefont{S\'olyom}(1979)}]{solyom79}
\bibinfo{author}{\bibfnamefont{J.}~\bibnamefont{S\'olyom}},
  \bibinfo{journal}{Adv. Phys.} \textbf{\bibinfo{volume}{28}},
  \bibinfo{pages}{201} (\bibinfo{year}{1979}).

\bibitem[{\citenamefont{Gogolin et~al.}(1998)\citenamefont{Gogolin, Nersesyan,
  and Tsvelik}}]{gogolin98}
\bibinfo{author}{\bibfnamefont{A.~O.} \bibnamefont{Gogolin}},
  \bibinfo{author}{\bibfnamefont{A.~A.} \bibnamefont{Nersesyan}},
  \bibnamefont{and} \bibinfo{author}{\bibfnamefont{A.~M.}
  \bibnamefont{Tsvelik}}, \emph{\bibinfo{title}{Bosonization and strongly
  correlated systems}} (\bibinfo{publisher}{Cambridge University Press},
  \bibinfo{year}{1998}).

\bibitem[{\citenamefont{Luther and Peschel}(1974)}]{luther74}
\bibinfo{author}{\bibfnamefont{A.}~\bibnamefont{Luther}} \bibnamefont{and}
  \bibinfo{author}{\bibfnamefont{I.}~\bibnamefont{Peschel}},
  \bibinfo{journal}{Phys. Rev. B} \textbf{\bibinfo{volume}{9}},
  \bibinfo{pages}{2911} (\bibinfo{year}{1974}).

\bibitem[{\citenamefont{Rozhkov}(2005)}]{rozhkov05}
\bibinfo{author}{\bibfnamefont{A.~V.} \bibnamefont{Rozhkov}},
  \bibinfo{journal}{Eur. Phys. J. B} \textbf{\bibinfo{volume}{47}},
  \bibinfo{pages}{193} (\bibinfo{year}{2005}).

\bibitem[{\citenamefont{Schotte and Schotte}(1969)}]{schotte69}
\bibinfo{author}{\bibfnamefont{K.~D.} \bibnamefont{Schotte}} \bibnamefont{and}
  \bibinfo{author}{\bibfnamefont{U.}~\bibnamefont{Schotte}},
  \bibinfo{journal}{Phys. Rev.} \textbf{\bibinfo{volume}{182}},
  \bibinfo{pages}{479} (\bibinfo{year}{1969}).

\bibitem[{\citenamefont{Luttinger}(1960)}]{luttinger60}
\bibinfo{author}{\bibfnamefont{J.~M.} \bibnamefont{Luttinger}},
  \bibinfo{journal}{Phys. Rev.} \textbf{\bibinfo{volume}{119}},
  \bibinfo{pages}{1153} (\bibinfo{year}{1960}).

\bibitem[{\citenamefont{Yamanaka et~al.}(1997)\citenamefont{Yamanaka, Oshikawa,
  and Affleck}}]{yamanaka97}
\bibinfo{author}{\bibfnamefont{M.}~\bibnamefont{Yamanaka}},
  \bibinfo{author}{\bibfnamefont{M.}~\bibnamefont{Oshikawa}}, \bibnamefont{and}
  \bibinfo{author}{\bibfnamefont{I.}~\bibnamefont{Affleck}},
  \bibinfo{journal}{Phys. Rev. Lett.} \textbf{\bibinfo{volume}{79}},
  \bibinfo{pages}{1110} (\bibinfo{year}{1997}).

\bibitem[{\citenamefont{Matveev
  et~al.}(2007{\natexlab{a}})\citenamefont{Matveev, Furusaki, and
  Glazman}}]{matveev07}
\bibinfo{author}{\bibfnamefont{K.~A.} \bibnamefont{Matveev}},
  \bibinfo{author}{\bibfnamefont{A.}~\bibnamefont{Furusaki}}, \bibnamefont{and}
  \bibinfo{author}{\bibfnamefont{L.~I.} \bibnamefont{Glazman}},
  \bibinfo{journal}{Phys. Rev. B} \textbf{\bibinfo{volume}{76}},
  \bibinfo{pages}{155440} (\bibinfo{year}{2007}{\natexlab{a}}).

\bibitem[{\citenamefont{Carmelo et~al.}(1991)\citenamefont{Carmelo, Horsch,
  Bares, and Ovchinnikov}}]{carmelo91}
\bibinfo{author}{\bibfnamefont{J.}~\bibnamefont{Carmelo}},
  \bibinfo{author}{\bibfnamefont{P.}~\bibnamefont{Horsch}},
  \bibinfo{author}{\bibfnamefont{P.~A.} \bibnamefont{Bares}}, \bibnamefont{and}
  \bibinfo{author}{\bibfnamefont{A.~A.} \bibnamefont{Ovchinnikov}},
  \bibinfo{journal}{Phys. Rev. B} \textbf{\bibinfo{volume}{44}},
  \bibinfo{pages}{9967} (\bibinfo{year}{1991}).

\bibitem[{\citenamefont{Essler et~al.}(2005)\citenamefont{Essler, Frahm,
  G\"ohmann, Kl\"umper, and Korepin}}]{essler05}
\bibinfo{author}{\bibfnamefont{F.~H.~L.} \bibnamefont{Essler}},
  \bibinfo{author}{\bibfnamefont{H.}~\bibnamefont{Frahm}},
  \bibinfo{author}{\bibfnamefont{F.}~\bibnamefont{G\"ohmann}},
  \bibinfo{author}{\bibfnamefont{A.}~\bibnamefont{Kl\"umper}},
  \bibnamefont{and} \bibinfo{author}{\bibfnamefont{V.~E.}
  \bibnamefont{Korepin}}, \emph{\bibinfo{title}{The One-Dimensional {H}ubbard
  Model}} (\bibinfo{publisher}{Cambridge University Press},
  \bibinfo{year}{2005}).

\bibitem[{\citenamefont{Yang}(1967)}]{yang67}
\bibinfo{author}{\bibfnamefont{C.~N.} \bibnamefont{Yang}},
  \bibinfo{journal}{Phys. Rev. Lett.} \textbf{\bibinfo{volume}{19}},
  \bibinfo{pages}{1312} (\bibinfo{year}{1967}).

\bibitem[{\citenamefont{Gaudin}(1967)}]{gaudin67}
\bibinfo{author}{\bibfnamefont{M.}~\bibnamefont{Gaudin}},
  \bibinfo{journal}{Phys. Lett. A} \textbf{\bibinfo{volume}{24}},
  \bibinfo{pages}{55} (\bibinfo{year}{1967}).

\bibitem[{\citenamefont{Essler}(2010)}]{essler10}
\bibinfo{author}{\bibfnamefont{F.~H.~L.} \bibnamefont{Essler}},
  \bibinfo{journal}{Phys. Rev. B} \textbf{\bibinfo{volume}{81}},
  \bibinfo{pages}{205120} (\bibinfo{year}{2010}).

\bibitem[{\citenamefont{Schmidt et~al.}(2010)\citenamefont{Schmidt, Imambekov,
  and Glazman}}]{schmidt09_3}
\bibinfo{author}{\bibfnamefont{T.~L.} \bibnamefont{Schmidt}},
  \bibinfo{author}{\bibfnamefont{A.}~\bibnamefont{Imambekov}},
  \bibnamefont{and} \bibinfo{author}{\bibfnamefont{L.~I.}
  \bibnamefont{Glazman}}, \bibinfo{journal}{Phys. Rev. Lett.}
  \textbf{\bibinfo{volume}{104}}, \bibinfo{pages}{116403}
  (\bibinfo{year}{2010}).

\bibitem[{\citenamefont{Pereira et~al.}(2006)\citenamefont{Pereira, Sirker,
  Caux, Hagemans, Maillet, White, and Affleck}}]{pereira06}
\bibinfo{author}{\bibfnamefont{R.~G.} \bibnamefont{Pereira}},
  \bibinfo{author}{\bibfnamefont{J.}~\bibnamefont{Sirker}},
  \bibinfo{author}{\bibfnamefont{J.-S.} \bibnamefont{Caux}},
  \bibinfo{author}{\bibfnamefont{R.}~\bibnamefont{Hagemans}},
  \bibinfo{author}{\bibfnamefont{J.~M.} \bibnamefont{Maillet}},
  \bibinfo{author}{\bibfnamefont{S.~R.} \bibnamefont{White}}, \bibnamefont{and}
  \bibinfo{author}{\bibfnamefont{I.}~\bibnamefont{Affleck}},
  \bibinfo{journal}{Phys. Rev. Lett.} \textbf{\bibinfo{volume}{96}},
  \bibinfo{pages}{257202} (\bibinfo{year}{2006}).

\bibitem[{\citenamefont{Matveev}(2004{\natexlab{a}})}]{matveev04_2}
\bibinfo{author}{\bibfnamefont{K.~A.} \bibnamefont{Matveev}},
  \bibinfo{journal}{Phys. Rev. Lett.} \textbf{\bibinfo{volume}{92}},
  \bibinfo{pages}{106801} (\bibinfo{year}{2004}{\natexlab{a}}).

\bibitem[{\citenamefont{Matveev}(2004{\natexlab{b}})}]{matveev04}
\bibinfo{author}{\bibfnamefont{K.~A.} \bibnamefont{Matveev}},
  \bibinfo{journal}{Phys. Rev. B} \textbf{\bibinfo{volume}{70}},
  \bibinfo{pages}{245319} (\bibinfo{year}{2004}{\natexlab{b}}).

\bibitem[{\citenamefont{Matveev
  et~al.}(2007{\natexlab{b}})\citenamefont{Matveev, Furusaki, and
  Glazman}}]{matveev07_2}
\bibinfo{author}{\bibfnamefont{K.~A.} \bibnamefont{Matveev}},
  \bibinfo{author}{\bibfnamefont{A.}~\bibnamefont{Furusaki}}, \bibnamefont{and}
  \bibinfo{author}{\bibfnamefont{L.~I.} \bibnamefont{Glazman}},
  \bibinfo{journal}{Phys. Rev. Lett.} \textbf{\bibinfo{volume}{98}},
  \bibinfo{pages}{096403} (\bibinfo{year}{2007}{\natexlab{b}}).

\bibitem[{\citenamefont{Fiete}(2007)}]{fiete07}
\bibinfo{author}{\bibfnamefont{G.~A.} \bibnamefont{Fiete}},
  \bibinfo{journal}{Rev. Mod. Phys.} \textbf{\bibinfo{volume}{79}},
  \bibinfo{pages}{801} (\bibinfo{year}{2007}).

\bibitem[{\citenamefont{Chao et~al.}(1978)\citenamefont{Chao, Spa\l{}ek, and
  Oles}}]{chao78}
\bibinfo{author}{\bibfnamefont{K.~A.} \bibnamefont{Chao}},
  \bibinfo{author}{\bibfnamefont{J.}~\bibnamefont{Spa\l{}ek}},
  \bibnamefont{and} \bibinfo{author}{\bibfnamefont{A.~M.} \bibnamefont{Oles}},
  \bibinfo{journal}{Phys. Rev. B} \textbf{\bibinfo{volume}{18}},
  \bibinfo{pages}{3453} (\bibinfo{year}{1978}).

\bibitem[{\citenamefont{Carmelo et~al.}(1992)\citenamefont{Carmelo, Horsch, and
  Ovchinnikov}}]{carmelo92}
\bibinfo{author}{\bibfnamefont{J.~M.~P.} \bibnamefont{Carmelo}},
  \bibinfo{author}{\bibfnamefont{P.}~\bibnamefont{Horsch}}, \bibnamefont{and}
  \bibinfo{author}{\bibfnamefont{A.~A.} \bibnamefont{Ovchinnikov}},
  \bibinfo{journal}{Phys. Rev. B} \textbf{\bibinfo{volume}{45}},
  \bibinfo{pages}{7899} (\bibinfo{year}{1992}).

\bibitem[{\citenamefont{Carmelo et~al.}(2004)\citenamefont{Carmelo, Penc,
  Martelo, Sacramento, {Lopes dos Santos}, Claessen, Sing, and
  Schwingenschl\"ogl}}]{carmelo04}
\bibinfo{author}{\bibfnamefont{J.~M.~P.} \bibnamefont{Carmelo}},
  \bibinfo{author}{\bibfnamefont{K.}~\bibnamefont{Penc}},
  \bibinfo{author}{\bibfnamefont{L.~M.} \bibnamefont{Martelo}},
  \bibinfo{author}{\bibfnamefont{P.~D.} \bibnamefont{Sacramento}},
  \bibinfo{author}{\bibfnamefont{J.~M.~B.} \bibnamefont{{Lopes dos Santos}}},
  \bibinfo{author}{\bibfnamefont{R.}~\bibnamefont{Claessen}},
  \bibinfo{author}{\bibfnamefont{M.}~\bibnamefont{Sing}}, \bibnamefont{and}
  \bibinfo{author}{\bibfnamefont{U.}~\bibnamefont{Schwingenschl\"ogl}},
  \bibinfo{journal}{Europhysics Letters} \textbf{\bibinfo{volume}{67}},
  \bibinfo{pages}{233} (\bibinfo{year}{2004}).

\bibitem[{\citenamefont{Carmelo et~al.}(2008)\citenamefont{Carmelo, Bozi, and
  Penc}}]{carmelo08}
\bibinfo{author}{\bibfnamefont{J.~M.~P.} \bibnamefont{Carmelo}},
  \bibinfo{author}{\bibfnamefont{D.}~\bibnamefont{Bozi}}, \bibnamefont{and}
  \bibinfo{author}{\bibfnamefont{K.}~\bibnamefont{Penc}},
  \bibinfo{journal}{Journal of Physics: Condensed Matter}
  \textbf{\bibinfo{volume}{20}}, \bibinfo{pages}{415103}
  (\bibinfo{year}{2008}).

\bibitem[{\citenamefont{Carmelo et~al.}(2006)\citenamefont{Carmelo, Penc,
  Sacramento, Sing, and Claessen}}]{carmelo08_2}
\bibinfo{author}{\bibfnamefont{J.~M.~P.} \bibnamefont{Carmelo}},
  \bibinfo{author}{\bibfnamefont{K.}~\bibnamefont{Penc}},
  \bibinfo{author}{\bibfnamefont{P.~D.} \bibnamefont{Sacramento}},
  \bibinfo{author}{\bibfnamefont{M.}~\bibnamefont{Sing}}, \bibnamefont{and}
  \bibinfo{author}{\bibfnamefont{R.}~\bibnamefont{Claessen}},
  \bibinfo{journal}{Journal of Physics: Condensed Matter}
  \textbf{\bibinfo{volume}{18}}, \bibinfo{pages}{5191} (\bibinfo{year}{2006}).

\bibitem[{\citenamefont{Penc et~al.}(1996)\citenamefont{Penc, Hallberg, Mila,
  and Shiba}}]{penc96}
\bibinfo{author}{\bibfnamefont{K.}~\bibnamefont{Penc}},
  \bibinfo{author}{\bibfnamefont{K.}~\bibnamefont{Hallberg}},
  \bibinfo{author}{\bibfnamefont{F.}~\bibnamefont{Mila}}, \bibnamefont{and}
  \bibinfo{author}{\bibfnamefont{H.}~\bibnamefont{Shiba}},
  \bibinfo{journal}{Phys. Rev. Lett.} \textbf{\bibinfo{volume}{77}},
  \bibinfo{pages}{1390} (\bibinfo{year}{1996}).

\bibitem[{\citenamefont{Mahan}(1990)}]{mahan90}
\bibinfo{author}{\bibfnamefont{G.}~\bibnamefont{Mahan}},
  \emph{\bibinfo{title}{Many-particle physics}} (\bibinfo{publisher}{Plenum},
  \bibinfo{address}{New York}, \bibinfo{year}{1990}).

\bibitem[{\citenamefont{Cheon and Shigehara}(1998)}]{cheon98}
\bibinfo{author}{\bibfnamefont{T.}~\bibnamefont{Cheon}} \bibnamefont{and}
  \bibinfo{author}{\bibfnamefont{T.}~\bibnamefont{Shigehara}},
  \bibinfo{journal}{Phys. Lett. A} \textbf{\bibinfo{volume}{243}},
  \bibinfo{pages}{111} (\bibinfo{year}{1998}).

\bibitem[{\citenamefont{Cheon and Shigehara}(1999)}]{cheon99}
\bibinfo{author}{\bibfnamefont{T.}~\bibnamefont{Cheon}} \bibnamefont{and}
  \bibinfo{author}{\bibfnamefont{T.}~\bibnamefont{Shigehara}},
  \bibinfo{journal}{Phys. Rev. Lett.} \textbf{\bibinfo{volume}{82}},
  \bibinfo{pages}{2536} (\bibinfo{year}{1999}).

\bibitem[{\citenamefont{Sutherland}(2004)}]{sutherland04}
\bibinfo{author}{\bibfnamefont{B.}~\bibnamefont{Sutherland}},
  \emph{\bibinfo{title}{Beautiful Models: 70 Years Of Exactly Solved Quantum
  Many-Body Problems}} (\bibinfo{publisher}{World Scientific},
  \bibinfo{year}{2004}).

\bibitem[{\citenamefont{Nayak et~al.}(2001)\citenamefont{Nayak, Shtengel,
  Orgad, Fisher, and Girvin}}]{nayak01}
\bibinfo{author}{\bibfnamefont{C.}~\bibnamefont{Nayak}},
  \bibinfo{author}{\bibfnamefont{K.}~\bibnamefont{Shtengel}},
  \bibinfo{author}{\bibfnamefont{D.}~\bibnamefont{Orgad}},
  \bibinfo{author}{\bibfnamefont{M.~P.~A.} \bibnamefont{Fisher}},
  \bibnamefont{and} \bibinfo{author}{\bibfnamefont{S.~M.}
  \bibnamefont{Girvin}}, \bibinfo{journal}{Phys. Rev. B}
  \textbf{\bibinfo{volume}{64}}, \bibinfo{pages}{235113}
  (\bibinfo{year}{2001}).

\bibitem[{\citenamefont{Karzig et~al.}()\citenamefont{Karzig, Glazman, and von
  Oppen}}]{karzig10}
\bibinfo{author}{\bibfnamefont{T.}~\bibnamefont{Karzig}},
  \bibinfo{author}{\bibfnamefont{L.~I.} \bibnamefont{Glazman}},
  \bibnamefont{and} \bibinfo{author}{\bibfnamefont{F.}~\bibnamefont{von
  Oppen}}, \bibinfo{note}{arXiv:1007.1152v1 [cond-mat.mes-hall]}.

\end{thebibliography}

\end{document}